\documentclass[11pt,a4paper]{article}
\usepackage{float}
\usepackage{multirow}
\usepackage[cp1252,utf8]{inputenc}
\usepackage{hyperref}
\usepackage{graphicx}
\usepackage{wrapfig}
\usepackage{amsmath}
\usepackage{xparse}
\usepackage{amsfonts}
\usepackage{amssymb}
\usepackage{relsize} 
\usepackage[top=1in, bottom=2cm, left=1.5cm, right=1.5cm]{geometry}
\usepackage{authblk}
\usepackage{ulem}
\usepackage{physics}
\usepackage{xcolor}
\NewDocumentCommand{\tens}{t_}
{%
	\IfBooleanTF{#1}
	{\tensop}
	{\otimes}%
}
\NewDocumentCommand{\tensop}{m}
{%
	\mathbin{\mathop{\otimes}\displaylimits_{#1}}%
}
\usepackage{changepage} 

\usepackage{afterpage}

\usepackage{placeins}

\usepackage{wrapfig}
\usepackage{cite} 
\usepackage[nottoc,notlof,notlot]{tocbibind} 

\title{\bf Mixed state information theoretic measures in boosted black brane}
\author[a]{\bf  Anirban Roy Chowdhury \thanks{iamanirban@bose.res.in}}
\author[b]{\bf  Ashis Saha \thanks{ashisphys18@klyuniv.ac.in}}
\author[c]{\bf Sunandan Gangopadhyay \thanks{sunandan.gangopadhyay@bose.res.in}}

\affil[a,c]{\textit{Department of Astrophysics and High Energy Physics\\}
\textit{S.N.~Bose National Centre for Basic Sciences,}
\textit{JD Block, Sector-III, Salt Lake, Kolkata 700106, India}}
\affil[b]{\textit{Department of Physics, University of Kalyani, Kalyani 741235, India}}

\date{}

\begin{document}
	\maketitle
	\begin{abstract}
		\noindent In this paper, we study various mixed state information theoretic quantities for a boosted black brane geometry. We have considered two setups, namely, a strip-like subsystem taken parallel and perpendicular to the direction of the boost. The quantities we calculate are the entanglement wedge cross-section, mutual information, entanglement negativity, and purification complexity. In the subsequent analysis, we have incorporated the thin-strip approximation and computed the leading order change (over pure AdS) in the concerned information theoretic quantities due to the boost parameter. We also show the relation of these quantities computed holographically to the energy and pressure of the boundary field theory. We then proceed to calculate the asymmetry ratios of these quantities, and observe that they are independent of the subsystem size. Finally, we proceed to study an interesting limit of the boosted black brane geometry, which is the so called AdS wave geometry. We once again compute all the mixed state information theoretic quantities for this geometry.
		\end{abstract}
	\section{Introduction}\label{sec1}
	In recent years the AdS/CFT correspondence \cite{Maldacena:1997re,Witten:1998qj,Aharony:1999ti} has emerged as a very powerful dictionary to study the strongly coupled ($\lambda\equiv {g_{\mathrm{YM}}}^2N_c \gg 1$) quantum field theories with the help of classical gravity. Due to this fascinating feature, this duality has its application from condensed matter physics to quantum chromodynamics. Furthermore, this conjecture has gifted us the `holographic' way to compute various information theoretic quantities in the conformal field theory side. In the context of quantum information theory, entanglement entropy (EE), a measure of quantum correlation, plays the role of the cardinal. It has the following simple definition, provided by von Neumann. Consider a bipartite quantum system with the Hamiltonian of the form $\mathcal{H}=\mathcal{H}_{A}\tens\mathcal{H}_{B}$, where $A\cup B$ is the full system. Now the entanglement entropy of the subsystem $A$ is defined as \cite{Chuang:2000}
	\begin{equation}
	S_{A}=-Tr(\rho_{A}\log\rho_{A})
	\end{equation} 
	where $\rho_{A}$ is the reduced density matrix of the subsystem $A$, which is obtained by tracing out the degrees of freedom of the subsystem $B=A^c$. This symbolically reads $\rho_{A}=Tr_{B}(\rho_{AB})$, $\rho_{AB}=\ket{\psi}\bra{\psi}$ where $\ket{\psi}\in\mathcal{H}$. The holographic counterpart of entanglement entropy is known as holographic entanglement entropy (HEE), which is computed by the Ryu-Takayanagi (RT) prescription \cite{Ryu:2006bv,Ryu:2006ef,Nishioka:2009un}. The RT-prescription relates the
	area of a co-dimension-$2$ static minimal surface $(\Gamma^{A}_{min})$ in the bulk, to the
	von Neumann entropy (EE) of the reduced density matrix at the boundary. Keeping in mind this statement, the HEE of a subsystem $A$ at the boundary is given by 
	\begin{equation}\label{RT}
	S_{EE}(A)=\frac{Area(\Gamma^{A}_{min})}{4G_N}
	\end{equation}
	where $G_N$ is the Newton's gravitational constant. Similar to EE for field theories, the HEE also contains an `area-like' divergent piece which is independent of the subsystem size (universal in nature) \cite{Srednicki:1993im}. With definition of entanglement entropy in hand, one can define another quantum mechanical quantity known as the mutual information $(I(A:B))$ \cite{Chuang:2000}
	\begin{equation}
	I(A:B)=S_{A}+S_{B}-S_{A\cup B}~.
	\end{equation}
	It measures the collective correlation between the two subsystems $A$ and $B$. Furthermore, it is positive and finite. The holographic counterpart of the mutual information can be obtained by using the RT-prescription to compute the relevant von Neumann entropies. However, the von Neumann definition of entanglement entropy is not a good measure of  quantum correlation if the state under consideration is a mixed one. Various research in this direction has suggested many different correlation measures for mixed states. Among these the entanglement of purification (EoP) \cite{Terhal_2002} is one of the most promising candidate. The process of purification suggests that one has to construct a pure state $\ket{\psi}$ from the mixed density matrix $\rho_{AB}$ by adding auxiliary degrees of freedom to the original Hilbert space $\mathcal{H}$ 
	\begin{eqnarray}
	\rho_{AB} = tr_{A^{\prime}B^{\prime}}\ket{\psi}\bra{\psi};~\psi \in \mathcal{H}_{AA^{\prime}BB^{\prime}}=\mathcal{H}_{AA^{\prime}} \tens \mathcal{H}_{BB^{\prime}}
	\end{eqnarray}
	The states $\ket{\psi}$ are denoted as the purifications of $\rho_{AB}$. In this set up, the EoP can be computed as \cite{Terhal_2002}
    \begin{eqnarray}\label{EoP}
    E_P(\rho_{AB})\equiv E_P(A,B) = \mathop{min}_{\ket{\psi}}S(\rho_{AA^{\prime}});~\rho_{AA^{\prime}} = tr_{BB^{\prime}}\ket{\psi}\bra{\psi}
    \end{eqnarray}
	where the minimization is taken over any state $\ket{\psi}$ with the property $\rho_{AB} = tr_{A^{\prime}B^{\prime}}\ket{\psi}\bra{\psi}$ being held constant. However, in the context of the QFT, it is a tricky task to compute the EoP. On the other hand, the holographic principle provides a quantity called the entanglement wedge cross-section (EWCS) \cite{Takayanagi:2017knl,Nguyen:2017yqw}, which can be treated as the holographic counterpart of the EoP. This observation has led to the $E_P=E_W$ duality. It has been observed that both $E_P(A,B)$ and $E_W(A,B)$ satisfy the following properties \cite{Takayanagi:2017knl}
	\begin{eqnarray}\label{prop}
	&&~E_P(A,B) = S_{EE}(A)=S_{EE}(B);~\rho_{AB}^2=\rho_{AB}\nonumber\\
	&&~\frac{1}{2} I(A:B) \leq E_P(A,B) \leq min\left[S_{EE}(A),S_{EE}(B)\right] \nonumber\\
	&& \frac{I(A:B)+I(A:C)}{2} \leq E_P(A,B\cup C)~~.
	\end{eqnarray} 
	However, it is to be noted that the direct proof of $E_P=E_W$ duality conjecture is yet to be discovered.\\
	As we have mentioned earlier, there are several different proposed candidates to measure quantum correlation for mixed states. Apart from the EoP, entanglement negativity or the logarithmic negativity \cite{Vidal:2002zz,PhysRevA.58.883,PhysRevA.60.3496} has emerged as other important correlation measures of mixed states. Entanglement negativity characterizes the upper bound on the distillable entanglement for a quantum system in a mixed state. Consider a tripartite scenario of a quantum mechanical set up, with the subsystems denoted as $A_{1}$,$A_{2}$ and $B$. The associated conditions are $A=A_{1}\cup A_{2}$ and $B=A^{c}$ \cite{Jain:2017aqk}. The Hilbert space of the bipartite system $A$ can be represented as $\mathcal{H}=\mathcal{H}_{1}\tens\mathcal{H}_{2}$, where $\mathcal{H}_{1}$ is the Hilbert space corresponding to  the subsystem $A_{1}$ and $\mathcal{H}_{2}$ is the Hilbert space associated with the subsystem $A_{2}$. Now the reduced density matrix of the subsystem $A$ can be obtained by taking the trace over the full density matrix of the system with respect to the rest of the system, that is, $A^{c}=B$. This reads
	\begin{equation}
	\rho_{A}=Tr_{A^{c}}(\rho_{AB})
	\end{equation}
	where $\rho_{AB}$ represents the total density matrix of the system in the mixed state. The computation involves a partial transpose of the reduced density matrix over one of the subsystems in the given bipartite system. This can be done in the following way. Let us consider $|e_{i}^{(1)}\rangle$ and $|e_{i}^{(2)}\rangle$ to be the basis of the Hilbert space associated with the subsystems $A_{1}$ and $A_{2}$ respectively. The partial transpose of the reduced density matrix with respect to $A_{2}$ is defined as
	\begin{eqnarray}
	\bra{e_{i}^{(1)}e_{j}^{(2)}}\rho_{A}^{T_{2}}\ket{e_{k}^{(1)}e_{l}^{(2)}}=\bra{e_{i}^{(1)}e_{l}^{(2)}}\rho_{A}\ket{e_{k}^{(1)}e_{j}^{(2)}}
	\end{eqnarray}
	where $\rho_{A}^{T_{2}}$ represents the partial transpose of the total density matrix $\rho$ with respect to $A^{c}$.  Entanglement negativity measures the degree to which  $\rho_{A}^{T_{2}}$ is not positive, this actually signifies the term `negativity'. If the trace norm of the partial transpose reduced density matrix is denoted by $\parallel\rho_{A}^{T_{2}}\parallel_{1}$, then we actually define two quantities, one is negativity and another one is the entanglement negativity or logarithmic negativity. The negativity between two subsystems $A_{1}$ and $A_{2}$ is defined in the following way \cite{Vidal:2002zz}
	\begin{equation}
	\mathcal{N}(\rho)=\frac{\parallel\rho_{A}^{T_{2}}\parallel_{1}-1}{2}~.
	\end{equation}
	The above corresponds to the absolute value of the sum of negative eigenvalues of $\rho_{A}^{T_{2}}$. One can show that for unentangled product state $\mathcal{N}(\rho)$ vanishes. On the other hand, entanglement negativity or logarithmic negativity between two subsystems $A_{1}$ and $A_{2}$ is defined as \cite{Vidal:2002zz}
	\begin{equation}
	E_{N}(\rho)=\ln(\parallel\rho_{A}^{T_{2}}\parallel_{1})~~.
	\end{equation}
	 In the context of quantum field theory it is very hard to compute the above quantity. However, gauge/gravity duality provides us a systematic way to calculate it by using the bulk/boundary dictionary. Furthermore, there are also some other quantities to probe the mixed state entanglement, namely, the odd entropy \cite{PhysRevLett.122.141601,Ghasemi:2021jiy,Basak:2022gcv}, reflected entropy \cite{Dutta:2019gen,Chu:2019etd,Basak:2022cjs}, etc.\\
	 Along with the above measures of quantum correlation, quantum complexity is one of the most fundamental quantity in context of quantum information theory. The computational complexity has its roots in the subject of computer science \cite{S.Arora:2009book,C.Moore:2011book}. In classical computation, an algorithm simply means a function which maps set of input bits to specific set of output bits. However, we are interested in quantum complexity. Here the function transform to the unitary operator $\hat{U}$ which acts on some input state for some number of qubits and gives an output quantum state on equal number of qubits. The unitary operation is constructed by set of elementary gates. If our reference state is $\ket{\psi_{R}}$ and the target state is $\ket{\psi_{T}}$, then one has to implement the following operation
	 \begin{equation}
	 \ket{\psi_{T}}=\hat{U}\ket{\psi_{R}}
	 \end{equation}
	 where $\hat{U}$ is an unitary operator which has been constructed with help of a set of elementary gates. Thus one can write the unitary operation as
	 \begin{equation}
	 \hat{U}=g_{1}g_{2}...g_{n}
	 \end{equation} 
	 $g_{i}$ denotes the elementary quantum gates. This in turn means that in order to reach the target state from the reference state one needs to implement the following operation in terms of the elementary gates
	 \begin{equation}
	 \ket{\psi_{T}}=\hat{U}\ket{\psi_{R}}=g_{n}g_{n-1}...g_{1}\ket{\psi_{R}}~.
	 \end{equation}
	 Now the circuit complexity of the target state, $\mathcal{C}(\ket{\psi_{T}})$ is defined as the minimum number of the gates which are needed to construct the unitary operation. This is the way by following which one can define the complexity in the state space. There are various different ways to compute the quantum complexity \cite{Nielsen1,Nielsen2,Susskind:2018pmk,Ali:2018fcz}.
	 In the context of guage/gravity duality, complexity plays an important role. In holography, it has been shown that the thermofield double state in the boundary field theory is dual to the two-sided eternal black hole in the bulk geometry. So in principle, by using the bulk geometry one can calculate the complexity of the $\ket{\mathrm{TFD}}$ (thermofield double state) state living at the boundary conformal field theory. It captures the information behind the horizon of the black hole. There are various proposals to compute the complexity holographically. We shall briefly discuss these proposals later in this paper. However, for mixed state the computation of complexity is a bit tricky. Observations in this direction has revealed that one can also introduce the concept of ``purification" in this context also. The purification complexity is defined as the minimal pure state complexity among all possible purifications available for a mixed state. If there two systems, say $A$ and $B$, the mutual complexity between them can be defined as \cite{Alishahiha:2018lfv,PhysRevD.99.086016,Caceres:2019pgf,Agon:2018zso,Ruan:2021wep}
	 \begin{eqnarray}
	 \Delta\mathcal{C} = \mathcal{C}(\rho_{A}) +\mathcal{C}(\rho_{B})-\mathcal{C}(\rho_{A\cup B})~.
	 \end{eqnarray}
	 In this paper, we holographically compute all the quantities discussed above, for the boosted black brane geometry. We have performed the calculations for both subsystems along boost and subsystems perpendicular to the direction of boost. We have employed the Hubeny-Ryu-Takayanagi (HRT) formalism to compute the entanglement entropy of the boundary theory holographically. We have also demonstrated that this matches with the result computed using the RT formalism in the appropriate limit. We have also computed the mentioned information theoretic quantities for the AdS wave geometry which is a special limit of the boosted black brane geometry.\\
	 	The motivation to study this type of system lies in the fact that it has been observed that the boosted brane provides a direct subtle connection between the boundary field theory and string theory. In the language of $AdS/CFT$ correspondence, when a $pp$ wave is propagating along a particular direction in the world volume of the classical $p$-brane configuration, two different cases can arise depending upon whether the Bogomo\'{l}nyi-Prasad-Sommerfield (BPS) bound is saturated or not. If the BPS bound is not saturated, then the effect of including $pp$ wave is equivalent to the inclusion of a locally Lorentz boost along the direction of propagation of the wave. However this equivalence is valid if the direction of propagation of the $pp$ wave is uncompactified. On the other hand, if the direction of propagation is compactified on a circle then this equivalence is just valid locally. Another motivation to study this system is that boosted AdS black brane background corresponds to a strongly coupled anisotropic thermal plasma uniformly boosted (with respect to an obsever seating in a static
	 	reference frame attached to the flat boundary spacetime) in a certain direction in the boundary theory  \cite{Mateos:2011ix,Bhatta:2019eog}. Such studies are important due to the experimental investigations of strongly coupled QCD plasma at RHIC and LHC \cite{SHURYAK200564,BACK200528,SHURYAK200948,Shuryak:2014zxa,PhysRevLett.103.232001}. Further, the boosted black brane geometry is the only regular solution of Einstein's field equations describing a dual field theory living on a flat spacetime encoding a homogeneous constant stress tensor \cite{Bhaseen:2013ypa}. These solutions are found to give a generalised first law of entanglement thermodynamics which contain the effects of chemical potential and charge density. Further, the effect of IR deformation (excitations in the CFT side) can be captured by computing the HEE of the boosted black brane. The rotational symmetry in the boundary theory breaks down due to the boost along a specific direction, in other words boost introduces anisotropy in the boundary theory. Hence, boosted black brane geometries in the bulk are important in investigating anisotropies of boundary field theories. It is also to be noted that the boost direction is compactified on a circle which results in Kaluza-Klein gauge charges.  Interestingly, boosted black brane compatified along one of its lightcone coordinates gives rise to Lifshitz theory \cite{Mishra:2018tzj}. Various physical quantities such as energy, momentum and pressure can be obtained by expanding the bulk $AdS$ geometry eq.(\ref{bbm})) in suitable Feffermann-Graham asymptotic coordinates \cite{Balasubramanian:1999re,Kraus:1999di}.\\
	 This paper is organized in the following way. In section (\ref{sec2}), we provide a brief discussion on the bulk geometry, that is, the boosted black brane geometry. 
	  In section (\ref{along}), we have computed various information theoretic quantities for subsystems with the orientation along the boost. In section (\ref{sec4}), we do the same for subsystems perpendicular to the direction of boost. The asymmetry ratios associated with the computed quantities has been obtained in section (\ref{sec5}). In section \eqref{AdSWave}, we consider the AdS wave geometry which arises from the boosted black brane, at the large boost limiting condition and compute the discussed information theoretic quantities holographically. We summarize our findings to conclude in section \eqref{conc}. We also provide an Appendix for the sake of completeness. 
	\section{Boosted black brane in $AdS_{d+1}$}\label{sec2}
	The $AdS_{d+1}$ boosted black brane can be represented by the following metric \cite{Mishra:2015cpa,Mishra:2016yor}
	\begin{eqnarray}\label{bbm}
	ds^{2}&=&\frac{1}{z^{2}}\left(-\frac{f(z)}{K(z)}dt^{2}+K(dy-\omega)^{2}+dx_{1}^{2}+...+dx_{d-2}^{2}+\frac{dz^{2}}{f(z)}\right)
	\end{eqnarray}
	where 
	\begin{eqnarray}
	f(z)=1-\frac{z^{d}}{z_{0}^{d}}~~, ~~ K(z)=1+\beta^{2}\gamma^{2}\frac{z^{d}}{z_{0}^{d}}~.
	\end{eqnarray}	
	In the above, $\beta$ is the boost parameter which is constrained as $0\le\beta\le 1$ and it is related to the Lorentz factor as $\gamma=\frac{1}{\sqrt{1-\beta^{2}}}$. Further, $z_{0}$ is the position of the event horizon. From the metric above, it is clear that the boost is along the $y$ direction which in turn introduces anisotropy in that particular direction. The Kaluza-Klein $1$-form $\omega$ reads 
	\begin{eqnarray}
	\omega=\beta^{-1}\left(1-\frac{1}{K(z)}\right)dt~~.
	\end{eqnarray}
In the following sections we will holographically compute various information theoretic measures by considering a strip like subsystem in both along the boost and perpendicular to the boost direction. We shall also investigate the effect of the boost parameter on these quantities.

\section{Strip like subsystem along the boost}\label{along}
In this section we start our analysis by considering a strip-like subsystem $A$. The geometry of this subsystem is specified by the volume $V_{sub}=L^{d-2}l$, with $-\frac{l}{2}\le y\le \frac{l}{2}$, and $x_{1},...,x_{d-2} \in \left[0,L\right]$ with $L\rightarrow\infty$. Further we assume that the length can vary only along the $y$-direction and the lengths in other directions are taken to be fixed. We choose the parametrization $z=z(y)$ and $t=t(y)$ in order to compute the surface area of the co-dimension two HRT surface $\Gamma^{min}_{A}(t)$ which shall eventually lead us to the holographic entanglement entropy.

\subsection{Holographic entanglement entropy}\label{sec3}
We now use the HRT prescription \cite{hubeny2007covariant} to holographically compute the entanglement entropy of subsystem $A$. According to the HRT prescription, the HEE reads
\begin{eqnarray}\label{hrt1}
	S_{EE}^{\parallel}&=&\frac{Area(\Gamma_A^{min}(t))_{\parallel}}{4G_{d+1}}\nonumber\\
	&=&\frac{V_{d-2}}{2G_{d+1}}\int_{-\frac{l}{2}}^{0}~\frac{dy}{z^{d-1}}\sqrt{K(z)+\frac{z^{\prime 2}}{f(z)}-t^{\prime 2}\left(\frac{f(z)}{K(z)}-\frac{K(z)}{\beta^{2}}\left[1-\frac{1}{K(z)}\right]^{2}\right)-\frac{2t^{\prime}(K(z)-1)}{\beta}}~.
\end{eqnarray}
We now identify the Lagrangian $\mathcal{L}=\mathcal{L}(z,z^{\prime},t,t^{\prime})$ from the above expression and note that $y$ is a cyclic coordinate. This cyclic coordinate leads to the following conserved quantity
\begin{eqnarray}
	\mathcal{H}=\frac{-K(z)\left(1-\frac{t^{\prime}(1-\frac{1}{K(z)})}{\beta}\right)}{z^{d-1}\sqrt{K+\frac{z^{\prime 2}}{f(z)}-t^{\prime 2}\left(\frac{f(z)}{K(z)}-\frac{K(z)}{\beta^{2}}\left[1-\frac{1}{K(z)}\right]^{2}\right)-\frac{2t^{\prime}(K(z)-1)}{\beta}}}=constant=c~~.
\end{eqnarray}
Now we introduce the turning point $(z_{t}^{\parallel},t_{t}^{\parallel})$ inside the bulk at which $z^{\prime}$ and $t^{\prime}$ vanishes. This fixes the above constant to be $c=-\frac{\sqrt{K(z_{t}^{\parallel})}}{z^{\parallel(d-1)}_{t}}$. This results in the following equation
\begin{eqnarray}\label{ph}
	K(z)+\frac{z^{\prime 2}}{f(z)}-t^{\prime 2}\left(\frac{f(z)}{K(z)}-\frac{K(z)}{\beta^{2}}\left[1-\frac{1}{K(z)}\right]^{2}\right)-\frac{2t^{\prime}(K(z)-1)}{\beta}=\frac{K^{2}(z)}{K(z_{t}^{\parallel})}\left(\frac{z_{t}^{\parallel}}{z}\right)^{2(d-1)}\left[1-\frac{t^{\prime}}{\beta}\left(1-\frac{1}{K(z)}\right)\right]^{2}~.\nonumber\\
\end{eqnarray}
Substituting the above expression in eq.(\ref{hrt1}), we obtain the HEE to be
\begin{eqnarray}\label{hrt2}
S_{EE}^{\parallel}=\frac{V_{d-2}}{2G_{d+1}}\int_{-\frac{l}{2}}^{0}\frac{dy}{z^{d-1}}\frac{K(z)}{\sqrt{K(z_{t}^{\parallel})}}\left(\frac{z_{t}^{\parallel}}{z}\right)^{d-1}\left(1-\frac{t^{\prime}}{\beta}\left[1-\frac{1}{K(z)}\right]\right).
\end{eqnarray}
Before we proceed further, we want to make some remarks regarding the above expression. We observe that the first term in the above expression is the HEE in the RT-prescription \cite{Ryu:2006bv} obtained by setting $dt=0$ in the metric (eq.(\ref{bbm})) and by choosing the parametrization $z=z(y)$. On the other hand the second term represents a correction term which arises due to the HRT formalism \cite{hubeny2007covariant}.\\
 To proceed further, we express $dy$ in terms of $dz$ by using eq.(\ref{ph}). This reads
\begin{eqnarray}\label{dy}
	\frac{dy}{dz}&=&\frac{\left(\frac{z}{z_{t}^{\parallel}}\right)^{d-1}\left[1-\frac{\left(\frac{z}{z_{t}^{\parallel}}\right)^{2d-2}\left[t^{\prime2}\left(\frac{f(z)}{K(z)}-\frac{K(z)}{\beta^{2}}(1-\frac{1}{K(z)})^{2}\right)+\frac{2t^{\prime}(K(z)-1)}{\beta}\right]}{K(z)\left[\frac{K(z)}{K(z_{t}^{\parallel})}\left(1-\frac{t^{\prime}}{\beta}\left[1-\frac{1}{K(z)}\right]\right)^{2}-\left(\frac{z}{z_{t}^{\parallel}}\right)^{2d-2}\right]}\right]^{\frac{1}{2}}}{\sqrt{K(z)f(z)}\left[\frac{K(z)}{K(z_{t}^{\parallel})}\left(1-\frac{t^{\prime}}{\beta}\left[1-\frac{1}{K(z)}\right]\right)^{2}-\left(\frac{z}{z_{t}^{\parallel}}\right)^{2d-2}\right]^{\frac{1}{2}}}~.
\end{eqnarray}
We now consider the thin strip approximation, that is, $\left(\frac{z_{t}^{\parallel}}{z_{0}}\right)^{d}<<1$. Keeping this approximation in mind, we only consider terms upto $\mathcal{O}\left(\left(\frac{z}{z_{t}^{\parallel}}\right)^{d}\right)$ and neglect the higher order terms in eq.\eqref{dy}. Eq.(\ref{dy}) now simplifies to the following expression 
\begin{eqnarray}\label{diffe}
dy&=&\frac{\left(\frac{z}{z_{t}^{\parallel}}\right)^{d-1}~dz}{\sqrt{K(z)f(z)}\left[\frac{K(z)}{K(z_{t}^{\parallel})}\left(1-\frac{t^{\prime}}{\beta}\left[1-\frac{1}{K(z)}\right]\right)^{2}-\left(\frac{z}{z_{t}^{\parallel}}\right)^{2d-2}\right]^{\frac{1}{2}}}
\end{eqnarray}
Now substituting eq.(\ref{diffe}) in eq.(\ref{hrt2}) and  using the conditions
$z(y=0)=z_{t}^{\parallel},z(y=\pm\frac{l}{2})=\epsilon$\footnote{$\epsilon$ is the UV cut-off which has been introduced to regularize the area functional.}, 
we can write down HEE in the following form 
\begin{eqnarray}\label{eepa}
S^{\parallel}_{EE}&=&\frac{V_{d-2}}{2G_{d+1}}\int_{\epsilon}^{z_{t}^{\parallel}}~dz\frac{\sqrt{K(z)/K(z_{t}^{\parallel})}}{z^{d-1}\sqrt{f(z)}}\frac{1}{\sqrt{\frac{K(z)}{K(z_{t}^{\parallel})}\left(1-\frac{t^{\prime}}{\beta}\left[1-\frac{1}{K(z)}\right]\right)^{2}-\left(\frac{z}{z_{t}^{\parallel}}\right)^{2d-2}}}\nonumber\\
&-&\frac{V_{d-2}}{2G_{d+1}}\int_{\epsilon}^{z_{t}^{\parallel}}~dz\frac{\sqrt{K(z)/K(z_{t}^{\parallel})}}{z^{d-1}\sqrt{f(z)}}\frac{\frac{t^{\prime}}{\beta}\left(1-\frac{1}{K(z)}\right)}{\sqrt{\frac{K(z)}{K(z_{t}^{\parallel})}\left(1-\frac{t^{\prime}}{\beta}\left[1-\frac{1}{K(z)}\right]\right)^{2}-\left(\frac{z}{z_{t}^{\parallel}}\right)^{2d-2}}}~.
\end{eqnarray}
In the subsequent discussion, we are interested in the leading order correction in the HEE due to the boost. In \cite{Mishra:2015cpa,Mishra:2016yor,mishra2018study}, it was pointed out that in the perturbative expansion for small strips, to obtain the leading order boost correction in the HEE, considering a constant time slice is sufficient.
In other words, when it comes to compute the leading order correction in the HEE (next to pure $AdS$), it is sufficient to work with the constant time slice co-dimension two RT-surface. The deviations in the geometry of the extremal surface away from the constant time slice contribute only to the second order terms in the perturbative expansion \cite{Mishra:2015cpa,Mishra:2016yor,mishra2018study}. To proceed further we will assume that all the computations are done in constant time slice, that is, we will set $t^{\prime}=0$. Therefore, under this approximation one can write down the subsystem size in terms of the bulk coordinate (by using eq.(\ref{diffe})) as \cite{Mishra:2015cpa,Mishra:2016yor}
\begin{eqnarray}\label{lpa}
\frac{l}{2}=\int_{0}^{z_{t}^{\parallel}}dz\frac{(z/z_{t}^{\parallel})^{d-1}}{\sqrt{f(z)K(z)}\left[\frac{K(z)}{K(z_{t}^{\parallel})}-(z/z_{t}^{\parallel})^{2d-2}\right]^{1/2}}~.
\end{eqnarray}
Keeping the above discussion in mind, one can obtain the following form of the HEE (by setting $t^{\prime}=0$) from eq.(\ref{eepa}) \cite{Mishra:2015cpa,Mishra:2016yor,mishra2018study}
\begin{eqnarray}
S^{\parallel}_{EE}&=&\frac{V_{d-2}}{2G_{d+1}}\int_{\epsilon}^{z_{t}^{\parallel}}~dz\frac{\sqrt{K(z)/K(z_{t}^{\parallel})}}{z^{d-1}\sqrt{f(z)}}\frac{1}{\sqrt{\frac{K(z)}{K(z_{t}^{\parallel})}-\left(\frac{z}{z_{t}^{\parallel}}\right)^{2d-2}}}~.	
\end{eqnarray}
As mentioned earlier, we will now use the thin strip approximation which implies
\begin{equation}\label{thinS}
\left(\frac{z_{t}^{\parallel}}{z_{0}}\right)^{d}<<1~~,~~\beta^{2}\gamma^{2}\left(\frac{z_{t}^{\parallel}}{z_{0}}\right)^{d}<<1
\end{equation}
 and we will keep terms upto $\mathcal{O}(\beta^{2}\gamma^{2})$. Under the thin strip approximation,  $\frac{K(z)}{K(z_{t}^{\parallel})}$ can be written down as \cite{Mishra:2015cpa,Mishra:2016yor} 
\begin{eqnarray}\label{tst}
\frac{K(z)}{K(z_{t}^{\parallel})}\approx 1+\beta^{2}\gamma^{2}\left(\frac{z_{t}^{\parallel}}{z_{0}}\right)^{d}\left[\left(\frac{z}{z_{t}^{\parallel}}\right)^{d}-1\right]~. 
\end{eqnarray}
Under the thin strip approximation and by using the expression given in eq.\eqref{tst}, the subsystem size is obtained to be \cite{Mishra:2015cpa,Mishra:2016yor}
\begin{eqnarray}\label{lpara}
l&=&2z_{t}^{\parallel}\left[b_{0}+\frac{1}{2}\left(\frac{z_{t}^{\parallel}}{z_{0}}\right)^{d}\left(b_{1}+\beta^{2}\gamma^{2}I_{l}\right)\right]
\end{eqnarray}
where the expressions corresponding to the constant terms $b_0$, $b_1$ and $I_l$ are given in the appendix. The subsystem size - turning point relation corresponding to the pure $AdS_{d+1}$ geometry can be obtained by taking the limits $\beta \rightarrow 0$ and $z_{0}\rightarrow \infty$. This reads \cite{Ryu:2006ef}
\begin{equation}\label{lpure}
l=2b_{0}\bar{z_{t}}
\end{equation}
where $\bar{z_{t}}$ is the turning point for pure $AdS_{d+1}$ geometry. To express $z_{t}^{\parallel}$ in terms $\bar{z_{t}}$, we make use of eq.(\ref{lpara}) and eq.(\ref{lpure}). This results in
\begin{eqnarray}\label{a}
z_{t}^{\parallel}&=&\frac{l/2}{\left[b_{0}+\frac{1}{2}\left(\frac{z_{t}^{\parallel}}{z_{0}}\right)^{d}\left(b_{1}+\beta^{2}\gamma^{2}I_{l}\right)\right]} \approx\frac{\bar{z_{t}}}{1+\frac{1}{2}\left(\frac{\bar{z_{t}}}{z_{0}}\right)^{d}\left(\frac{b_{1}}{b_{0}}+\frac{\beta^{2}\gamma^{2}}{b_{0}}I_{l}\right)}~.
\end{eqnarray}
Now we proceed to compute the explicit form of HEE by using eq.(s)(\ref{eepa},\ref{tst}). Under the thin strip approximation, HEE is obtained to be \cite{Mishra:2015cpa,Mishra:2016yor}
\begin{eqnarray}\label{Spara}
S^{\parallel}_{EE}=S_{div}+\frac{V_{d-2}}{2G_{d+1}}\frac{a_{0}}{(z_{t}^{\parallel})^{d-2}}\left[1+\frac{p^{d}}{2}\frac{a_{1}}{a_{0}}+\frac{q^{d}}{2}\left(\frac{d+1}{d-1}\frac{b_{1}}{a_{0}}-\frac{1}{d-1}\frac{b_{0}}{a_{0}}\right)\right]
\end{eqnarray} 
where $S_{div}$ represents the subsystem independent divergent term $S_{div}=\frac{V_{d-2}}{2G_{d+1}(d-2)}\frac{1}{\epsilon^{d-2}}$ and $a_0$ is a constant term, given in the appendix. Furthermore,  $p^{d}=\left(\frac{z_{t}^{\parallel}}{z_{0}}\right)^{d}$ and $q^{d}=\beta^{2}\gamma^{2}\left(\frac{z_{t}^{\parallel}}{z_{0}}\right)^{d}$. The finite part of the HEE can be written in terms of the subsystem size by using eq.(\ref{lpure}), eq.(\ref{a}) and eq.(\ref{Spara}). This reads \cite{Mishra:2015cpa,Mishra:2016yor}
\begin{eqnarray}\label{sfin}
S^{\parallel}_{EE}&=&S_{div}+\frac{V_{d-2}a_{0}}{2G_{d+1}}\Bigg[\left(\frac{2b_{0}}{l}\right)^{d-2}+\frac{1}{2z_{0}^{d}}\left(\frac{l}{2b_{0}}\right)^{2}\frac{b_{1}}{a_{0}}\left(\frac{d-1}{2}+\beta^{2}\gamma^{2}\right)\Bigg]~.
\end{eqnarray}
The higher order terms in the above expresiion of entanglement entropy is obtained in \cite{Maulik:2020tzm}. For the pure $AdS_{d+1}$ background, the HEE is known to be\footnote{This can also be obtained from eq.(\ref{sfin}) by taking the limits $\beta \rightarrow 0$ and $z_{0}\rightarrow \infty$.}  \cite{Ryu:2006ef}
\begin{equation}\label{spure}
S^{pure}_{EE}=S_{div}+\frac{V_{d-2}a_{0}}{2G_{d+1}}\left(\frac{2b_{0}}{l}\right)^{d-2}
\end{equation}
We now define a quantity $\delta S^{\parallel}_{EE}$, in order to remove divergent piece and to represent the change in HEE due to the boost. This reads \cite{Mishra:2015cpa,Mishra:2016yor}
\begin{eqnarray}\label{cspara}
\delta S^{\parallel}_{EE}\equiv S^{\parallel}_{EE}-S^{pure}_{EE}=\frac{V_{d-2}(d+1)b_{1}l^{2}}{32G_{d+1}b_{0}^{2}z_{0}^{d}}\left[\frac{d-1}{d+1}+\frac{2\beta^{2}\gamma^{2}}{d+1}\right]~.
\end{eqnarray} 
The expression of HEE in eq.(\ref{sfin}) suggests that the entanglement entropy increases in the presence of the boost. A possible reason for this increase in the entanglement entropy is an increase in the area of the strip residing in the CFT side. Increase in the boost also leads to more excitations in the CFT side, and this may also be a reason for the increase in the entanglement entropy from the CFT point of view.\\ 
\subsection{EWCS and Holographic mutual information}
 As we have mentioned earlier, the EE is a good way to quantify quantum entanglement as long as the system under consideration is in a pure state. We now holographically compute EoP, which is one of the promising candidates to quantify entanglement for mixed states. This we do by computing the EWCS for the boosted black brane. This computation holographically probes EoP on the basis of $E_P = E_W$ duality \cite{Takayanagi:2017knl,Jokela:2019ebz,BabaeiVelni:2019pkw,Nguyen:2017yqw}.\\
  We consider two strip like subsystems on the boundary $\partial M$ ($\partial M$ is the boundary of the canonical time-slice $M$ that has been considered in the gravity dual). We denote these subsystems as $A$ and $B$ with both of them having the same length $l$. Further we consider that $A$ and $B$ are separated by a distance $x$ with the condition $A\cap B =0$. The Ryu-Takayanagi surfaces corresponding to $A$, $B$ and $AB$ can be denoted as $\Gamma_A^{min}$, $\Gamma_B^{min}$ and $\Gamma_{AB}^{min}$ respectively. The co-dimension-0 domain of entanglement wedge $M_{AB}$ is characterized by the following boundary 
\begin{eqnarray}\label{16}
	\partial M_{AB} = A \cup B \cup \Gamma_{AB}^{min} =\bar{\Gamma}_A \cup \bar{\Gamma}_B
\end{eqnarray}
where $\bar{\Gamma}_A = A \cup \Gamma_{AB}^{A}$, $\bar{\Gamma}_B = B \cup \Gamma_{AB}^{B}$. In the above equation we have used the condition $\Gamma_{AB}^{min} = \Gamma_{AB}^{A} \cup \Gamma_{AB}^{B}$. In this set up, one can define the holographic entanglement entropies $S(\rho_{A \cup \Gamma_{AB}^{A}})$ and $S(\rho_{B \cup \Gamma_{AB}^{B}})$ and compute them by finding a static RT surface $\Sigma^{min}_{AB}$ with the following condition
\begin{eqnarray}
	\partial \Sigma^{min}_{AB} = \partial \bar{\Gamma}_A =  \partial \bar{\Gamma}_B~.
\end{eqnarray}
The splitting condition $\Gamma_{AB}^{min} = \Gamma_{AB}^{A} \cup \Gamma_{AB}^{B}$ which has been incorporated in not unique and there can be infinite number of possible choices. Further, this means that there can be infinite number of choices for the surface $\Sigma^{min}_{AB}$. The EWCS is computed by minimizing the area of $\Sigma^{min}_{AB}$ over all possible choices for $\Sigma^{min}_{AB}$. This reads
\begin{eqnarray}\label{18}
	E_W(\rho_{AB}) = \mathop{min}_{\bar{\Gamma}_A \subset \partial M_{AB}}\left[\frac{A\left(\Sigma^{min}_{AB}\right)}{4G_{d+1}}\right]~.
\end{eqnarray}
This means that to compute EWCS we have to calculate the vertical constant $y$ hypersurface with minimal area which splits $M_{AB}$ into two domains corresponding to $A$ and $B$. The time induced metric on this constant $y$ hypersurface reads
\begin{eqnarray}
	ds^{2}\mid_{ind}=\frac{1}{z^{2}}\left(dx_{1}^{2}+...+dx_{d-2}^{2}+\frac{dz^{2}}{f(z)}\right)~~.
\end{eqnarray}
By using this above mentioned induced metric and the formula given in eq.\eqref{18}, the minimal cross-section of the entanglement wedge ($M_{AB}$) is found to be
\begin{eqnarray}
	E_{W}^{\parallel}&=&\frac{V_{d-2}}{4G_{d+1}}\int_{z_{t}^{\parallel}(x)}^{z_{t}^{\parallel}(2l+x)}\frac{dz}{z^{d-1}\sqrt{f(z)}}\nonumber\\
	&\approx&\frac{V_{d-2}}{4G_{d+1}}\int_{z_{t}^{\parallel}(x)}^{z_{t}^{\parallel}(2l+x)}\frac{dz}{z^{d-1}}\left[1+\frac{1}{2}\left(\frac{z_{t}^{\parallel}}{z_{0}}\right)^{d}\left(\frac{z}{z_{t}^{\parallel}}\right)^{d}\right]\nonumber\\
	&=&\frac{V_{d-2}}{4G_{d+1}(d-2)}\left[\frac{1}{(z_{t}^{\parallel}(x))^{d-2}}-\frac{1}{(z_{t}^{\parallel}(2l+x))^{d-2}}\right]
	+\frac{V_{d-2}}{16G_{d+1}z_{0}^{d}}\left[(z_{t}^{\parallel}(2l+x))^{2}-(z_{t}^{\parallel}(x))^{2}\right]
\end{eqnarray}
where we have used the thin strip approximation. By using eq.(\ref{a}), we can recast the above computed expression can in terms of the subsystem size $l$ and the separation distance $x$. This reads
\begin{eqnarray}\label{ewpara}
	E_{W}^{\parallel}=\frac{V_{d-2}}{4(2b_{0})^{2}G_{d+1}}\Bigg[\frac{(2b_{0})^{d}}{d-2}\left(\frac{1}{x^{d-2}}-\frac{1}{(2l+x)^{d-2}}\right)
	+\Bigg(\frac{1}{2z_{0}^{d}}\left[\left(1+\frac{2\beta^{2}\gamma^{2}}{d-1}\right)\frac{b_{1}}{b_{0}}-\frac{\beta^{2}\gamma^{2}}{d-1}\right]-\frac{1}{2}\Bigg)\left(x^{2}-(2l+x)^{2}\right)\Bigg]~.\nonumber\\
\end{eqnarray}
The EWCS for the pure $AdS_{d+1}$ geometry can be obtained by taking the limit $\beta \rightarrow 0$, $z_0\rightarrow\infty$ of the above expression, which gives   
\begin{eqnarray}\label{ewpure}
	E_{W}^{pure}=\frac{V_{d-2}(2b_{0})^{d-2}}{4G_{d+1}(d-2)}\left[\frac{1}{x^{d-2}}-\frac{1}{(2l+x)^{d-2}}\right]~.
\end{eqnarray}
We now introduce a scaling for the sake of simplification. This reads  
\begin{eqnarray}
	\bar{E}_{W}^{para}= \frac{2G_{d+1}}{V_{d-2}} E_{W}^{\parallel}~~~\mathrm{and}~~~\bar{E}_{W}^{pure}=\frac{2G_{d+1}}{V_{d-2}}E_{W}^{pure}~.
\end{eqnarray}
It has been observed that both $E_P$ and $E_W$ satisfies the bound $E_W(A:B)\geq \frac{1}{2}I(A:B)$. In order to verify this for the case in hand, firstly we need to compute the explicit expression of the mutual information.\\
\noindent The holographic mutual information (HMI) has the following form in this set up
\begin{eqnarray}\label{hmi}
	I(A:B)=S^{\parallel}_{EE}(A)+S^{\parallel}_{EE}(B)-S^{\parallel}_{EE}(A\cup B)\equiv2S^{\parallel}_{EE}(l)-S^{\parallel}_{EE}(x)-S^{\parallel}_{EE}(2l+x)~.
\end{eqnarray}
In the above we have used $S^{\parallel}_{EE}(A\cup B)=S^{\parallel}_{EE}(2l+x)+S^{\parallel}_{EE}(x)$ (for $x/l$ ``small"). With this definition in hand, we now use eq.(\ref{sfin}) to compute the HMI for this setup. This turns out to be
\begin{eqnarray}\label{ipara}
	I^{\parallel}(A:B)&=&\frac{V_{d-2}}{2G_{d+1}}\Bigg[\frac{2^{d-2}b_{0}^{d-1}}{d-2}\left(-\frac{2}{l^{d-2}}+\frac{1}{x^{d-2}}+\frac{1}{(2l+x)^{d-2}}\right)\nonumber\\
	&+&\frac{b_{1}}{(2b_{0})^{2}z_{0}^{d}}\left(\frac{d-1}{2}+\beta^{2}\gamma^{2}\right)\left(l^{2}-\frac{x^{2}}{2}-\frac{(2l+x)^{2}}{2}\right)\Bigg]~~.
\end{eqnarray}
By taking the limit $\beta \rightarrow 0$, $z_0\rightarrow\infty$ of the above expression, we get the HMI for pure $AdS_{d+1}$ background
\begin{eqnarray}\label{ipure}
	I^{pure}(A:B)=\frac{V_{d-2}2^{d-2}b_{0}^{d-1}}{2G_{d+1}(d-2)}\left(-\frac{2}{l^{d-2}}+\frac{1}{x^{d-2}}+\frac{1}{(2l+x)^{d-2}}\right)~.
\end{eqnarray} 
Similar to EWCS, we now introduce the scaling $\bar{I}(A:B) =\frac{2G_{d+1}}{V_{d-2}} I(A:B)$. By incorporating this, we define
\begin{eqnarray}
	\bar{I}^{para}(A:B)=\frac{2G_{d+1}}{V_{d-2}} I^{\parallel}(A:B)~~\mathrm{and}~~\bar{I}^{pure}(A:B)= \frac{2G_{d+1}}{V_{d-2}} I^{pure}(A:B)~.
\end{eqnarray}
With the explicit expressions of EWCS and HMI in hand, we shall now compute the critical separation length $x_c$ in this set up and try to represent it in terms of the subsystem length $l$ and boost parameter $\beta$. The critical separation length $x_c$ is defined as the point at which the mutual correlation between subsystems $A$ and $B$, that is, $I^{\parallel}(A:B)$ vanishes and the EWCS shows a phase transition.\\
For the sake of simplicity, we firstly fix the spacetime dimension, namely, we perform the computations for $d=3$ and $d=4$. In $d=3$, the condition $I^{\parallel}(A:B)=0$ at $x=x_c$ gives the following equation
\begin{eqnarray}
2\left(\frac{\sqrt{\pi}\Gamma(3/4)}{\Gamma(1/5)}\right)^{2}\left[-\frac{2}{l}+\frac{1}{x_c}+\frac{1}{2l+x_c}\right]+\frac{\sqrt{\pi}\frac{\Gamma(3/4)}{4}(1+\beta^{2}\gamma^{2})}{\left(\frac{2\sqrt{\pi}\Gamma(3/4)}{\Gamma(1/5)}\right)^{2}(10)^{3}}\left[l^{2}-\frac{x_c^{2}}{2}-\frac{(2l+x_c)^{2}}{2}\right]&=&0~.
\end{eqnarray} 
On the other hand, for $d=4$ we get
\begin{eqnarray}
2\left(\sqrt{\pi}\frac{\Gamma(4/6)}{\Gamma(1/7))}\right)^{3}\left[-\frac{2}{l^{2}}+\frac{1}{x_c^{2}}+\frac{1}{(2l+x_c)^{2}}\right]+\frac{\frac{\sqrt{\pi}\Gamma(4/3)}{5\Gamma(5/6)}(\frac{3}{2}+\beta^{2}\gamma^{2})}{\left(2\sqrt{\pi}\frac{\Gamma(4/6)}{\Gamma(1/7)}\right)^{2}(10)^{4}}\left[l^{2}-\frac{x_c^{2}}{2}-\frac{(2l+x_c)^{2}}{2}\right]&=&0~.
\end{eqnarray}
One needs to solve the above expressions in order represent $x_c$ in terms of the subsystem size $l$. We now graphically represent our observations.\\
\begin{figure}
	\centering
	\includegraphics[width=0.5\textwidth]{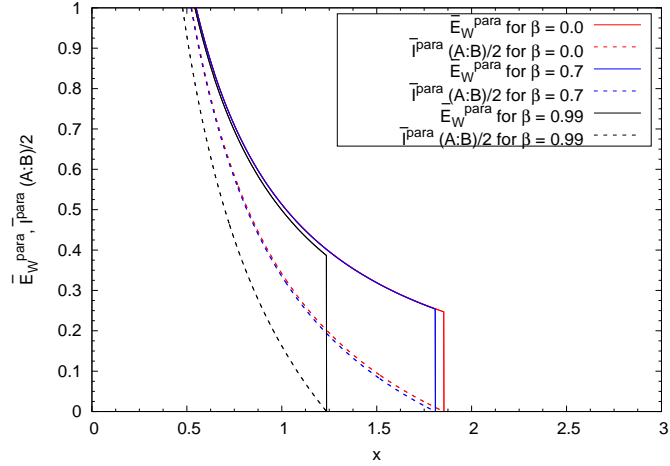}
	\caption{Variation of EWCS and HMI with respect to the separation distance for different values of the boost parameter. The solid curves represent the EWCS and the dotted curves represent the HMI.}
	\label{fig1}
\end{figure}\\	
It has been argued in \cite{Takayanagi:2017knl} that mutual information in eq.(\ref{ipara}) vanishes at a  critical separation length, namely, $x_{c}$ for a fixed subsystem size. This in turn means that at $x=x_{c}$, EWCS shows a discontinuity which represent a phase transition between the connected phase and disconnected phase of the entanglement wedge. Upto $x=x_{c}$, connected phase is the physical whereas beyond $x > x_{c}$ disconnected phase is physical. Some recent works in this direction can be found in \cite{Saha:2021kwq,Chowdhury:2021idy,Sahraei:2021wqn,Liu:2021rks,Basu:2021awn}. In order to compute the critical length $x_c$ for the case in hand, we firstly fix the parameters $d=3$, $l=3$ and $z_{0}=10$. We then plot the expressions of EWCS and HMI, namely $\bar{E}_{W}^{para}$ and $\bar{I}^{para}(A:B)$, for various values of the boost parameter $\beta$. We observe that increasing the value of the boost parameter causes earlier saturation of the HMI and in turn means earlier phase transition for the corresponding entanglement wedge. This has been shown in Fig.(\ref{fig1}).\\
Similar to the HEE, we now compute the change in HMI and EWCS due to inclusion of the anisotropy (boost) in the boundary field theory. This has been given below 
	\begin{eqnarray}\label{cewpara}
		\delta E_{W}^{\parallel}\equiv E_{W}^{\parallel}-E_{W}^{pure}=\frac{V_{d-2}}{32G_{d+1}z_{0}^{d}b_{0}^{2}}\left[\frac{1}{2}-\left(1+\frac{2\beta^{2}\gamma^{2}}{d-1}\right)\frac{b_{1}}{b_{0}}+\frac{\beta^{2}\gamma^{2}}{d-1}\right]\left[(2l+x)^{2}-x^{2}\right]
	\end{eqnarray}
and by using eq.(s)(\ref{ipara},\ref{ipure})
\begin{eqnarray}\label{cipara}
	\delta I^{\parallel}(A:B)\equiv I^{\parallel}(A:B)-I^{pure}(A:B)=\frac{V_{d-2}b_{1}(d+1)}{32G_{d+1}b_{0}^{2}z_{0}^{d}}\left(\frac{d-1}{d+1}+\frac{2\beta^{2}\gamma^{2}}{d+1}\right)\bigg(2l^{2}-x^{2}-(2l+x)^{2}\bigg)~.
\end{eqnarray}
Further, if we take the adjacent subsystem limit, that is, $x\rightarrow0$ in eq.\eqref{cewpara} and compare the obtained expression with eq.\eqref{cspara}, we observe
\begin{eqnarray}\label{newex3}
	\delta E_{W}^{\parallel} =\lambda_1~ \delta S_{EE}^{\parallel}~.
\end{eqnarray}
This in turn means that the change in EWCS due to the presence of boost is proportional to the boost initiated change in HEE upto a constant factor $\lambda_1$\footnote{The explicit expression of $\lambda_1$ has been given in Appendix B.}. It is to be kept in mind that we have shown this under the thin strip approximation. The above relation can be used to relate the leading order change in EWCS to the excitation energy and pressure of the boundary CFT. This can be done by using the relation for the change in HEE ($\delta S^{\parallel}_{EE}$) with the boundary field theoretic quantities like excitation energy and pressure. This relation reads \cite{Mishra:2015cpa,Mishra:2016yor}
\begin{eqnarray}\label{para1st}
	\delta S^{\parallel}_{EE}&=&\frac{1}{T_{E}}\left(\Delta E_{\parallel}-\frac{d-1}{d+1}\mathcal{V}\Delta P_{\parallel}\right)	
\end{eqnarray}
where $\Delta E_{\parallel}$, $\Delta P_{\parallel}$, $T_{E}$, $\mathcal{V}=V_{d-2}l$ are the excitation energy, pressure, entanglement temperature and volume of the subsystem respectively. The expressions for these quantities read \cite{Mishra:2015cpa,Mishra:2016yor}
\begin{eqnarray}\label{bq}
	\Delta E_{\parallel}&=&\frac{V_{d-2}l}{16\pi G_{d+1}}\left(\frac{d-1}{d}+\beta^{2}\gamma^{2}\right)\frac{d}{z_{0}^{d}}\nonumber\\
	\Delta P_{\parallel}&=&\frac{d}{16\pi G_{d+1}}\left(\frac{1}{d}+\beta^{2}\gamma^{2}\right)\frac{1}{z_{0}^{d}}\nonumber\\
	T_{E}&=&\frac{b_{0}^{2}d}{a_{1}\pi l}~.
\end{eqnarray}
Now by substituting eq.\eqref{para1st} in eq.\eqref{newex3}, we obtain
\begin{eqnarray}\label{newex8}
	\delta E^{\parallel}_{W}&=&\frac{\lambda_1}{T_{E}}\left(\Delta E_{\parallel}-\frac{d-1}{d+1}\mathcal{V}\Delta P_{\parallel}\right)~.	
\end{eqnarray}
\subsection{Entanglement negativity}
We now consider another entanglement measure known as the entanglement negativity (also known as the logarithmic negativity) $(E_{N})$ which is also a promising candidate to quantify entanglement for mixed states. In the introduction, we have briefly discussed the concept of entanglement negativity and its significance in context of quantum information theory. However, to proceed further we now qualitatively discuss how one can compute entanglement negativity holographically.\\
Two different proposals have been suggested by various literatures in this direction.\\
In one proposal, $E_{N}$ is given by the area of an extremal cosmic brane that terminates at the boundary of the entanglement wedge \cite{Kudler-Flam:2018qjo,Kusuki:2019zsp}. This proposal is motivated by the quantum error correcting codes and states that the logarithmic negativity is equivalent to the cross-sectional area of the entanglement wedge with a bulk correction term. However, for a general entangling surface this is difficult to compute due to the backreaction of the cosmic brane. This calculation simplifies a lot for a ball shaped subregion. In this set up, the backreaction is accounted for by an overall constant to the area of the entanglement wedge cross-section. Then it is conjectured that \cite{Kudler-Flam:2018qjo,Kusuki:2019zsp,Blanco:2013joa}
\begin{eqnarray}
	E_{N}&=&\chi_{d}\frac{E_{W}}{4G_{N}}+E_{bulk}
\end{eqnarray}
where $E_{W}$ is the minimal cross-sectional area of the entanglement wedge associated with the concerned boundary region and $\chi_{d}$ is a constant which depends on the dimension of the spacetime. $E_{bulk}$ is the quantum correction term corresponding to the logarithmic negativity between the bulk fields on either sides of the entanglement wedge cross-section.\\
Another proposal suggests that the entanglement negativity is given by certain combinations of co-dimension-two static minimal bulk surfaces \cite{Chaturvedi:2016rft,Chaturvedi:2016rcn,Jain:2017xsu,Jain:2017uhe,Malvimat:2018izs,Malvimat:2018cfe}. Both of these proposals reproduce the exact known result of entanglement negativity in CFT. In this paper, we follow the second proposal where entanglement negativity is given by the a certain combination of the area of co-dimension-two static minimal surfaces in the bulk. These combinations can be obtained from the dual CFT correlators. Some recent works in this directions can be found in \cite{Rogerson:2022yim,Matsumura:2022ide,Bertini:2022fnr,Roik:2022gbb,Dong:2021oad,Bhattacharya:2021dnd,Hejazi:2021yhz,Afrasiar:2021hld,Basu:2022nds}.\\ 
We noew compute $E_N$ by considering two different set ups, namely, for two adjacent strip like subsystems and for two disjoint strip like subsystems. At first we consider adjacent scenario. Let us consider two strip like subsystems $A$ and $B$ with lengths $l_{1}$ and $l_{2}$ with zero-overlapping. In the case of such adjacent subsystems the entanglement negativity ($E_N$) is defined as \cite{Chaturvedi:2016rft,Chaturvedi:2016rcn,Jain:2017xsu,Jain:2017uhe,Malvimat:2018izs,Malvimat:2018cfe}
\begin{equation}
E^{\parallel}_{N_{adj}}=\frac{3}{4}\left[S_{EE}^{\parallel}(l_{1})+S_{EE}^{\parallel}(l_{2})-S_{EE}^{\parallel}(l_{1}+l_{2})\right]~.
\end{equation}
By using eq.(\ref{sfin}), one can find the entanglement negativity for adjacent subsystems is  
\begin{eqnarray}\label{eepara}
E^{\parallel}_{N_{adj}}&=&\frac{3V_{d-2}a_{0}}{8G_{d+1}}\Bigg[(2b_{0})^{d-2}\left[\frac{1}{l_{1}^{d-2}}+\frac{1}{l_{2}^{d-2}}-\frac{1}{(l_{1}+l_{2})^{d-2}}\right]\nonumber\\
&+&\frac{1}{2z_{0}^{d}(2b_{0})^{2}}\frac{b_{1}}{a_{0}}\left(\frac{d-1}{2}+\beta^{2}\gamma^{2}\right)\left[l_{1}^{2}+l_{2}^{2}-(l_{1}+l_{2})^{2}\right]\Bigg]+\frac{3}{4}S_{div}~~.
\end{eqnarray}
For pure $AdS_{d+1}$ background, the entanglement negativity for two adjacent subsystem can be obtained from the above expression by taking the limit $\beta\rightarrow0$, $z_0\rightarrow\infty$. This reads
\begin{eqnarray}\label{eepure}
E^{pure}_{N_{adj}}=\frac{3V_{d-2}a_{0}(2b_{0})^{d-2}}{8G_{d+1}}\left[\frac{1}{l_{1}^{d-2}}+\frac{1}{l_{2}^{d-2}}-\frac{1}{(l_{1}+l_{2})^{d-2}}\right]+\frac{3}{4}S_{div}~~.
\end{eqnarray} 
 One can take note from eq.(s)(\ref{eepara},\ref{eepure}) that the divergent piece of the entanglement entropy contributes to the entanglement negativity. We now compute the change in $E_N$ due to the inclusion of boost, this we define as
\begin{eqnarray}\label{newex6}
\delta E^{\parallel}_{N_{adj}}\equiv E^{\parallel}_{N_{adj}}-E^{pure}_{N_{adj}}=\frac{3V_{d-2}b_{1}}{16z_{0}^{d}G_{d+1}(2b_{0})^{2}}\left[\frac{d-1}{2}+\beta^{2}\gamma^{2}\right]\left(l_{1}^{2}+l_{2}^{2}-(l_{1}+l_{2})^{2}\right)~.
\end{eqnarray}
Furthermore, if we consider the subsystems to be of same length, that is, $l_{1}=l_{2}=l$, we then obtain the following expression of $\delta E^{\parallel}_{N_{adjacent}}$
\begin{eqnarray}\label{newex}
	\delta E^{\parallel}_{N_{adj}}|_{l_{1}=l_{2}}= - \frac{3V_{d-2}b_{1}l^2}{8z_{0}^{d}G_{d+1}(2b_{0})^{2}}\left[\frac{d-1}{2}+\beta^{2}\gamma^{2}\right]~.
\end{eqnarray}
As we have mentioned earlier, we now consider two disjoint strip like subsystems $A$ and $B$ with length $l_{1}$ and $l_{2}$. The subsystems under consideration are separated by a length $x$. In this case the entanglement negativity reads \cite{Malvimat:2018ood,KumarBasak:2020viv}
\begin{eqnarray}
E^{\parallel}_{N_{dis}}=\frac{3}{4}\left[S_{EE}^{\parallel}(l_{1}+x)+S_{EE}^{\parallel}(l_{2}+x)-S_{EE}^{\parallel}(l_{1}+l_{2}+x)-S_{EE}^{\parallel}(x)\right]~~.
\end{eqnarray}
We now make use of eq.(\ref{sfin}) to find the explicit expression for $(E_{N})$ in the disjoint set up. This is obtained to be 
\begin{eqnarray}\label{disjointBBB}
E^{\parallel}_{N_{dis}}&=&\frac{3V_{d-2}a_{0}}{8G_{d+1}}\Bigg[(2b_{0})^{d-2}\left(\frac{1}{(l_{1}+x)^{d-2}}+\frac{1}{(l_{2}+x)^{d-2}}-\frac{1}{(l_{1}+l_{2}+x)^{d-2}}-\frac{1}{x^{d-2}}\right)\nonumber\\
&+&\frac{b_{1}/a_{0}}{2z_{0}^{d}(2b_{0})^{2}}\left(\frac{d-1}{2}+\beta^{2}\gamma^{2}\right)\Bigg((l_{1}+x)^{2}+(l_{2}+x)^{2}-(l_{1}+l_{2}+x)^{2}-x^{2}\Bigg)\Bigg]~.
\end{eqnarray}
Similar to the adjacent case, in the limit $\beta \rightarrow0$ and $z_0\rightarrow\infty$ of the above expression, we get the the pure $AdS_{d+1}$ result. This reads
\begin{eqnarray}\label{disjointPure}
E^{pure}_{N_{dis}}&=&\frac{3V_{d-2}a_{0}}{8G_{d+1}}(2b_{0})^{d-2}\left(\frac{1}{(l_{1}+x)^{d-2}}+\frac{1}{(l_{2}+x)^{d-2}}-\frac{1}{(l_{1}+l_{2}+x)^{d-2}}-\frac{1}{x^{d-2}}\right)~.
\end{eqnarray}
Furthermore, it is to be noted that in the disjoint set up, the entanglement negativity does not receive any contribution from the divergent piece. On the other hand, the change in entanglement negativity (disjoint case) due to the boost can be defined as
\begin{eqnarray}
\delta E^{\parallel}_{N_{dis}}\equiv E^{\parallel}_{N_{dis}}-E^{pure}_{N_{dis}}=\frac{3V_{d-2}b_{1}}{16z_{0}^{d}(2b_{0})^{2}}\left(\frac{d-1}{2}+\beta^{2}\gamma^{2}\right)\Bigg((l_{1}+x)^{2}+(l_{2}+x)^{2}-(l_{1}+l_{2}+x)^{2}-x^{2}\Bigg)~.
\end{eqnarray}
So far we have considered the subsystems have lengths $l_1$ and $l_2$. We now consider a special scenario where the lengths of the two strip like subsystems $A$ and $B$ are equal, that is, $l_{1}=l_{2}=l$. At this condition, the expressions given in eq.(s)(\ref{disjointBBB},\ref{disjointPure}) get simplified to the following forms 
\begin{eqnarray}
\bar{E}^{para}_{N_{dis}}|_{l_{1}=l_{2}}&=&\frac{3}{4}\Bigg[a_{0}(2b_{0})^{d-2}\left(\frac{2}{(l+x)^{d-2}}-\frac{1}{(2l+x)^{d-2}}-\frac{1}{x^{d-2}}\right)\nonumber\\
&+&\frac{b_{1}/a_{0}}{2z_{0}^{d}(2b_{0})^{2}}\left(\frac{d-1}{2}+\beta^{2}\gamma^{2}\right)\Bigg(2(l+x)^{2}-(2l+x)^{2}-x^{2}\Bigg)\Bigg]\label{enpara}\\
\bar{E}^{pure}_{N_{dis}}|_{l_{1}=l_{2}}&=&\frac{3a_{0}}{4}(2b_{0})^{d-2}\left(\frac{2}{(l+x)^{d-2}}-\frac{1}{(2l+x)^{d-2}}-\frac{1}{x^{d-2}}\right)~.\label{enpuer}
\end{eqnarray}
where we have used the scaling $\bar{E}_{N}=\frac{2G_{d+1}}{V_{d-2}}E_{N}$. On the similar note, the change in entanglement negativity with the condition $l_{1}=l_{2}=l$ reads 
\begin{eqnarray}\label{aa}
\delta E^{\parallel}_{N_{dis}}|_{l_{1}=l_{2}}&=&\frac{3V_{d-2}b_{1}}{16z_{0}^{d}(2b_{0})^{2}G_{d+1}}\left(\frac{d-1}{2}+\beta^{2}\gamma^{2}\right)\Bigg(2(l+x)^{2}-(2l+x)^{2}-x^{2}\Bigg)~.
\end{eqnarray}
It is also reassuring to observe that if we consider the adjacent subsystem limit $x\rightarrow0$ in eq.\eqref{aa}, we obtain eq.\eqref{newex}. On the other hand, if we compare the expression given in eq.\eqref{newex} with eq.\eqref{cspara}, we obtain the following relation
\begin{eqnarray}\label{newex8}
\delta E^{\parallel}_{N_{adj}}|_{l_{1}=l_{2}}= \lambda_2~ \delta S^{\parallel}_{EE}~.	
\end{eqnarray}
The above relation suggests that the change in entanglement negativity due to the presence of boost is proportional to the change in HEE upto a constant factor $\lambda_2$\footnote{The explicit expression of $\lambda_2$ has been given in Appendix B.}. This has also been observed in case of EWCS.\\
\begin{figure}
	\centering
	\includegraphics[width=0.5\textwidth]{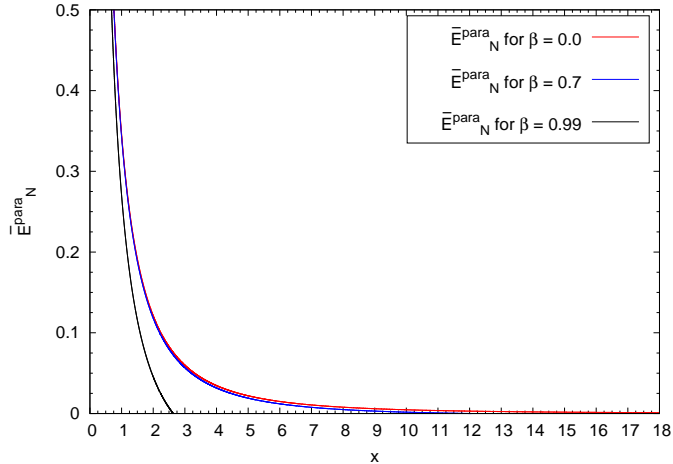}
	\caption{In the above figure, we have showed the variation of  entanglement negativity for two disjoint subsystems (with length $l=3$) with the separation distance between the two subsystems. We have set $d=3$ and $z_0=10$.}
	\label{ENpara}
\end{figure}
Now in order to understand the effect of the boost parameter on the entanglement negativity, we introduce a simple plot in fig.(\eqref{ENpara}). We plot the expression of $E_N$ which has been obtained for two disjoint subsystems with equal lengths $l$ (given in eq.(\ref{enpara})). We consider different values of the boost parameter $\beta$. We find that for $\beta\rightarrow0$ and $z_{0}\rightarrow \infty$, the entanglement negativity does not vanish at any value of the separation distance between two subsystems. However, for a non-zero, finite value of the boost parameter $\beta$, we observe that entanglement negativity vanishes at a critical separation distance $x_c^{\prime}$. Furthermore the value of this critical separation length $x_c^{\prime}$ decreases with the increase in the value of the boost parameter $\beta$. Thus, one can say that entanglement negativity measures correlation between two disjoint subsystems even when they are not in the connected phase because entanglement negativity vanishes at a larger value of the separtion compared to mutual information and EWCS.\\

\subsection{Holographic subregion complexity}
The basic concept of circuit complexity has already been discussed earlier. Now to proceed further, we shall describe various holographic proposals to compute the quantum complexity of the boundary states by using the bulk AdS geometry. There are various different proposals to compute the complexity holographically. Initially, it was suggested that the complexity of the boundary state is related to the volume of the Einstein-Rosen bridge (ERB) connecting the boundaries of a two-sided eternal black hole \cite{Maldacena:2013xja,Susskind:2014moa,Susskind:2018pmk}. This has been dubbed as the `Complexity = Volume' (CV) conjecture. This conjecture gives the following relation
 \begin{eqnarray}\label{CV1}
\mathcal{C}_V(t_L,t_R)=\frac{V^{ERB}(t_L,t_R)}{8\pi R G_{d+1}}
\end{eqnarray}\\
where $R$ is the AdS radius and $V^{ERB}(t_L,t_R)$ is the co-dimension one extremal volume of the ERB. This is bounded by the two spatial slices at times $t_L$ and $t_R$ of two CFTs, living on the two boundaries of the eternal black hole.\\
Another proposal relates the gravitational action computed in a certain bulk region known as the Wheeler-Dewitt (WdW) patch (which is bounded by the light sheets) to the quantum complexity of the boundary state. This proposal is denoted as the `Complexity = Action' (CA) conjecture \cite{Brown:2015lvg,Brown:2015bva,Cai:2016xho,Goto:2018iay,Alishahiha:2018lfv,Alishahiha:2018tep,PhysRevD.100.086004,Carmi:2017jqz}. Mathematically it can be represented as 
\begin{equation}
\mathcal{C}_{A}=\frac{\mathcal{I}_{WdW}}{\pi\hbar}
\end{equation}
where $\mathcal{I}_{WdW}$ is the gravitational action evaluated on the WdW patch.\footnote{There is also another proposal suggested for the holographic complexity which is known as the `Complexity = Spacetime volume' (CV 2.0) \cite{Couch:2016exn} which states that complexity is related to the spacetime volume ($\hat{V}$) of the WdW patch 
$\mathcal{C}_{V 2.0}=\frac{\hat{V}}{GR^{2}}$.}\\
These conjectures depend on the whole state of the boundary system. However, there is another conjecture which depends on the reduced state of the physical system at the boundary. This is known as the holographic subregion complexity (HSC) conjecture. It relates co-dimension one volume ($V(\gamma)$) enclosed by the extremal RT surface to the complexity \cite{Alishahiha:2015rta}
\begin{eqnarray}\label{HSC}
\mathcal{C}_V=\frac{V(\gamma)}{8\pi R G_{d+1}}~.
\end{eqnarray}
Some interesting works related to the holographic computation of complexity can be found in \cite{Emparan:2021hyr,Bhattacharyya:2022ren,Alishahiha:2021thv,Alishahiha:2019cib,Engelhardt:2021mju,Mounim:2021bba,Ghanbarian:2020cdv,Borvayeh:2020yip,HosseiniMansoori:2017tsm,Akhavan:2019zax,Omidi:2020oit,Hashemi:2019aop,Doroudiani:2019llj,Akhavan:2018wla,Alishahiha:2017hwg}. Now in this set up (where the subsystem is taken to be along the direction of the boost), if $V(\gamma)^{\parallel}$ denotes the maximal
co-dimension-one volume enclosed by the co-dimension-two static,
minimal Ryu-Takayanagi (RT) surface in the bulk, then
the holographic subregion complexity (HSC) reads
\begin{equation}\label{cparad}
\mathcal{C}_{V}^{\parallel}=\frac{V(\gamma)^{\parallel}}{8\pi R G_{d+1}}~.
\end{equation}
It is to be kept in mind that, in this case where the subsystem is placed parallel to the boost direction, the volume of the corresponding RT surface can be obtained by using eq.(\ref{bbm}) and eq.(\ref{diffe}). This in turn leads to the following \cite{Karar:2019bwy}
\begin{eqnarray}\label{cpara}
\mathcal{C}_{V}^{\parallel}&=&\frac{V_{d-2}}{8\pi G_{d+1}(d-1)}\Bigg[\frac{l}{\epsilon^{d-1}}-\frac{2^{d-2}\pi^{\frac{d-1}{2}}}{(d-1)}\left(\frac{\Gamma(\frac{d}{2d-2})}{\Gamma(\frac{1}{(2d-2)})}\right)^{d-3}\frac{1}{l^{d-2}}\nonumber\\
&-&\frac{l^{2}}{4b^{2}_{0}z_{0}^{d}}\Bigg[\Bigg(\frac{(d-2)\pi b_{1}}{2(d-1)b_{0}^{2}}+(2-d)c_{0}\Bigg)
+\beta^{2}\gamma^{2}\Bigg(\frac{(d-2)\pi}{2(d-1)^{2}b_{0}}\left(\frac{2b_{1}}{b_{0}}-1\right)+c_{2}-c_{0}d\Bigg)\Bigg]\Bigg]~.\nonumber\\
\end{eqnarray}
where we have set AdS radius $R=1$. Similarly the holographic subregion complexity for a subsystem of length $l$ for pure $AdS_{d+1}$ geometry reads 
\begin{eqnarray}
\mathcal{C}_{V}^{pure}=\frac{V^{pure}}{8\pi G_{d+1}}=\frac{1}{8\pi G_{d+1}}\Bigg[\frac{V_{d-2}}{d-1}\frac{l}{\epsilon^{d-1}}-\frac{2^{d-2}\pi^{\frac{d-1}{2}}}{(d-1)^{2}}\left(\frac{\Gamma(\frac{d}{2d-2})}{\Gamma(\frac{1}{(2d-2)})}\right)^{d-3}\frac{V_{d-2}}{l^{d-2}}\Bigg]~.
\end{eqnarray}
The above can be obtained by taking the limits $\beta\rightarrow0$ and $z_0\rightarrow\infty$ of eq.(\ref{cpara}). We compute the change in complexity due to the presence of the boost parameter. This we define as $\delta \mathcal{C}_{V}^{\parallel}\equiv \mathcal{C}_{V}^{\parallel}-\mathcal{C}_{V}^{pure}$. By substituting the corresponding expressions, we get   
\begin{eqnarray}\label{dcv}
\delta C_{V}^{\parallel}&=&-\frac{1}{8\pi G_{d+1}}\frac{V_{d-2}l^{2}}{4b^{2}_{0}z_{0}^{d}(d-1)}\Bigg[\Bigg(\frac{(d-2)\pi b_{1}}{2(d-1)b_{0}^{2}}+(2-d)c_{0}\Bigg)
+\beta^{2}\gamma^{2}\Bigg(\frac{(d-2)\pi}{2(d-1)^{2}b_{0}}\left(\frac{2b_{1}}{b_{0}}-1\right)+c_{2}-c_{0}d\Bigg)\Bigg]~.\nonumber\\
\end{eqnarray}
\subsection{Mutual complexity}
\begin{figure}[!h]
	\centering
	\includegraphics[width=0.5\textwidth]{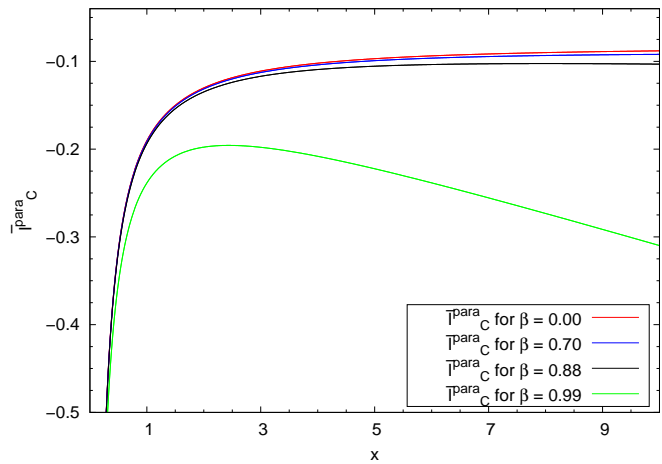}
	\caption{Variation of the mutual complexity with respect to the separation distance $x$ between the two subsystems. We have set $l=3$, $d=3$ and $z_0=10$.}
	\label{mcpara}
\end{figure}

In order to find the complexity for mixed states, we follow the purification complexity (CoP) protocol. It is defined as the minimal complexity among all possible purifications of the mixed state. This implies optimization over the circuits which take the reference state (unentangled product state) to a certain target state $\ket{\psi_{AB}}$ (which is a purification of the desired mixed state $\rho_A$) and also optimization over the possible purifications of $\rho_A$. This is expressed as
$\mathcal{C}(\rho_{A})= \mathrm{min}_{B}~ \mathcal{C}(\ket{\psi_{AB}});~ \rho_A = \mathrm{Tr}_B \ket{\psi_{AB}}\bra{\psi_{AB}}$
where $A^c=B$. In recent times, a quantity known as the `mutual complexity' has been defined in order to compute the above mentioned mixed state complexity \cite{PhysRevD.99.086016,Alishahiha:2018lfv,Caceres:2019pgf,Ghodrati:2019hnn}. One starts with a state $\rho_{AB}$ in the extended Hilbert space (original Hilbert space with auxiliary degrees of freedom), then by tracing out the degrees of freedom of $B$, one obtains $\rho_{A}$. Similarly, by tracing out the degrees of freedom of $A$ leads to $\rho_B$. These computed results along with the following formula leads to the mutual complexity \cite{Alishahiha:2018lfv}
\begin{eqnarray}\label{26}
\mathcal{C} = \mathcal{C}(\rho_{A}) +\mathcal{C}(\rho_{B})-\mathcal{C}(\rho_{A\cup B})~.
\end{eqnarray}
Now we will calculate the mutual complexity by using HSC conjecture. We consider two subsystems $A$ and $B$ of equal length $l$ on the boundary Cauchy slice, separated by a distance $x$. For $\frac{x}{l}<1$, the mutual complexity is given by \cite{Alishahiha:2018lfv,PhysRevD.99.086016,Caceres:2019pgf,Agon:2018zso,Ruan:2021wep,Saha:2021kwq}
\begin{eqnarray}\label{hmc}
\Delta\mathcal{C}^{\parallel}&=&\mathcal{C}_{V}^{\parallel}(A)+\mathcal{C}_{V}^{\parallel}(B)-\mathcal{C}_{V}^{\parallel}(A\cup B)\nonumber\\ &=&2\mathcal{C}_{V}^{\parallel}(l)-\mathcal{C}_{V}^{\parallel}(2l+x)+\mathcal{C}_{V}^{\parallel}(x)
\end{eqnarray}
where in the last line we have used $C_{V}^{\parallel}(A\cup B)=C_{V}^{\parallel}(2l+x)-C_{V}^{\parallel}(x)$, for $\frac{x}{l}<1$. This relation is valid in general for two subsystems of equal length and with separation $x$ \cite{Ghodrati:2019hnn}. Now by using eq.(\ref{cpara}), we find the mutual complexity as
\begin{eqnarray}\label{mcpa}
\Delta\mathcal{C}^{\parallel}&=&\frac{V_{d-2}}{8\pi G_{d+1}(d-1)}\Bigg[\frac{2^{d-2}\pi^{\frac{d-1}{2}}}{(d-1)}\left(\frac{\Gamma(\frac{d}{2d-2})}{\Gamma(\frac{1}{(2d-2)})}\right)^{d-3}\Bigg(\frac{1}{(2l+x)^{d-2}}-\frac{1}{x^{d-2}}-\frac{2}{l^{d-2}}\Bigg)\nonumber\\
&-&\frac{1}{4b^{2}_{0}z_{0}^{d}}\Bigg[\Bigg(\frac{(d-2)\pi b_{1}}{2(d-1)b_{0}^{2}}+(2-d)c_{0}\Bigg)+\beta^{2}\gamma^{2}\Bigg(\frac{(d-2)\pi}{2(d-1)^{2}b_{0}}\left(\frac{2b_{1}}{b_{0}}-1\right)+c_{2}-c_{0}d\Bigg)\Bigg]\left(2l^{2}+x^{2}-(2l+x)^{2}\right)\Bigg]~\nonumber\\
\end{eqnarray}
where the expressions corresponding to the constant terms $c_0$, $c_1$ and $c_2$ are given in the appendix. On the other hand, the mutual complexity for pure $AdS_{d+1}$ background is obtained to be 
\begin{eqnarray}\label{hmcp}
\Delta\mathcal{C}^{pure}&=&\frac{V_{d-2}}{8\pi G_{d+1}(d-1)}\frac{2^{d-2}\pi^{\frac{d-1}{2}}}{(d-1)}\left(\frac{\Gamma(\frac{d}{2d-2})}{\Gamma(\frac{1}{(2d-2)})}\right)^{d-3}\Bigg(\frac{1}{(2l+x)^{d-2}}-\frac{1}{x^{d-2}}-\frac{2}{l^{d-2}}\Bigg)~.
\end{eqnarray}
With the above expressions in hand, we can compute the change in mutual complexity, that is, $\delta\mathcal{C}^{\parallel}\equiv \Delta\mathcal{C}^{\parallel}-\Delta\mathcal{C}^{pure}$ as
\begin{eqnarray}\label{bbb}
\delta\mathcal{C}^{\parallel}&=&-\frac{V_{d-2}}{8\pi G_{d+1}(d-1)}\frac{1}{4b^{2}_{0}z_{0}^{d}}\Bigg[\Bigg(\frac{(d-2)\pi b_{1}}{2(d-1)b_{0}^{2}}+(2-d)c_{0}\Bigg)\nonumber\\
&+&\beta^{2}\gamma^{2}\Bigg(\frac{(d-2)\pi}{2(d-1)^{2}b_{0}}\left(\frac{2b_{1}}{b_{0}}-1\right)+c_{2}-c_{0}d\Bigg)\Bigg]\left(2l^{2}+x^{2}-(2l+x)^{2}\right)~~.
\end{eqnarray}
Similar to the previously computed quantities, we scale the mutual complexity in the following way
\begin{eqnarray}
\bar{I}^{para}_{C}=\frac{2G_{d+1}}{V_{d-2}}\Delta\mathcal{C}^{\parallel}~\mathrm{and}~\bar{I}^{pure}_{C}=\frac{2G_{d+1}}{V_{d-2}}\Delta\mathcal{C}^{pure}~.
\end{eqnarray}
With these expressions in hand, we proceed to represent our results graphically. It has been noted that there can be two different natures of the mutual complexity, that is, $\Delta\mathcal{C}$ is said to be subadditive if $\Delta\mathcal{C}>0$ and superadditive if $\Delta\mathcal{C}<0$. In order to observe this for the case in hand, we take the help of fig.(\ref{mcpara}). We observe that for every chosen value of the boost parameter $\beta$, the mutual complexity is super-additive Furthermore, the effect of the boost parameter on the mutual complexity can be noted firmly from the mentioned figure.
\section{Strip perpendicular to the boost}\label{sec4}
In this section we consider a strip like subsystem having orientation perpendicular to the boost direction and compute various information theoretic measures as we have done in the previous section. Further, we assume that the subsystem is placed along the $x_{1}$ direction while boost is along the $y$ direction. The subsystem is specified by the $-\frac{l}{2}\le x_{1}\le\frac{l}{2}$, $0\le y\le l_{y}$, and $0\le x_{i} \le l_{i}$. $l_{y}$ and $l_{i}$ is taken to be very large, that is $l_{y},l_{i}>>l$. Furthermore we assume that that the lengths along the directions other than $x_1$ are fixed. We now choose the parametrisation $z=z(x_{1})$ and \textbf{$t=t(x_{1})$} to compute the the area of the co-dimension two static minimal surface $\Gamma_A^{min}$. Similar to the previous section, we have taken the strip to be thin.
\subsection{Holographic entanglement entropy}
We now again use the HRT prescription to obtain the HEE of a strip perpendicular to the boost 
\begin{eqnarray}\label{hrtper}
	S_{EE}^{\perp}&=&\frac{V_{d-2}}{2G_{d+1}}\int_{-\frac{l}{2}}^{0}\frac{dx_{1}}{z^{d-1}}\sqrt{K(z)}\left[1+\frac{z^{\prime2}}{f(z)}-t^{\prime2}\left(\frac{f(z)}{K(z)}-\frac{K(z)}{\beta^{2}}\left(1-\frac{1}{K(z)}\right)^{2}\right)\right]^{\frac{1}{2}}~.
\end{eqnarray}
Identifying the Lagrangian $\mathcal{L}=\mathcal{L}(z,z^{\prime},t,t^{\prime})$ from the above expression, we note that $x_1$ is a cyclic coordinate. This results in the following conserved quantity
\begin{eqnarray}\label{newex1}
	\mathcal{H}=-\frac{\sqrt{K(z)}}{z^{d-1}\sqrt{1+\frac{z^{\prime2}}{f(z)}-t^{\prime2}\left(\frac{f(z)}{K(z)}-\frac{K(z)}{\beta^{2}}\left(1-\frac{1}{K(z)}\right)^{2}\right)}}=constant=c~.
\end{eqnarray}
Similar to the parallel case, we introduce the turning point $(z_{t}^{\perp},t_{t}^{\perp})$ inside the bulk at which $z^{\prime}$ and $t^{\prime}$ vanishes. This fixes the above constant to be $c=-\frac{\sqrt{K(z_{t}^{\perp})}}{z^{\perp(d-1)}_{t}}$. By using this value of the constant $c$ in eq.\eqref{newex1}, we obtain the following equation
\begin{eqnarray}\label{perhrt}
	1+\frac{z^{\prime2}}{f(z)}-t^{\prime2}\left(\frac{f(z)}{K(z)}-\frac{K(z)}{\beta^{2}}\left(1-\frac{1}{K(z)}\right)^{2}\right)=\frac{K(z)}{K(z_{t}^{\perp})}\left(\frac{z_{t}^{\perp}}{z}\right)^{2d-2}~~.
\end{eqnarray}
Using this above expression in eq.(\ref{hrtper}), the entanglement entropy of the strip is given by
\begin{eqnarray}\label{HRT}
	S_{EE}^{\perp}&=&\frac{V_{d-2}}{2G_{d+1}}\int_{-\frac{l}{2}}^{0}\frac{dx_{1}}{z^{d-1}}\frac{K(z)}{\sqrt{K(z_{t}^{\perp})}}\left(\frac{z_{t}^{\perp}}{z}\right)^{d-1}~.
\end{eqnarray}
As we have mentioned in the parallel scenario, the above expression can also be obtained by following the RT prescription. To proceed further, we now write the boundary coordinate in terms of the bulk coordinate. This can be done by solving eq.(\ref{perhrt}). From eq.(\ref{perhrt}) we can write $dx_{1}$ in terms of $dz$ in the following way
\begin{equation}
	dx=\frac{dz(z/z_{t}^{\perp})^{d-1}\left[1-\frac{t^{\prime2}(z/z_{t}^{\perp})^{2d-2}\left(\frac{f(z)}{K(z)}-\frac{K(z)}{\beta^{2}}\left(1-\frac{1}{K(z)}\right)^{2}\right)}{\left[\frac{K(z)}{K(z_{t}^{\perp})}-\left(\frac{z}{z_{t}^{\perp}}\right)^{2d-2}\right]}\right]^{\frac{1}{2}}}{\sqrt{f(z)}\left[\frac{K(z)}{K(z_{t}^{\perp})}-\left(\frac{z}{z_{t}^{\perp}}\right)^{2d-2}\right]^{\frac{1}{2}}}~~.
\end{equation}
Again we consider the thin strip approximation, that is, in the above expression we will keep terms upto $\mathcal{O}\left((z/z_{t}^{\perp})^{d}\right)$ and neglect the higher order terms. Using this fact we recast the above expression in the following way
\begin{eqnarray}
	dx_{1}=\frac{dz\left(\frac{z}{z_{t}^{\perp}}\right)^{d-1}}{\sqrt{f(z)}\left[\frac{K(z)}{K(z_{t}^{\perp})}-\left(\frac{z}{z_{t}^{\perp}}\right)^{2d-2}\right]^{\frac{1}{2}}}~~.
\end{eqnarray}
The HEE is therefore obtained to be \cite{Mishra:2015cpa,Mishra:2016yor}
\begin{equation}
	S_{EE}^{\perp}=\frac{V_{d-2}}{2G_{d+1}}\int_{\epsilon}^{z_{t}^{\perp}}\frac{dz}{z^{d-1}\sqrt{f(z)}\left[1-\left(\frac{z}{z_{t}^{\perp}}\right)^{2d-2}\frac{K(z)}{K(z_{t}^{\perp})}\right]^{\frac{1}{2}}}~.
\end{equation}
Using the above expression, the HEE is obtained to be \cite{Mishra:2015cpa,Mishra:2016yor}
\begin{eqnarray}\label{sper}
S_{EE}^{\perp}&=&\frac{Area(\Gamma^{min}_{A})_{\perp}}{4G_{(d+1)}}=S_{div}+\frac{V_{d-2}a_{0}}{z_{t}^{\perp(d-2)}}\Bigg(1+\left(\frac{p^{\prime d}+q^{\prime d}}{2}\right)\frac{a_{1}}{a_{0}}+\frac{q^{\prime d}}{2}\Bigg[\frac{d+1}{d-1}\frac{b_{1}}{a_{0}}-\frac{1}{d-1}\frac{b_{0}}{a_{0}}\Bigg]\Bigg)
\end{eqnarray}
where $G_{d+1}$ is the $(d+1)$-dimensional Newton's constant and $V_{d-2}=l_{y}l_{2}...l_{d-2}$ is the spatial volume of the boundary and $p^{\prime d}=\left(\frac{z_{t}^{\perp}}{z_{0}}\right)^{d}~,~q^{\prime d}=\beta^{2}\gamma^{2}\left(\frac{z_{t}^{\perp}}{z_{0}}\right)^{d}~$.  $S_{div}$ represents the subsystem independent divergent term $S_{div}=\frac{V_{d-2}}{2G_{d+1}(d-2)}\frac{1}{\epsilon^{d-2}}$. In obtaining the above expression, we use the thin strip approximation (eq.(\ref{thinS})) as we have done in previous section. This reads 
\begin{eqnarray}\label{thS}
\frac{K(z_{t}^{\perp})}{K(z)}\approx1+\beta^{2}\gamma^{2}\left[1-\left(\frac{z}{z_{0}}\right)^{d}\right]\left(\frac{z_{t}^{\perp}}{z_{0}}\right)^{d}~.
\end{eqnarray}
On the other hand, under the thin strip approximation, the relationship between the subsystem size $l$ and the turning point $z_{t}^{\perp}$ stands to be \cite{Mishra:2015cpa,Mishra:2016yor}
\begin{eqnarray}
l\approx 2z_{t}^{\perp}\left[b_{0}+\frac{1}{2}\left(\frac{z_{t}^{\perp}}{z_{0}}\right)^{d}\left(b_{1}+\beta^{2}\gamma^{2}I_{l}\right)\right]~~.
\end{eqnarray}
On the other hand, the relationship between $z_{t}^{\perp}$ and $\bar{z_{t}}$ is obtained to be
\begin{eqnarray}\label{lper}
z_{t}^{\perp}&=&\frac{l/2}{b_{0}+\frac{1}{2}\left(\frac{z_{t}^{\perp}}{z_{0}}\right)^{d}\left[b_{1}+\beta^{2}\gamma^{2}I_{l}\right]}\approx\frac{\bar{z_{t}}}{1+\frac{1}{2}\left(\frac{\bar{z_{t}}}{z_{0}}\right)^{d}\Bigg(\frac{b_{1}}{b_{0}}+\frac{\beta^{2}\gamma^{2}}{b_{0}}I_{l}\Bigg)}~.
\end{eqnarray}
Now the HEE in terms of the subsystem size can be written down as \cite{Mishra:2015cpa,Mishra:2016yor}
\begin{eqnarray}\label{seper}
S_{EE}^{\perp}=S_{div}+\frac{V_{d-2}a_{0}}{2G_{d+1}}\left(\frac{2b_{0}}{l}\right)^{d-2}+\frac{V_{d-2}b_{1}(d+1)l^{2}}{32G_{d+1}b_{0}^{2}z_{0}^{d}}\Bigg(\frac{d-1}{d+1}+\beta^{2}\gamma^{2}\Bigg)~.
\end{eqnarray}
With the help of eq.(\ref{seper}) and eq.(\ref{spure}), we can compute the change in HEE due to the presence of boost. This reads
\begin{equation}\label{csper}
\delta S_{EE}^{\perp}\equiv S^{\perp}_{EE}-S^{pure}_{EE}=\frac{V_{d-2}b_{1}(d+1)l^{2}}{32G_{d+1}b_{0}^{2}z_{0}^{d}}\Bigg(\frac{d-1}{d+1}+\beta^{2}\gamma^{2}\Bigg)~~.
\end{equation}
Furthermore, the above expression and eq.(\ref{cspara}) leads to the following relation between $\delta S_{EE}^{\perp}$ and $\delta S_{EE}^{\parallel}$ 
\begin{equation}\label{newex4}
\delta S_{EE}^{\parallel}=\left[\frac{1+\frac{2}{d-1}\beta^{2}\gamma^{2}}{1+\frac{d+1}{d-1}\beta^{2}\gamma^{2}}\right]\delta S_{EE}^{\perp}~.
\end{equation}
From eq.(\ref{cspara}) and eq.(\ref{csper}), one can note that $\delta S_{EE}^{\perp} \ge \delta S_{EE}^{\parallel}$, that is, the entanglement entropy of the strip lying perpendicular to the boost is larger than the entanglement entropy for the strip which is parallel to the boost. This in turn highlights boost induced anisotropic nature of the boundary theory, as the boost breaks the rotational symmetry in boundary field theory \cite{Bhatta:2019eog}. The reason for this enhanced entanglement entropy in the perpendicular direction compared to the parallel case is due to the pressure asymmetry on the CFT side. The CFT `pressure' is a quantity that is different in the perpendicular and the parallel case. In the parallel case, the pressure is more than the perpendicular case and this difference is solely due to the boost \cite{mishra2018study}. Due to this unequal pressure, the excitations in the CFT side consumes more energy in the parallel strip in comparison to the perpendicular strip. This leads to an enhancement in the entanglement entropy in the perpendicular case compared to the parallel case \cite{Mishra:2015cpa,Mishra:2016yor}. The argument suggests that the CFT `pressure' in the boundary plays a crucial role in determining the entanglement entropy of the subsystems living in the boundary.
\subsection{EWCS and Holographic mutual information}
We now compute the minimal cross-section of the entanglement wedge, where the subsystems are perpendicular to the direction of boost. To do this let us consider two subsystems $A$ and $B$ with equal length $l$ and they are separated by a distance $x$. In the set up, we have to calculate the vertical constant $x_{1}$ hyper-surface with minimal area which splits the domain of the entanglement wedge $M_{AB}$ into two domains corresponding to $A$ and $B$. The time induced metric on this constant $x_{1}$ hyper-surface reads
\begin{eqnarray}
ds^{2}|_{ind}&=&\frac{1}{z^{2}}\Bigg(Kdy^{2}+...+dx_{d-2}^{2}+\frac{dz^{2}}{f(z)}\Bigg)~~.
\end{eqnarray}
The above mentioned induced metric leads to the following expression for EWCS
\begin{eqnarray}\label{iperex}
E_{W}^{\perp}&=&\frac{V_{d-2}}{4G_{d+1}}\int_{z_{t}^{\perp}(x)}^{z_{t}^{\perp}(2l+x)}\frac{dz}{z^{d-1}}\sqrt{\frac{K(z)}{f(z)}}~~.
\end{eqnarray}
Under the thin strip approximation, the explicit expression of $E_{W}^{\perp}$ is found to be
\begin{eqnarray}
E_{W}^{\perp}
&=&\frac{V_{d-2}}{4G_{d+1}(d-2)}\left[\frac{1}{(z_{t}^{\perp}(x))^{d-2}}-\frac{1}{(z_{t}^{\perp}(2l+x))^{d-2}}\right]+\frac{V_{d-2}(1+\beta^{2}\gamma^{2})}{16G_{d+1}z_{0}^{d}}\left[(z_{t}^{\perp}(2l+x))^{2}-(z_{t}^{\perp}(x))^{2}\right]~.\nonumber\\
\end{eqnarray}
By using eq.(\ref{lper}) in the above expression, we get EWCS in terms of the subsystem size $l$ and distance of separation $x$. This reads 
\begin{eqnarray}\label{ewper}
E_{W}^{\perp}&=&\frac{V_{d-2}}{2G_{d+1}}\Bigg[\frac{(2b_{0})^{d-2}}{2(d-2)}\left[\frac{1}{x^{d-2}}-\frac{1}{(2l+x)^{d-2}}\right]+\frac{\left[(2l+x)^{2}-x^{2}\right]}{4z_{0}^{d}(2b_{0})^{2}}\left(\frac{1+\beta^{2}\gamma^{2}}{2}-\left(\frac{b_{1}}{b_{0}}+\frac{\beta^{2}\gamma^{2}I_{l}}{b_{0}}\right)\right)\Bigg]~.~~~~
\end{eqnarray}
Similar to the previous section, we now introduce the following scaling
\begin{eqnarray}\label{ScalingPerpEW}
\bar{E}_{W}^{perp}= \frac{2G_{d+1}}{V_{d-2}} E_{W}^{\perp}~.
\end{eqnarray}
Furthermore, the change in EWCS in this set up stands to be
\begin{eqnarray}\label{cewper}
\delta E_{W}^{\perp}\equiv E_{W}^{\perp}-E_{W}^{pure}=\frac{V_{d-2}}{32G_{d+1}z_{0}^{d}b_{0}^{2}}\left[\frac{1}{2}\left(1+\beta^{2}\gamma^{2}\right)-\left(\frac{b_{1}}{b_{0}}+\beta^{2}\gamma^{2}\frac{I_{l}}{b_{0}}\right)\right]\left[(2l+x)^{2}-x^{2}\right]~~.
\end{eqnarray}
We note that there exists a relation between $\delta E_{W}^{\perp}$ and $\delta E_{W}^{\parallel}$. This can be shown by using the eq(s).(\ref{cewpara},\ref{cewper})
\begin{equation}\label{newex2}
\delta E_{W}^{\perp}=\left[\frac{\frac{1}{2}(1+\beta^{2}\gamma^{2})-\left(\frac{b_{1}}{b_{0}}+\frac{\beta^{2}\gamma^{2}I_{l}}{b_{0}}\right)}{\frac{1}{2}-\left(1+\frac{2\beta^{2}\gamma^{2}}{d-1}\right)\frac{b_{1}}{b_{0}}+\frac{\beta^{2}\gamma^{2}}{d-1}}\right]\delta E_{W}^{\parallel}~.
\end{equation}
Further, if we again consider the adjacent subsystem limit ($x\rightarrow0$), the obtained expression of $\delta E_{W}^{\perp}$ then reads
\begin{eqnarray}
\delta E_{W}^{\perp}=\frac{V_{d-2}l^2}{8G_{d+1}z_{0}^{d}b_{0}^{2}}\left[\frac{1}{2}\left(1+\beta^{2}\gamma^{2}\right)-\left(\frac{b_{1}}{b_{0}}+\beta^{2}\gamma^{2}\frac{I_{l}}{b_{0}}\right)\right]~.
\end{eqnarray}
In this limit (adjacent subsystem limit), the relation given in eq.\eqref{newex2} stands to be true. Now if we compare the above expression with the one given in eq.\eqref{csper}, we get
\begin{eqnarray}\label{newex9}
	\delta E_{W}^{\perp}=\lambda_3 ~\delta S_{EE}^{\perp}~.
\end{eqnarray}
In order to relate $\delta E_{W}^{\perp}$ to the boundary CFT quantities, we recall the following relation \cite{Mishra:2015cpa,Mishra:2016yor}
\begin{eqnarray}\label{per1st1}
	\delta S_{EE}^{\perp}&=&\frac{1}{T_{E}}\left(\Delta E_{\perp}-\frac{d-1}{d+1}\mathcal{V}\Delta P_{\perp}\right)
\end{eqnarray}
where $\Delta E_{\perp}=\Delta E_{\parallel}$, $\Delta P_{\perp}=\frac{1}{z_{0}^{d}}\frac{1}{16\pi G_{d+1}}$. The expression of entanglement temperature $T_{E}$ is given in eq.(\ref{bq}). Now by using the abve relation in eq.\eqref{newex9}, we get
\begin{eqnarray}\label{per1st}
	\delta E_{W}^{\perp}&=&\frac{\lambda_3}{T_{E}}\left(\Delta E_{\perp}-\frac{d-1}{d+1}\mathcal{V}\Delta P_{\perp}\right)~.
\end{eqnarray}


On the other hand, the HMI in this set up is obtained to be
\begin{eqnarray}\label{iper2}
I^{\perp}(A:B)&=&\frac{V_{d-2}}{2G_{d+1}}\Bigg[\frac{2^{d-2}b_{0}^{d-1}}{(d-2)}\left(-\frac{2}{l^{d-2}}+\frac{1}{x^{d-2}}+\frac{1}{(2l+x)^{d-2}}\right)\nonumber\\
&+&\frac{b_{1}(d+1)}{16b_{0}^{2}z_{0}^{d}}\Bigg(\frac{d-1}{d+1}+\beta^{2}\gamma^{2}\Bigg)\left[2l^{2}-x^{2}-(2l+x)^{2}\right]\Bigg]~~.
\end{eqnarray}
Similar to the EWCS, we now compute the change in HMI due to the boost. This reads
\begin{eqnarray}\label{ciper}
\delta I^{\perp}(A:B)=I^{\perp}(A:B)-I^{pure}(A:B)=\frac{V_{d-2}b_{1}(d+1)}{32G_{d+1}b_{0}^{2}z_{0}^{d}}\Bigg(\frac{d-1}{d+1}+\beta^{2}\gamma^{2}\Bigg)\left[2l^{2}-x^{2}-(2l+x)^{2}\right]
\end{eqnarray}
where $I^{pure}(A:B)$ is the HMI with pure $AdS_{d+1}$ geometry in the bulk, given in eq.(\ref{ipure}).
Further, the relation between $\delta I^{\perp}(A:B)$ and $\delta I^{\parallel}(A:B)$ can be obtained by using eq.(\ref{cipara}) and eq.(\ref{ciper})
\begin{eqnarray}
\delta I^{\perp}(A:B)=\left[\frac{\frac{d-1}{d+1}+\beta^{2}\gamma^{2}}{\frac{d-1}{d+1}+\frac{2\beta^{2}\gamma^{2}}{d+1}}\right]\delta I^{\parallel}(A:B)~.
\end{eqnarray} 
We again introduce the following scaling for the sake of simplicity
\begin{eqnarray}\label{HMIPerpScaled}
\bar{I}^{perp}(A:B)=\frac{2G_{d+1}}{V_{d-2}}I^{\perp}(A:B)~.
\end{eqnarray}  
\begin{figure}[!h]
	\begin{minipage}[t]{0.48\textwidth}
		\centering\includegraphics[width=\textwidth]{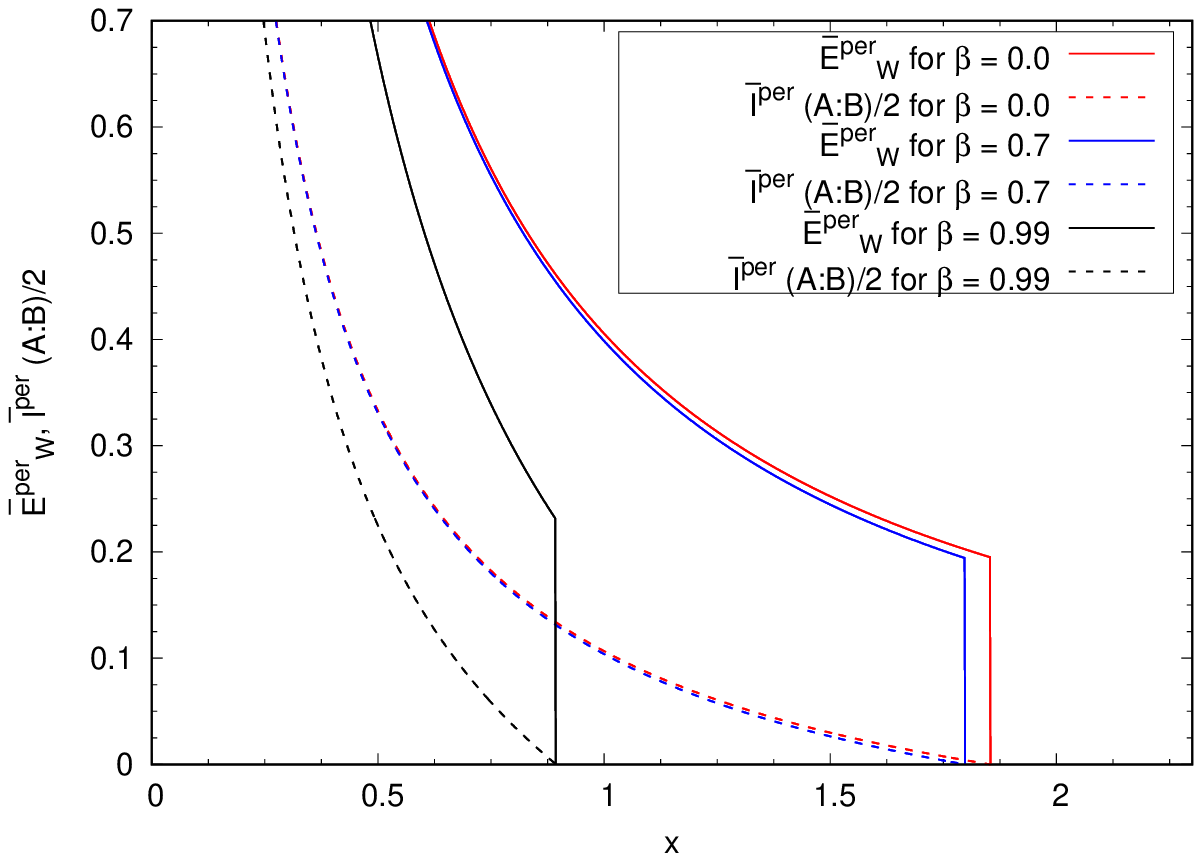}\\
	\end{minipage}\hfill
	\begin{minipage}[t]{0.48\textwidth}
		\centering\includegraphics[width=\textwidth]{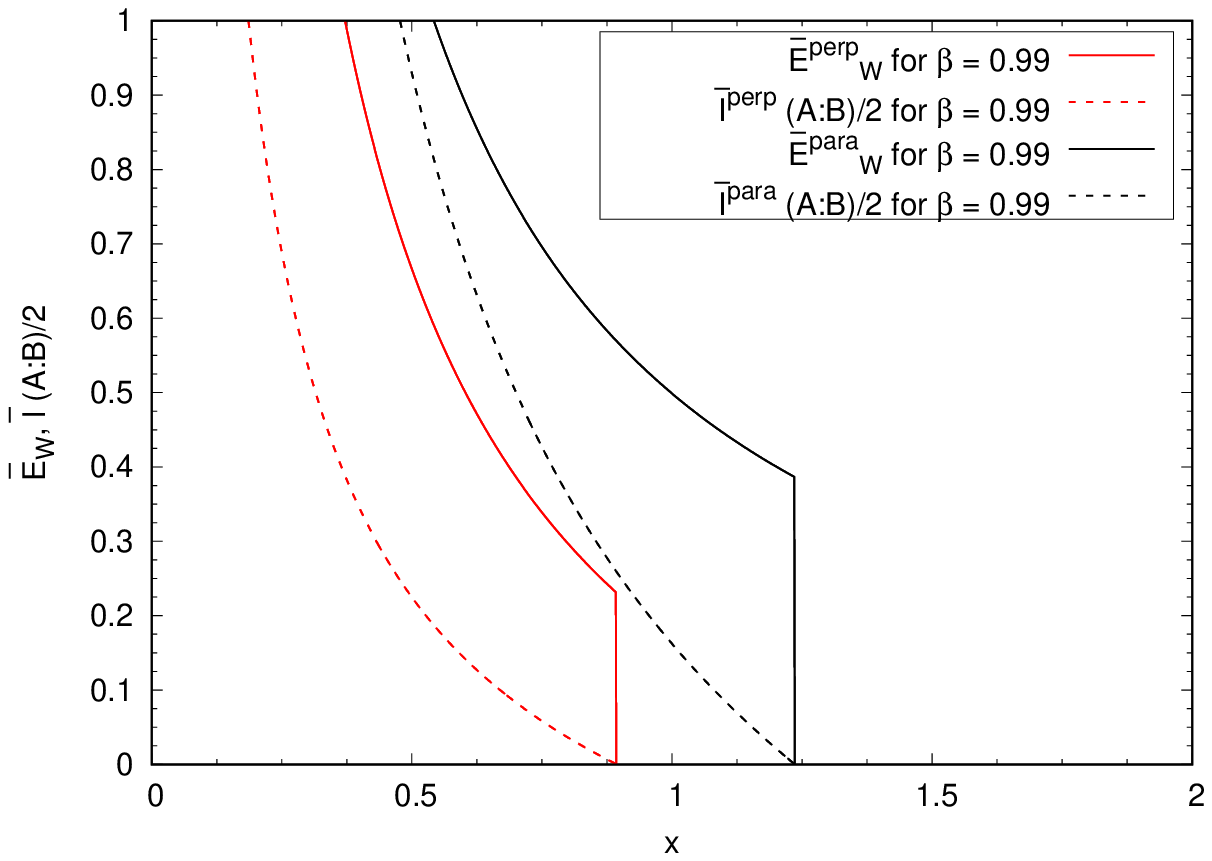}\\
	\end{minipage}
	\caption{In this figure we have plotted EWCS and HMI as a function of separation distance $x$ between two subsystems of fixed length $l$. In the left panel, we have plotted the EWCS and HMI when the subsystems are taken perpendicular to the boost direction. In the right panel, we compare the EWCS and HMI computed in both perpendicular and parallel set up. This we show for $\beta=0.99$. We have set $l=3$, $d=3$ and $z_0=10$.}
	\label{EWCompare}
\end{figure}
Similar to the parallel case, we now graphically represent our results to show the effect of boost parameter $\beta$. This we have done in fig.\eqref{EWCompare}. In the left panel, we have plotted the obtained expressions of EWCS and HMI. These are given in eq.\eqref{ScalingPerpEW} and eq.\eqref{HMIPerpScaled}. We observe that the value of the critical separation length $x_c$ decreases with the increase in the value of the boost parameter. This observation is similar to the parallel case. In the right panel, we qualitatively compare EWCS and HMI computed in both perpendicular and parallel set up. This we show for a particular value of the boost parameter, namely, $\beta=0.99$. We observe that for the same value of the $\beta$, the discontinuity in EWCS occurs at a early value of $x$, for the perpendicular case.\\
As we have shown earlier, the critical separation distance $x_c$ can be represented in terms of the subsystem length and boost parameter. In order to do this we use the fact that mutual information vanishes at $x=x_c$. For $d=3$, $I^{\perp}(A:B)=0$ at $x=x_c$ condition gives
\begin{eqnarray}
2\left(\frac{\sqrt{\pi}\Gamma(3/4)}{\Gamma(1/5)}\right)^{2}\left[-\frac{2}{l}+\frac{1}{x_c}+\frac{1}{2l+x_c}\right]+\frac{4\sqrt{\pi}\frac{\Gamma(3/4)}{4}(\frac{1}{2}+\beta^{2}\gamma^{2})}{16(10)^{3}\left(\frac{\sqrt{\pi}\Gamma(3/4)}{\Gamma(1/5)}\right)^{2}}\left[2l^{2}-x_c^{2}-(2l+x_c)^{2}\right]&=&0~.
\end{eqnarray}
Similarly, for $d=4$ we get
\begin{eqnarray}
2\left(\sqrt{\pi}\frac{\Gamma(4/6)}{\Gamma(1/7))}\right)^{3}\left[-\frac{2}{l^{2}}+\frac{1}{x_c^{2}}+\frac{1}{(2l+x_c)^{2}}\right]+\frac{4\frac{\sqrt{\pi}\Gamma(4/3)}{5\Gamma(5/6)}(\frac{3}{5}+\beta^{2}\gamma^{2})}{16(10)^{4}\left(\sqrt{\pi}\frac{\Gamma(4/6)}{\Gamma(1/7)}\right)^{2}}\left[2l^{2}-x_c^{2}-(2l+x_c)^{2}\right]&=&0~.
\end{eqnarray}
In order to obtain the analytical expression corresponding to $x_c$, one needs to solve the above equation ($d=3$). As we have mentioned earlier, the point $x_c$ is defined as the critical point at which the mutual information vanishes and the EWCS shows the phase transition from the connected phase to the disconnected phase. With the expressions of $\bar{I}^{perp}(A:B)$ and $\bar{I}^{para}(A:B)$ in hand, we can get a qualitative comparison between critical points computed in the parallel and perpendicular to the boost scenario. This we show in fig.\eqref{xcComp}. We observe that the deviation between the critical separation lengths is prominent for relatively larger value of the boost parameter $\beta$. \\
\begin{figure}
	\centering
	\includegraphics[width=0.5\textwidth]{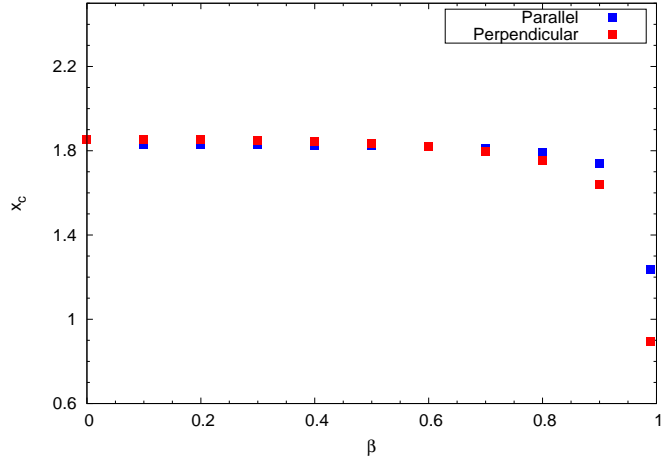}
	\caption{Comparison between the critical separation length $x_c$computed for both parallel and perpendicular case.}
	\label{xcComp}
\end{figure}
\subsection{Entanglement negativity}
We now proceed to compute the entanglement negativity $(E_{N})$ for the perpendicular case. Similar to the parallel case, we consider two different configurations, firstly two adjacent subsystems of length $l_{1}$ and $l_{2}$ and then two disjoint subsystems of length $l_{1}$ and $l_{2}$, separated by a distance $x$.\\
For the adjacent set up, entanglement negativity reads
\begin{eqnarray}
E^{\perp}_{N_{adj}}&=&\frac{3}{4}S_{div}+\frac{3V_{d-2}a_{0}(2b_{0})^{d-2}}{8G_{d+1}}\Bigg(\frac{1}{l_{1}^{d-2}}+\frac{1}{l_{2}^{d-2}}-\frac{1}{(l_{1}+l_{2})^{d-2}}\Bigg)\nonumber\\
&+&\frac{3V_{d-2}b_{1}(d+1)}{128G_{d+1}b_{0}^{2}z_{0}^{d}}\Bigg(\frac{d-1}{d+1}+\beta^{2}\gamma^{2}\Bigg)\left[l_{1}^{2}+l_{2}^{2}-(l_{1}+l_{2})^{2}\right]~~.
\end{eqnarray}
On the other hand, the change in $E_N$ reads
\begin{eqnarray}\label{newex5}
\delta E^{\perp}_{N_{adj}}&=&E^{\perp}_{N_{adj}}-E^{pure}_{N_{adj}}=\frac{3V_{d-2}b_{1}(d+1)}{128G_{d+1}b_{0}^{2}z_{0}^{d}}\Bigg(\frac{d-1}{d+1}+\beta^{2}\gamma^{2}\Bigg)\left[l_{1}^{2}+l_{2}^{2}-(l_{1}+l_{2})^{2}\right]
\end{eqnarray}
where $E^{pure}_{N_{adj}}$ is given in eq.(\ref{eepure}). One can now observe a relation between $\delta E^{\parallel}_{N_{adj}}$ and $\delta E^{\perp}_{N_{adj}}$ by using eq.(s)\eqref{newex6}, \eqref{newex5}. This reads
\begin{eqnarray}
\delta E^{\perp}_{N_{adj}} = \left[\frac{d+1}{2}\right]	\left[\frac{\frac{d-1}{d+1}+\beta^{2}\gamma^{2}}{\frac{d-1}{2}+\beta^{2}\gamma^{2}}\right] \delta E^{\parallel}_{N_{adj}}~.
\end{eqnarray}	
Now if we consider that the lengths corresponding to both of the subsystems are equal, that is, $l_1=l_2=l$, we then obtain
\begin{eqnarray}\label{newex7}
	\delta E^{\perp}_{N_{adj}}|_{l_1=l_2}=-\frac{3V_{d-2}b_{1}(d+1)l^2}{64G_{d+1}b_{0}^{2}z_{0}^{d}}\Bigg(\frac{d-1}{d+1}+\beta^{2}\gamma^{2}\Bigg)~.
\end{eqnarray}
Now we consider two disjoint subsystems $A$ and $B$ of length $l_{1}$ and $l_{2}$ perpendicular to the boost direction, with a separation of $x$. In this case, the entanglement negativity reads
\begin{eqnarray}\label{disjointEN}
E^{\perp}_{N_{dis}}&=&\frac{3}{8}\frac{V_{d-2}a_{0}(2b_{0})^{d-2}}{G_{d+1}}\Bigg[\frac{1}{(l_{1}+x)^{d-2}}+\frac{1}{(l_{2}+x)^{d-2}}-\frac{1}{(l_{1}+l_{2}+x)^{d-2}}-\frac{1}{x^{d-2}}\Bigg]\nonumber\\
&+&\frac{3V_{d-2}b_{1}(d+1)}{128b_{0}^{2}G_{d+1}z_{0}^{d}}\left(\frac{d-1}{d+1}+\beta^{2}\gamma^{2}\right)\Bigg[(l_{1}+x)^{2}+(l_{2}+x)^{2}-(l_{1}+l_{2}+x)^{2}-x^{2}\Bigg]~~.
\end{eqnarray}
Similar to the parallel case, if we take the length of subsystems to be equal, that is $l_{1}=l_{2}=l$, then the entanglement of negativity (disjoint configuration) reads
\begin{eqnarray}\label{enper}
\bar{E}^{per}_{N_{dis}}|_{l_{1}=l_{2}}&=&\frac{3}{4}\Bigg[a_{0}(2b_{0})^{d-2}\Bigg(\frac{2}{(l+x)^{d-2}}-\frac{1}{(2l+x)^{d-2}}-\frac{1}{x^{d-2}}\Bigg)\nonumber\\
&+&\frac{b_{1}(d+1)}{32b_{0}^{2}z_{0}^{d}}\left(\frac{d-1}{d+1}+\beta^{2}\gamma^{2}\right)\Bigg[2(l+x)^{2}-(2l+x)^{2}-x^{2}\Bigg]\Bigg]~.
\end{eqnarray}
Where $\bar{E}^{perp}_{N}=\frac{2G_{d+1}}{V_{d-2}}E^{\perp}_{N}$. 
\begin{figure}[!h]
	\begin{minipage}[t]{0.48\textwidth}
		\centering\includegraphics[width=\textwidth]{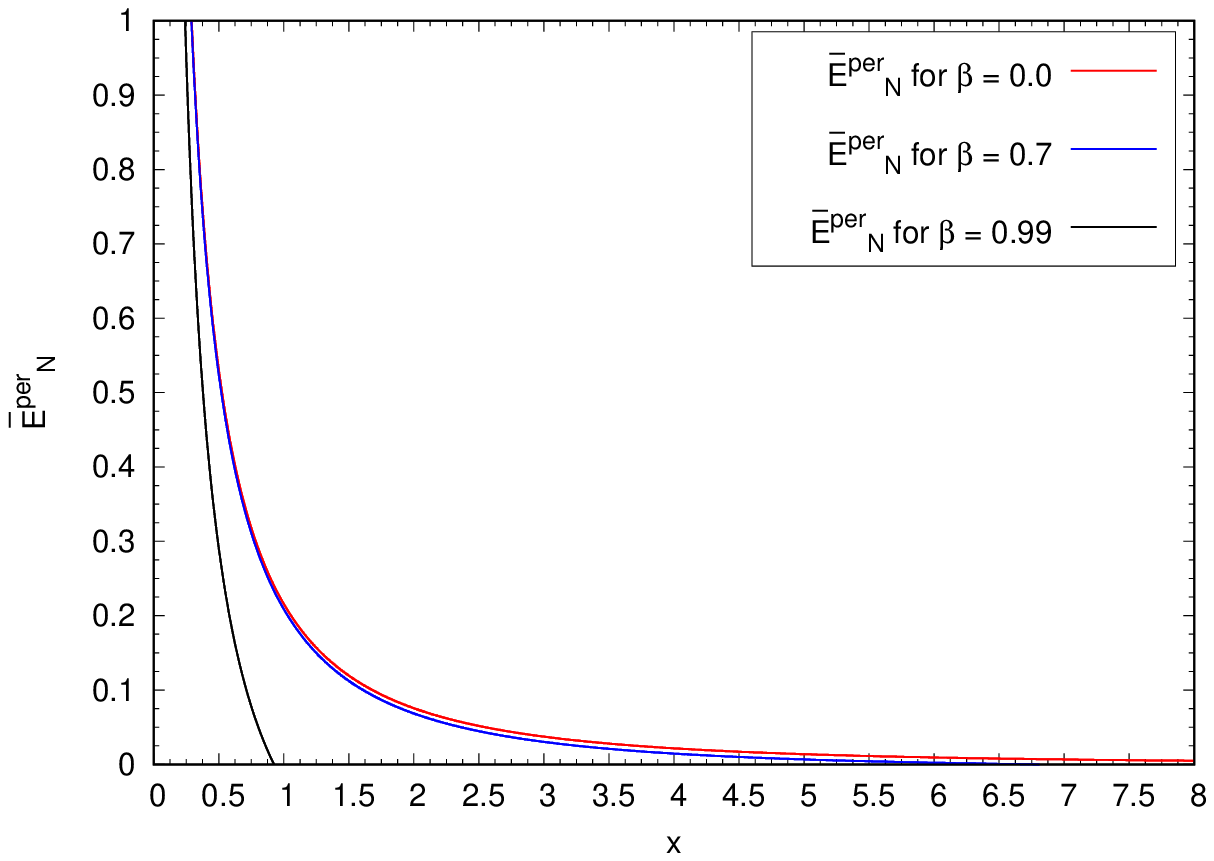}\\
	\end{minipage}\hfill
	\begin{minipage}[t]{0.48\textwidth}
		\centering\includegraphics[width=\textwidth]{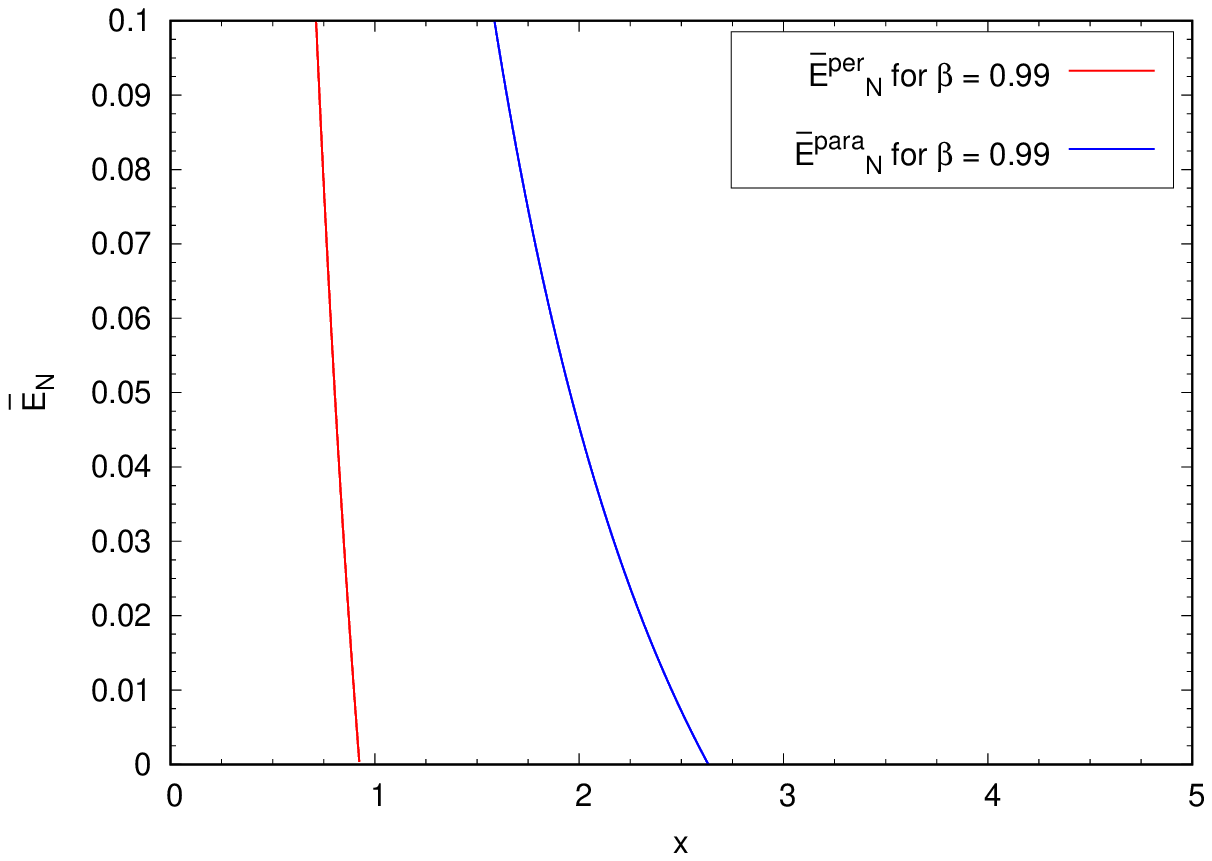}\\
	\end{minipage}
	\caption{In this above figure we have plotted entanglement negativity for two disjoint subsystems (when subsystems are perpendicular to the boost) with same length $l$, as function of the separation distance $x$. In the left panel, we probe the effect of the boost parameter on the entanglement negativity, computed in the perpendicular set up. In the right panel, we have compared the results of two different configurations The blue curve shows the result when the subsystems are along the boost and the red curve represents the result when the subsystems are perpendicular to the boost. In both these cases we have taken the boost parameter $\beta=0.99$. We have set $d=3$, $l=3$ and $z_0=10$.}
	\label{neg}
\end{figure}
We now graphically represent the entanglement negativity of two disjoint subsystems (given in eq.(\ref{enper})) with different values of the boost parameter. We find that $E_N$ computed for $\beta\rightarrow0$ and $z_0\rightarrow\infty$ does not vanishes at any value of the separation distance. However, for a non-zero, finite value of the $\beta$ parameter, $E_N$ vanishes at a finite value of the separation distance. Further, as we increase the $\beta$, entanglement negativity vanishes st a smaller value of the separation length. This is very much clear from the fig.(\ref{neg}). Furthermore, from the right panel of fig.(\ref{neg}) we observe that, entanglement negativity vanishes earlier for the perpendicular case, at the same value of the boost parameter (in this case $\beta=0.99$).
\begin{figure}[!h]
	\centering
	\includegraphics[width=0.5\textwidth]{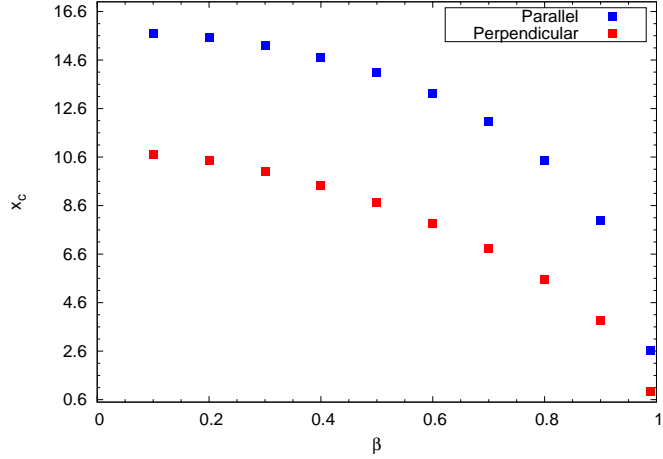}
	\caption{Comparison between the critical separation length $x_c$computed for both parallel and perpendicular case, at which the entanglement negativity vanishes.}
	\label{xcComp1}
\end{figure}
\noindent This we have observed in the parallel scenario also. We denote this length scale (at which $E_N$ vanishes) as the critical separation length $x_c$ for the entanglement negativity. We now compare the values of this length scale, computed in both parallel and perpendicular set up. This we have shown in fig.\eqref{xcComp1}. We observe that in the perpendicular set up, the entanglement negativity vanishes for a smaller value of $x_c$, for the same value of the boost parameter. Furthermore, $x_c$ decreases with the increase in the boost parameter.\\
\noindent With this computed result (given in eq.\eqref{disjointEN}) in hand, we can now compute the quantity $\delta E^{\perp}_{N_{dis}}$. This reads
\begin{eqnarray}
\delta E^{\perp}_{N_{dis}}&=&\frac{3V_{d-2}b_{1}(d+1)}{128b_{0}^{2}G_{d+1}z_{0}^{d}}\left(\frac{d-1}{d+1}+\beta^{2}\gamma^{2}\right)\Bigg[(l_{1}+x)^{2}+(l_{2}+x)^{2}-(l_{1}+l_{2}+x)^{2}-x^{2}\Bigg]~~.
\end{eqnarray}
It is clear from the above result that the divergent part in the entanglement entropy does not contribute to the entanglement negativity in case of disjoint subsystem which we have observed in the previous section also. Furthermore, when $l_1=l_2=l$, the change in entanglement negativity reads
 \begin{eqnarray}\label{bb}
 \delta E^{\perp}_{N_{dis}}|_{l_{1}=l_{2}}&=&\frac{3V_{d-2}b_{1}(d+1)}{128b_{0}^{2}G_{d+1}z_{0}^{d}}\left(\frac{d-1}{d+1}+\beta^{2}\gamma^{2}\right)\Bigg[2(l+x)^{2}-(2l+x)^{2}-x^{2}\Bigg]~.
 \end{eqnarray}
 With the help of the above expression and eq.(\ref{aa}), we can find the following relation
\begin{equation}
 \delta E^{\perp}_{N_{dis}}=\left[\frac{d+1}{2}\right]	\left[\frac{\frac{d-1}{d+1}+\beta^{2}\gamma^{2}}{\frac{d-1}{2}+\beta^{2}\gamma^{2}}\right]\delta E^{\parallel}_{N_{dis}}~.
 \end{equation}
By looking at the above relation and the one given in eq.\eqref{newex7}, we can conclude that the relation between $\delta E^{\perp}_{N}$ and $\delta E^{\parallel}_{N}$ stands to be unchanged irrespective of the adjacent or disjoint subsystem choice. Further, it is also reassuring to observe the fact that in the limit $x\rightarrow0$, eq.\eqref{bb} reproduces eq.\eqref{newex7}. Similar to the parallel set up, once again we construct a relation between $ \delta E^{\perp}_{N_{adj}}|_{l_{1}=l_{2}}$ and $ \delta S^{\perp}_{EE}$. This reads
\begin{eqnarray}
	\delta E^{\perp}_{N_{adj}}|_{l_{1}=l_{2}} = \lambda_4~\delta S^{\perp}_{EE}~.
\end{eqnarray}
\subsection{Holographic subregion complexity}
\begin{figure}[!h]
	\centering
	\includegraphics[width=0.5\textwidth]{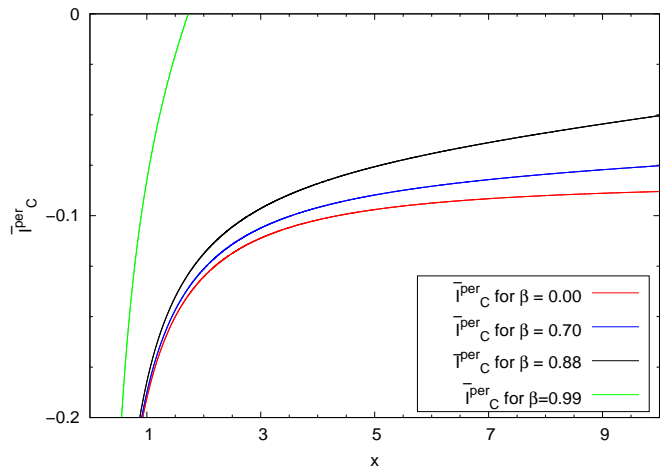}
	\caption{Variation of mutual complexity with respect to the separation distance between two subsystems taken perpendicular to the direction of boost.}
	\label{mcper}
\end{figure}

We now proceed to calculate the holographic subregion complexity of a  strip like subsystem say $A$ which is perpendicular to the direction of boost. In this set up, the HSC is obtained to be
\begin{eqnarray}\label{cper}
C^{\perp}_{V}&=&\frac{V_{\perp}({\Gamma})}{8\pi G_{d+1}}
\end{eqnarray}  
where $V_{\perp}(\Gamma)$ denotes the maximal
co-dimension one volume enclosed by the co-dimension two
minimal Ryu-Takayanagi (RT) surfaces in the bulk. Following similar procedure as we have done in case of the subsystem parallel to boost scenario, we compute the co-dimension one volume $V_{\perp}(\Gamma)$ and substitute in eq.\eqref{cper}. This leads to the following \cite{Karar:2019bwy}
\begin{eqnarray}\label{hscper}
C_{V}^{\perp}&=&\frac{1}{8\pi G_{d+1}}\Bigg[\frac{V_{d-2}}{d-1}\frac{l}{\epsilon^{d-1}}-\frac{2^{d-2}\pi^{\frac{d-1}{2}}}{(d-1)^{2}}\left(\frac{\Gamma(\frac{d}{2d-2})}{\Gamma(\frac{1}{(2d-2)})}\right)^{d-3}\frac{V_{d-2}}{l^{d-2}}-\frac{V_{d-2}l^{2}}{4b_{0}^{2}(d-1)z_{0}^{d}}\Bigg[\Bigg(\frac{(d-2)\pi b_{1}}{2(d-1)b_{0}^{2}}+(2-d)c_{0}\Bigg)\nonumber\\
&+&\beta^{2}\gamma^{2}\left(\frac{(d-2)\pi I_{l}}{2b_{0}^{2}(d-1)}+c_{2}-(d-1)c_{0}\right)\Bigg]\Bigg]~.
\end{eqnarray}
On the other hand, the change in HSC is obtained to be
\begin{eqnarray}
\delta C_{V}^{\perp}=-\frac{V_{d-2}l^{2}}{32G_{d+1}b_{0}^{2}(d-1)z_{0}^{d}}\Bigg[\Bigg(\frac{(d-2)\pi b_{1}}{2(d-1)b_{0}^{2}}+(2-d)c_{0}\Bigg)
+\beta^{2}\gamma^{2}\left(\frac{(d-2)\pi I_{l}}{2b_{0}^{2}(d-1)}+c_{2}-(d-1)c_{0}\right)\Bigg]~.
\end{eqnarray}
Now we will relate this change in the  HSC with the boundary field theoretic quatities. This provides a thermodynamics like law for HSC. The change in the holographic subregion complexity $(\delta C_{V}^{\perp})$ related to $\delta S^{\parallel}_{EE}$ (given in eq.(\ref{cspara})) and $\delta S^{\perp}_{EE}$ (given in eq.(\ref{csper})) reads \cite{Karar:2019bwy}
\begin{eqnarray}
	\delta C^{\perp}_{V}&=&\frac{1}{2(d-1)^{3}}\left[\frac{\delta S^{\parallel}_{EE}}{(d+1)b_{1}^{2}}-\left(\frac{d-2}{b_{0}^{2}}-\frac{d-3}{(d+1)b_{1}^{2}}\right)\delta S^{\perp}_{EE}\right]~.
\end{eqnarray}
Now using eqs.(\ref{para1st},\ref{per1st1}), one can relate the change in the holographic subregion complexity with the boundary field theoretic quantities like excitation energy, pressure in the following way.
\begin{eqnarray}\label{c1st}
	\delta C^{\perp}_{V}&=&\frac{1}{2(d-1)^{3}T_{E}}\left[\Delta E\left(\frac{d-2}{(d+1)b_{1}^{2}}-\frac{d-2}{b_{0}^{2}}\right)-\mathcal{V}\left(\frac{d-1}{d+1}\right)\left(\frac{\Delta P_{\parallel}}{b_{1}^{2}}-\Delta P_{\perp}\left(\frac{d-2}{b_{0}^{2}}-\frac{d-3}{(d+1)b_{1}^{2}}\right)\right)\right]\nonumber\\.
\end{eqnarray}
Similarly following \cite{Karar:2019bwy}, we can write $\delta C_{V}^{\parallel}$ (given in eq.(\ref{dcv})) in the following way
\begin{eqnarray}\label{c2nd}
	\delta C_{V}^{\parallel}&=&\frac{1}{2(d-1)^{3}T_{E}}\left[\Delta E\left(\frac{d-2}{(d+1)b_{1}^{2}}-\frac{d-2}{b_{0}^{2}}\right)-\mathcal{V}\left(\frac{d-1}{d+1}\right)\left(\frac{\Delta P_{\parallel}}{b_{1}^{2}}-\Delta P_{\perp}\left(\frac{d-2}{b_{0}^{2}}-\frac{d-3}{(d+1)b_{1}^{2}}\right)\right)\right]\nonumber\\
	&+&\frac{\mathcal{V}l\beta^{2}\gamma^{2}c_{0}}{32\pi G_{d+1}(d-1)b_{0}^{2}z_{0}^{d}}\left[1+(d-2)(d+1)\frac{b_{1}^{2}}{b_{0}^{2}}\right]~~.
\end{eqnarray}
From both the eqs.(\ref{c1st},\ref{c2nd}), we observe an interesting fact that the change in complexity for strip placed along the perpendicular to the boost direction is related to the CFT pressure perendicular to boost and also on the pressure parallel to the boost.  The same is true for the change in HSC for parallel strip (placed along $y$-direction).\\
With the computed result of HSC (given in eq.(\ref{hscper})) in hand, we now proceed to calculate the mutual complexity between two subsystems $A$ and $B$ which are in perpendicular orientation with respect to the direction of boost. For that we will consider two subsystems of equal length $l$ and with a separation of $x$. In this setup, the mutual complexity reads
\begin{eqnarray}
\Delta\mathcal{C}^{\perp}=\mathcal{C}_{V}^{\perp}(A)+\mathcal{C}_{V}^{\perp}(B)-\mathcal{C}_{V}^{\perp}(A\cup B)\equiv
2\mathcal{C}_{V}^{\perp}(l)-\mathcal{C}_{V}^{\perp}(2l+x)+\mathcal{C}_{V}^{\perp}(x)~.
\end{eqnarray}
For the above mentioned configuration, mutual complexity reads
\begin{eqnarray}
\Delta\mathcal{C}^{\perp}&=&\frac{V_{d-2}}{8\pi G_{d+1}(d-1)}\Bigg[\frac{2^{d-2}\pi^{\frac{d-1}{2}}}{(d-1)}\left(\frac{\Gamma(\frac{d}{2d-2})}{\Gamma(\frac{1}{(2d-2)})}\right)^{d-3}\Bigg(\frac{1}{(2l+x)^{d-2}}-\frac{1}{x^{d-2}}-\frac{2}{l^{d-2}}\Bigg)\nonumber\\
&-&\frac{1}{4b_{0}^{2}z_{0}^{d}}\Bigg[\Bigg(\frac{(d-2)\pi b_{1}}{2(d-1)b_{0}^{2}}+(2-d)c_{0}\Bigg)+\beta^{2}\gamma^{2}\left(\frac{(d-2)\pi I_{l}}{2b_{0}^{2}(d-1)}+c_{2}-(d-1)c_{0}\right)\Bigg]\left[2l^{2}+x^{2}-(2l+x)^{2}\right]\Bigg]~~.\nonumber\\
\end{eqnarray}
Furthermore, the change in mutual complexity $\delta\mathcal{C}^{\perp}\equiv \Delta\mathcal{C}^{\perp}-\Delta\mathcal{C}^{pure}$ is obtained to be
\begin{eqnarray}\label{a1}
	\delta\mathcal{C}^{\perp}=-\frac{V_{d-2}}{8\pi G_{d+1}(d-1)}\frac{1}{4b_{0}^{2}z_{0}^{d}}\Bigg[\Bigg(\frac{(d-2)\pi b_{1}}{2(d-1)b_{0}^{2}}+(2-d)c_{0}\Bigg)&+&\beta^{2}\gamma^{2}\left(\frac{(d-2)\pi I_{l}}{2b_{0}^{2}(d-1)}+c_{2}-(d-1)c_{0}\right)\Bigg]\nonumber\\
	&&\times\left[2l^{2}+x^{2}-(2l+x)^{2}\right].
\end{eqnarray}
The relation between $\delta\mathcal{C}^{\parallel}$ and $\delta\mathcal{C}^{\perp}$ reads
\begin{eqnarray}
\frac{\delta\mathcal{C}^{\parallel}}{\delta\mathcal{C}^{\perp}}&=&\frac{\Bigg[\Bigg(\frac{(d-2)\pi b_{1}}{2(d-1)b_{0}^{2}}+(2-d)c_{0}\Bigg)
	+\beta^{2}\gamma^{2}\Bigg(\frac{(d-2)\pi}{2(d-1)^{2}b_{0}}\left(\frac{2b_{1}}{b_{0}}-1\right)+c_{2}-c_{0}d\Bigg)\Bigg]}{\Bigg[\Bigg(\frac{(d-2)\pi b_{1}}{2(d-1)b_{0}^{2}}+(2-d)c_{0}\Bigg)+\beta^{2}\gamma^{2}\left(\frac{(d-2)\pi I_{l}}{2b_{0}^{2}(d-1)}+c_{2}-(d-1)c_{0}\right)\Bigg]}~.
\end{eqnarray}
\noindent To obtain the above expression we have used eqs.(\ref{bbb},\ref{a1}).
\noindent Similar to parallel case, we now introduce the scaling $\bar{I}^{per}_{C}=\frac{2G_{d+1}}{V_{d-2}}\Delta\mathcal{C}^{\perp}$ and proceed to represent our findings graphically. This we show in fig.\eqref{mcper}, where we set $d=3$, $l=3$ and $z_0=10$. Similar to the parallel case, the mutual complexity in this scenario is also superadditive.\\ 
\section{Asymmetry ratio of different holographic measures}\label{sec5}
In this section we compute asymmetry ratio of different information theoretic measures. For entanglement entropy the asymmetry ratio is defined as
\begin{eqnarray}
\mathcal{A}_{S_{EE}}&=&\frac{\delta S^{\perp}_{EE}-\delta S^{\parallel}_{EE}}{\delta S^{\perp}_{EE}+\delta S^{\parallel}_{EE}}~~.
\end{eqnarray}
By using eq.(\ref{cspara}) and eq.(\ref{csper}), $\mathcal{A}_{S_{EE}}$ can be computed. This reads
\begin{eqnarray}\label{sea}
\mathcal{A}_{S_{EE}}=\frac{\beta^{2}\gamma^{2}}{\left(2+\frac{d+3}{d-1}\beta^{2}\gamma^{2}\right)}~.
\end{eqnarray}
It is clear from the above result that the asymmetry ratio for HEE is independent of the subsystem size. It depends only on the boost parameter and the value of the spacetime dimension. In the large boost limit ($\beta\rightarrow1$), eq.(\ref{sea}) can be recast to the form
\begin{eqnarray}
\mathcal{A}_{S_{EE}}|_{\beta\rightarrow 1}=\frac{d-1}{d+3}~~.
\end{eqnarray}
Similarly, the asymmetry ratio for mutual information is obtained to be
\begin{eqnarray}
\mathcal{A}_{I}=\frac{\delta I^{\perp}-\delta I^{\parallel}}{\delta I^{\perp}+\delta I^{\parallel}}=\frac{\beta^{2}\gamma^{2}}{\left(2+\frac{d+3}{d-1}\beta^{2}\gamma^{2}\right)}~.
\label{AI}
\end{eqnarray}
In the above computation, we have used eq.(\ref{cipara}) and eq.(\ref{ciper}) to get the explicit expression. The above result again implies that $\mathcal{A}_{I}$ is independent of the subsystem size. Again in the large boost limit ($\beta\rightarrow1$), eq.(\ref{AI}) reads
\begin{eqnarray}
\mathcal{A}_{I}|_{\beta\rightarrow 1}=\frac{d-1}{d+3}~~.
\label{AIB}
\end{eqnarray}
The results given in eq.(s)(\ref{AI},\ref{AIB}) are similar to the ones we have obtained for HEE. This is quite natural as HMI is nothing but a certain combination of the HEE. On the other hand, the asymmetry ratio for the entanglement wedge cross-section reads (we use eq(s).(\ref{cewpara},\ref{cewper}))
\begin{eqnarray}\label{aew}
\mathcal{A}_{E_{W}}=\frac{\delta E_{W}^{\perp}-\delta E_{W}^{\parallel}}{\delta E_{W}^{\perp}+\delta E_{W}^{\parallel}}=\frac{\beta^{2}\gamma^{2}\left[1-2\frac{b_{1}}{b_{0}}\right]}{(1-2\frac{b_{1}}{b_{0}})+\frac{\beta^{2}\gamma^{2}}{2}\left[1+2\frac{b_{1}}{b_{0}}-4\frac{I_{l}}{b_{0}}\right]}~~.
\end{eqnarray}
In the limit $\beta\rightarrow1$, the above expression simplifies to the following form
\begin{eqnarray}\label{aew2}
\mathcal{A}_{E_{W}}|_{\beta\rightarrow 1}=\frac{2(1-2\frac{b_{1}}{b_{0}})}{1+2\frac{b_{1}}{b_{0}}-4\frac{I_{l}}{b_{0}}}~.
\end{eqnarray}
From the obtained expression, given in the eq(s).(\ref{aew},\ref{aew2}), it is clear that the asymmetry ratio for EWCS is independent of shape and size of the subsystem. 
Asymmetry ratio for the HSC reads \cite{Karar:2019bwy}
\begin{eqnarray}\label{assc}
\mathcal{A}_{C_{V}}=\frac{\delta C_{V}^{\perp}-\delta C_{V}^{\parallel}}{\delta C_{V}^{\perp}+\delta C_{V}^{\parallel}}=\left[\frac{\frac{2-d}{2(d-1)^{3}b_{0}^{2}}-\frac{1}{2(d-1)^{3}(d+1)b_{1}^{2}}}{\frac{2-d}{2(d-1)^{3}b_{0}^{2}}+\frac{\mathcal{R}+2d-5}{2(d-1)^{3}(d+1)b_{1}^{2}}}\right]\mathcal{A}_{S_{EE}}
\end{eqnarray} 
where $\mathcal{A}_{S_{EE}}$ is given in eq.(\ref{sea}) and $\mathcal{R}$ is 
\begin{eqnarray}
\mathcal{R}=\frac{\delta S_{EE}^{\parallel}}{\delta S_{EE}^{\perp}}=\frac{1+\frac{2}{d-1}\beta^{2}\gamma^{2}}{1+\frac{d+1}{d-1}\beta^{2}\gamma^{2}}~.
\end{eqnarray}
The above result corresponding to HSC, matches with the ones given in \cite{Karar:2019bwy}. In the limit $\beta\rightarrow 1$ eq.(\ref{assc}) reduces to 
\begin{eqnarray}
\mathcal{A}_{C_{V}}|_{\beta\rightarrow 1}=\left[\frac{\frac{2-d}{2(d-1)^{3}b_{0}^{2}}-\frac{1}{2(d-1)^{3}(d+1)b_{1}^{2}}}{\frac{2-d}{2(d-1)^{3}b_{0}^{2}}+\frac{\frac{2}{d+1}+2d-5}{2(d-1)^{3}(d+1)b_{1}^{2}}}\right]\frac{d-1}{d+3}~.
\end{eqnarray}
 Now by using eq(s).(\ref{aa},\ref{bb}) the asymmetry ratio for entanglement negativity for disjoint intervals is obtained to be
\begin{eqnarray}\label{dise}
\mathcal{A}_{E_{N_{dis}}}=\frac{\delta E^{\perp}_{N_{dis}}-\delta E^{\parallel}_{N_{dis}}}{\delta E^{\perp}_{N_{dis}}+\delta E^{\parallel}_{N_{dis}}}=\frac{\beta^{2}\gamma^{2}\frac{d-1}{2}}{(d+1)+\beta^{2}\gamma^{2}\frac{d+3}{2}}~~.
\end{eqnarray}
On the other hand, for adjacent intervals the asymmetry ratio for the entanglement negativity reads
\begin{eqnarray}\label{adje}
\mathcal{A}_{E_{N_{adj}}}=\frac{\delta E^{\perp}_{N_{adj}}-\delta E^{\parallel}_{N_{adj}}}{\delta E^{\perp}_{N_{adj}}+\delta E^{\parallel}_{N_{adj}}}=\frac{\beta^{2}\gamma^{2}\frac{d-1}{2}}{(d+1)+\beta^{2}\gamma^{2}\frac{d+3}{2}}~~.
\end{eqnarray}
It is interesting to observe that for both the adjacent and disjoint interval case, the asymmetry ratio for entanglement negativity is same and independent of the subsystem size. In the large boost limit ($\beta\rightarrow 1$), both the eqs.(\ref{dise},\ref{adje}) reduces to the following expression 
\begin{eqnarray}
\mathcal{A}_{E_{N_{dis}}}|_{\beta\rightarrow 1}=\mathcal{A}_{E_{N_{adj}}}|_{\beta\rightarrow 1}=\frac{d-1}{d-3}~.
\end{eqnarray}
Further, the asymmetry ratio for mutual complexity reads
\begin{eqnarray}
\mathcal{A}_{\Delta \mathcal{C}}=\frac{\beta^{2}\gamma^{2}\left(\frac{(d-2)\pi I_{l}}{2b_{0}^{2}(d-1)}+c_{2}-(d-1)c_{0}-\frac{(d-2)\pi}{2(d-1)^{2}b_{0}}\left(\frac{2b_{1}}{b_{0}}-1\right)-c_{2}+c_{0}d\right)}{2\Bigg(\frac{(d-2)\pi b_{1}}{2(d-1)b_{0}^{2}}+(2-d)c_{0}\Bigg)+\beta^{2}\gamma^{2}\left(\frac{(d-2)\pi I_{l}}{2b_{0}^{2}(d-1)}+c_{2}-(d-1)c_{0}+\frac{(d-2)\pi}{2(d-1)^{2}b_{0}}\left(\frac{2b_{1}}{b_{0}}-1\right)+c_{2}-c_{0}d\right)}~.
\end{eqnarray}
In the large boost limit ($\beta\rightarrow 1$), the above expression simplifies to the following form
\begin{eqnarray}
\mathcal{A}_{\Delta \mathcal{C}}|_{\beta\rightarrow 1}=\frac{\left(\frac{(d-2)\pi I_{l}}{2b_{0}^{2}(d-1)}+c_{2}-(d-1)c_{0}-\frac{(d-2)\pi}{2(d-1)^{2}b_{0}}\left(\frac{2b_{1}}{b_{0}}-1\right)-c_{2}+c_{0}d\right)}{\left(\frac{(d-2)\pi I_{l}}{2b_{0}^{2}(d-1)}+c_{2}-(d-1)c_{0}+\frac{(d-2)\pi}{2(d-1)^{2}b_{0}}\left(\frac{2b_{1}}{b_{0}}-1\right)+c_{2}-c_{0}d\right)}
\end{eqnarray}
 We would like to mention that the asymmetry ratios computed in this section are bounded from above \cite{Mishra:2016yor} and the bound saturates in the large boost limit, that is, $\beta\rightarrow 1$.
 
\section{$AdS$ wave geometry}\label{AdSWave}
The $AdS$ wave geometry can be obtained by taking the limits $\beta\rightarrow 1, z_{0}\rightarrow\infty$ in eq.(\ref{bbm}), with the following condition
\begin{equation}
\frac{\beta^{2}\gamma^{2}}{z_{0}^{d}}=\frac{1}{z_{I}^{d}}=\mathrm{fixed}~
\end{equation}
where $z=z_{I}$ is an scale which determines momentum
of the wave traveling in the $y$ direction.
In this limit, the boosted black brane metric (given in eq.(\ref{bbm}) reduces to following \cite{Mishra:2016yor}
\begin{eqnarray}
ds^{2}&=&\frac{1}{z^{2}}\left(-\frac{dt^{2}}{K}+K(dy-\omega)^{2}+dx_{1}^{2}+...+dx_{2}^{2}+dz^{2}\right)
\end{eqnarray}
with 
\begin{equation}
K(z)=1+\frac{z^{d}}{z_{I}^{d}}~,~\omega=\left(1-\frac{1}{K(z)}\right)dt~~.
\end{equation}
We shall now compute various information theoretic measures of the conformal field theory living at the boundary using the above geometry holographically. Here we will again consider two different cases, first, the subsystem is taken along the boost and then the subsystem is perpendicular to the boost. We follow the same approach as we have shown in case of the boosted black brane.
\subsection{Holographic entanglement entropy}
\subsubsection{Subsystem along the boost}
Similar to the parallel subsystem scenario of boosted black brane, we consider a subsystem which is specified by the volume $V_{sub}=L^{d-2}l$, with $-\frac{l}{2}\le y\le \frac{l}{2}$, and $x_{1},...,x_{d-2} \in \left[0,L\right]$ with $L\rightarrow\infty$. Further we assume that in this case the length can vary only along the $y$-direction and the lengths in other directions are taken to be fixed. Following the same procedure as we have shown for the boosted black brane and by incorporating the thin strip approximation we compute the HEE in this context also. This is obtained to be
\begin{eqnarray}\label{hee}
	S_{EE}^{\parallel}&=&S_{div}+\frac{V_{d-2}}{2G_{d+1}}\left[a_{0}\left(\frac{2b_{0}}{l}\right)^{d-2}+\left(\frac{l}{2b_{0}}\right)^{2}\frac{b_{1}}{2z_{I}^{d}}\right]
\end{eqnarray}
where we have used the relation
\begin{eqnarray}\label{1}
	z_{t}^{\parallel}\approx\frac{l}{2\left[b_{0}-\frac{1}{2}\left(\frac{z_{t}}{z_{I}}\right)^{d}(b_{1}-I_{l})\right]}\approx\frac{\bar{z}_{t}}{1-\frac{1}{2}\left(\frac{\bar{z_{t}}}{z_{I}}\right)^{d}\left(\frac{b_{1}}{b_{0}}-\frac{I_{l}}{b_{0}}\right)}~.
\end{eqnarray}
In the above $\bar{z_{t}}$ is the turning point corresponding to the pure $AdS_{d+1}$ geometry in the bulk. In the second line we have use eq.(\ref{lpure}) to write $z_{t}^{\parallel}$ in terms of $\bar{z_{t}}$. The change in entanglement entropy $(\delta S^{\parallel}_{EE}=S^{\parallel}_{EE}-S^{pure}_{EE})$ is found to be
\begin{eqnarray}
\delta S^{\parallel}_{EE}&=&\frac{V_{d-2}}{2G_{d+1}}\left(\frac{l}{2b_{0}}\right)^{2}\frac{b_{1}}{2z_{I}^{d}}
\end{eqnarray}
\subsubsection{Subsystem perpendicular to the boost}
We now consider a strip-like subsystem perpendicular to the boost. The system is specified by volume $V=ll_{y}...l_{d-2}$, with $-\frac{l}{2}\le x_{1}\le\frac{l}{2}$,$0\le y\le l_{y}$, and $0\le x_{i} \le l_{i}$. $l_{y}$ and $l_{i}$ is taken to be very very large, that is $l_{y},l_{i}>>l$. The length along $x_{1}$ direction changes and length along other directions is fixed. In this set up, the HEE in terms of the subsystem size is found to be
\begin{eqnarray}\label{Sper}
	S^{\perp}_{EE}&=&S_{div}+\frac{V_{d-2}}{2G_{d+1}}\left[a_{0}\left(\frac{2b_{0}}{l}\right)^{d-2}+\frac{a_{1}}{z_{I}^{d}}\left(\frac{l}{2b_{0}}\right)^{2}\right]~.
\end{eqnarray}
where we have used the following relation
\begin{eqnarray}\label{L}
	z_{t}^{\perp}\approx\frac{l}{2b_{0}\left[1+\left(\frac{z_{t}}{z_{I}}\right)^{d}I_{l}\right]}\approx\frac{\bar{z_{t}}}{\left[1+\left(\frac{\bar{z}_{t}}{z_{I}}\right)^{d}I_{l}\right]}~.
\end{eqnarray}
Now we will compute change in the entanglement entropy $(\delta S^{\perp}_{EE}=S^{\perp}_{EE}-S^{pure}_{EE})$ using eqs.(\ref{Sper},\ref{spure}). The expression of $\delta S^{\perp}_{EE}$ reads 
\begin{eqnarray}
\delta S^{\perp}_{EE}&=&\frac{V_{d-2}}{2G_{d+1}}\frac{a_{1}}{z_{I}^{d}}\left(\frac{l}{2b_{0}}\right)^{2}~.
\end{eqnarray} 
Furthermore, we observe that the following relation can be pointed out in context of the AdS wave geometry
\begin{eqnarray}
\delta S^{\perp}_{EE} = \left(\frac{2a_1}{b_1}\right)\delta S^{\parallel}_{EE}~.	
\end{eqnarray}
\subsection{EWCS and Holographic mutual information}
We now compute the minimal cross-section of the entanglement wedge and HMI by following the same procedure we have shown for the boosted brane, that is, two disjoint subsystems of equal length $l$, separated by a distance $x$.
\subsubsection{Subsystems along the boost}
When the subsystems are along the boost, the EWCS is found to be
\begin{eqnarray}\label{ewp1}
	E_{W}^{\parallel}=\frac{V_{d-2}}{2G_{d+1}}\Bigg[\frac{(2b_{0})^{d-2}}{2(d-2)}\left(\frac{1}{x^{d-2}}-\frac{1}{(2l+x)^{d-2}}\right)+\frac{\left(\frac{I_{l}}{b_{0}}-\frac{b_{1}}{b_{0}}\right)}{4(2b_{0})^{2}z_{I}^{d}}\left(x^{2}-(2l+x)^{2}\right)\Bigg]~.
\end{eqnarray}
On the other hand, HMI corresponding to this set up reads
\begin{eqnarray}
	I^{\parallel}(A:B)=\frac{V_{d-2}}{2G_{d+1}}\Bigg[a_{0}(2b_{0})^{d-2}\left(\frac{2}{l^{d-2}}-\frac{1}{x^{d-2}}-\frac{1}{(2l+x)^{d-2}}\right)+\frac{b_{1}}{(2b_{0})^{2}z_{I}^{d}}\left(l^{2}-\frac{x^{2}}{2}-\frac{(2l+x)^{2}}{2}\right)\Bigg]~.
\end{eqnarray}
The changes in EWCS and HMI due to the presence of boost read
\begin{eqnarray}
	\delta E_{W}^{\parallel}&=&\frac{V_{d-2}}{2G_{d+1}}\frac{\left(\frac{I_{l}}{b_{0}}-\frac{b_{1}}{b_{0}}\right)}{4(2b_{0})^{2}z_{I}^{d}}\left(x^{2}-(2l+x)^{2}\right)\\
	\delta I^{\parallel}(A:B)&=&\frac{V_{d-2}}{2G_{d+1}}\frac{b_{1}}{(2b_{0})^{2}z_{I}^{d}}\left(l^{2}-\frac{x^{2}}{2}-\frac{(2l+x)^{2}}{2}\right)~~.
\end{eqnarray}
	
\subsubsection{Subsystems perpendicular to the boost}
When the subsystems are perpendicular to the boost, the EWCS is found to be
\begin{eqnarray}\label{EWper}
	E^{\perp}_{W}=\frac{V_{d-2}}{2G_{d+1}}\Bigg[\frac{(2b_{0})^{d-2}}{2(d-2)}\left(\frac{1}{x^{d-2}}-\frac{1}{(2l+x)^{d-2}}\right)+\frac{1}{4z_{I}^{d}(2b_{0})^{2}}\left(1-\frac{I_{l}}{b_{0}}\right)\left[(2l+x)^{2}-x^{2}\right]\Bigg]~.
\end{eqnarray}
In this set up, HMI reads
\begin{eqnarray}\label{Iper}
	I^{\perp}(A:B)=\frac{V_{d-2}}{2G_{d+1}}\Bigg[a_{0}(2b_{0})^{d-2}\left(\frac{2}{l^{d-2}}-\frac{1}{x^{d-2}}-\frac{1}{(2l+x)^{d-2}}\right)+\frac{a_{1}}{z_{I}^{d}(2b_{0})^{2}}\left(2l^{2}-x^{2}-(2l+x)^{2}\right)\Bigg]~~.
\end{eqnarray}
We now compute the changes in EWCS and HMI due to the presence of boost. This reads
\begin{eqnarray}
	\delta E_{W}^{\perp}&=&\frac{V_{d-2}}{2G_{d+1}}\frac{1}{4z_{I}^{d}(2b_{0})^{2}}\left(1-\frac{I_{l}}{b_{0}}\right)\left[(2l+x)^{2}-x^{2}\right]\\
	\delta I^{\perp}(A:B)&=&\frac{V_{d-2}}{2G_{d+1}}\frac{a_{1}}{z_{I}^{d}(2b_{0})^{2}}\left(2l^{2}-x^{2}-(2l+x)^{2}\right)~.
\end{eqnarray}
\subsection{Entanglement negativity}
We now holographically compute the entanglement negativity for the AdS wave geometry. Firstly, we consider two adjacent subsystems with lengths $l_1$ and $l_2$ and compute the entanglement negativity for this configuration. We then consider two disjoint subsystems with lengths $l_1$ and $l_2$, separated by a distance $x$.
\subsubsection{Subsystems along the boost}
When two adjacent subsystems of length $l_1$ and $l_2$ are taken along the boost, the entanglement negativity is reads
\begin{eqnarray}
	E^{\parallel}_{N_{adj}}=\frac{3}{4}S_{div}+\frac{3}{4}\frac{V_{d-2}}{2G_{d+1}}\Bigg[a_{0}(2b_{0})^{d-2}\left(\frac{1}{l_{1}^{d-2}}+\frac{1}{l_{2}^{d-2}}-\frac{1}{(l_{1}+l_{2})^{d-2}}\right)+\frac{b_{1}}{2z_{I}^{d}(2b_{0})^{2}}\left(l_{1}^{2}+l_{2}^{2}-(l_{1}+l_{2})^{2}\right)\Bigg]~.~~
\end{eqnarray}
On the other hand, when we consider two disjoint subsystems (with length $l_1$ and $l_2$), separated by a distance $x$, the entanglement negativity is obtained to be
\begin{eqnarray}\label{enn1}
	E^{\parallel}_{N_{dis}}
	&=&\frac{3}{4}\frac{V_{d-2}}{2G_{d+1}}\Bigg[a_{0}(2b_{0})^{d-2}\left(\frac{1}{(l_{1}+x)^{d-2}}+\frac{1}{(l_{2}+x)^{d-2}}-\frac{1}{(l_{1}+l_{2}+x)^{d-2}}-\frac{1}{x^{d-2}}\right)\nonumber\\
	&+&\frac{b_{1}}{2(2b_{0})^{2}z_{I}^{d}}\left((l_{1}+x)^{2}+(l_{2}+x)^{2}-(l_{1}+l_{2}+x)^{2}-x^{2}\right)\Bigg]~.
\end{eqnarray}
With the above expressions in hand, we can now proceed to compute the change in entanglement negativity due to the involvement of the boost. This we do for both adjacent and disjoint case. The findings are given below
\begin{eqnarray}
	\delta E^{\parallel}_{N_{adj}}&=&\frac{3V_{d-2}b_{1}}{16(2b_{0})^{2}G_{d+1}z_{I}^{d}}\left[l_{1}^{2}+l_{2}^{2}-(l_{1}+l_{2})^{2}\right]\\
	\delta E^{\parallel}_{N_{dis}}
	&=&\frac{3}{4}\frac{V_{d-2}}{2G_{d+1}}\frac{b_{1}}{2(2b_{0})^{2}z_{I}^{d}}\left((l_{1}+x)^{2}+(l_{2}+x)^{2}-(l_{1}+l_{2}+x)^{2}-x^{2}\right)~.
\end{eqnarray}
\subsubsection{Subsystems perpendicular to the boost}
Once again we consider two adjacent subsystems of length $l_1$ and $l_2$ in perpendicular orientation with respect to the direction of boost. This leads to the following expression for the entanglement negativity
\begin{eqnarray}\label{ENPER}
	E^{\perp}_{N_{adj}}=\frac{3}{4}S_{div}+\frac{V_{d-2}}{2G_{d+1}}\frac{3}{4}\Bigg[a_{0}(2b_{0})^{d-2}\left(\frac{1}{l_{1}^{d-2}}+\frac{1}{l_{2}^{d-2}}-\frac{1}{(l_{1}+l_{2})^{d-2}}\right)+\frac{a_{1}}{z_{I}^{d}(2b_{0})^{2}}\left(l_{1}^{2}+l_{2}^{2}-(l_{1}+l_{2})^{2}\right)\Bigg]~.~~
\end{eqnarray}
Furthermore, when we consider two disjoint subsystems with length $l_1$ and $l_2$ in the perpendicular direction (with respect to the boost), separated by a distance $x$, we get the following result for the entanglement negativity
\begin{eqnarray}\label{ENp}
	E^{\perp}_{N_{dis}}&=&\frac{3}{4}\frac{V_{d-2}}{2G_{d+1}}\Bigg[a_{0}(2b_{0})^{d-2}\left(\frac{1}{(l_{1}+x)^{d-2}}+\frac{1}{(l_{2}+x)^{d-2}}-\frac{1}{(l_{1}+l_{2}+x)^{d-2}}-\frac{1}{x^{d-2}}\right)\nonumber\\
	&+&\frac{a_{1}}{(2b_{0})^{2}z_{I}^{d}}\left((l_{1}+x)^{2}+(l_{2}+x)^{2}-(l_{1}+l_{2}+x)^{2}-x^{2}\right)\Bigg]~.
\end{eqnarray}
Similar to the parallel case, we now compute the change in entanglement negativity (for both adjacent and disjoint case) due to the boost. This reads
\begin{eqnarray}
	\delta E^{\perp}_{N_{adj}}
	&=&\frac{3}{4}\frac{V_{d-2}}{2G_{d+1}}\frac{a_{1}}{z_{I}^{d}(2b_{0})^{2}}\left(l_{1}^{2}+l_{2}^{2}-(l_{1}+l_{2})^{2}\right)~.\nonumber\\
	\delta E^{\perp}_{N_{dis}}&=&\frac{3}{4}\frac{V_{d-2}}{2G_{d+1}}\frac{a_{1}}{z_{I}^{d}(2b_{0})^{2}}\left((l_{1}+x)^{2}+(l_{2}+x)^{2}-(l_{1}+l_{2}+x)^{2}-x^{2}\right)
\end{eqnarray}
\subsection{Holographic subregion complexity}
We now proceed to compute the HSC for the AdS wave geometry. Once again we will use the formula given in eq.(\ref{cparad}).
\subsubsection{Subsystem along the boost}
When we consider a subsystem of length $l$ in the direction along the boost, the HSC is found to be
\begin{eqnarray}\label{CC}
	C^{\parallel}_{V}&=&\frac{V_{d-2}}{8\pi G_{d+1}}\Bigg[\frac{1}{d-1}\frac{l}{\epsilon^{d-1}}-\frac{2^{d-2}\pi^{\frac{d-1}{2}}}{(d-1)^{2}}\left(\frac{\Gamma(\frac{d}{2d-2})}{\Gamma(\frac{1}{(2d-2)})}\right)^{d-3}\frac{1}{l^{d-2}}\nonumber\\
	&-&\frac{1}{z_{I}^{d}(2b_{0})^{2}(d-1)}\Bigg[c_{2}-c_{0}d+\frac{(d-2)\pi}{2(d-1)b_{0}}\left(\frac{I_{l}}{b_{0}}-\frac{b_{1}}{b_{0}}\right)\Bigg]l^{2}~\Bigg]
\end{eqnarray}
and the change in HSC due to the presence of boost reads
\begin{eqnarray}
	\delta C^{\parallel}_{V}&=&-\frac{V_{d-2}}{8\pi G_{d+1}}\frac{1}{z_{I}^{d}(2b_{0})^{2}(d-1)}\Bigg[c_{2}-c_{0}d+\frac{(d-2)\pi}{2(d-1)b_{0}}\left(\frac{I_{l}}{b_{0}}-\frac{b_{1}}{b_{0}}\right)\Bigg]l^{2}~.
\end{eqnarray}
This change in the subregion complexity can be related to the boundary field theoretic quantities in the following way
\begin{eqnarray}
	\delta C^{\parallel}_{V}&=&-\frac{\left[c_{2}-c_{0}d+\frac{(d-2)\pi}{2(d-1)b_{0}^{2}}\left(I_{l}-b_{1}\right)\right]}{4\pi b_{1}(d-1)T_{E}}\left[\Delta E-\frac{d-1}{d+}\mathcal{V}\Delta P_{\parallel}\right]
\end{eqnarray}
where $T_{E}$, $\Delta E$, $\Delta P_{\parallel}$ are given in eq.(\ref{bq}).
\subsubsection{Subsystem perpendicular to the boost}
On the other hand, when we consider a subsystem of length $l$ in the direction perpendicular to the boost, the computed expression of HSC reads
\begin{eqnarray}
	C_{V}^{\perp}&=&\frac{V_{d-2}}{8\pi G_{d+1}}\Bigg[\frac{1}{d-1}\frac{l}{\epsilon^{d-1}}-\frac{2^{d-2}\pi^{\frac{d-1}{2}}}{(d-1)^{2}}\left(\frac{\Gamma(\frac{d}{2d-2})}{\Gamma(\frac{1}{(2d-2)})}\right)^{d-3}\frac{1}{l^{d-2}}\nonumber\\
	&-&\frac{1}{z_{I}^{d}(d-1)(2b_{0})^{2}}\left[c_{2}-c_{0}(d-1)+\frac{I_{l}(d-2)}{2(d-1)b_{0}^{2}}\right]l^{2}\Bigg]~.
\end{eqnarray}
With the above expression in hand, we can compute the change in HSC due to the boost , in the perpendicular set up. This reads
\begin{eqnarray}
	\delta C^{\perp}_{V}&=&-\frac{V_{d-2}}{8\pi G_{d+1}}\frac{1}{z_{I}^{d}(d-1)(2b_{0})^{2}}\left[c_{2}-c_{0}(d-1)+\frac{I_{l}(d-2)}{2(d-1)b_{0}^{2}}\right]l^{2}~.
\end{eqnarray}
The change in the HSC $(\delta C^{\perp}_{V})$ can be written in terms of the CFT excitation energy and entanglement temperature in the following way
\begin{equation}
	\delta C^{\perp}_{V}=-\frac{\left[c_{2}-c_{0}(d-1)+\frac{I_{l}(d-2)}{2(d-1)b_{0}^{2}}\right]}{4\pi a_{1}T_{E}}\Delta E~~.
\end{equation}
\subsection{Mutual complexity}
In this section, we compute the mutual complexity by incorporating the HSC conjecture. We do this by considering two subsystems of equal length $l$, separated by a distance $x$.
\subsubsection{Subsystems along the boost}
In this set up, we consider the equal length subsystems (separated by the distance $x$) are along the direction of the boost. This configuration leads to the following expression of mutual complexity
\begin{eqnarray}\label{mcpw}
	\Delta\mathcal{C}^{\parallel}&=&\frac{V_{d-2}}{2G_{d+1}}\frac{1}{4\pi}\Bigg[-\frac{2^{d-2}\pi^{\frac{d-1}{2}}}{(d-1)^{2}}\left(\frac{\Gamma(\frac{d}{2d-2})}{\Gamma(\frac{1}{(2d-2)})}\right)^{d-3}\left(\frac{2}{l^{d-2}}+\frac{1}{x^{d-2}}-\frac{1}{(2l+x)^{d-2}}\right)\nonumber\\
	&-&\frac{1}{z_{I}^{d}(2b_{0})^{2}(d-1)}\Bigg[c_{2}-c_{0}d+\frac{(d-2)\pi}{2(d-1)b_{0}}\left(\frac{I_{l}}{b_{0}}-\frac{b_{1}}{b_{0}}\right)\Bigg]\left(2l^{2}+x^{2}-(2l+x)^{2}\right)\Bigg]~.
\end{eqnarray}
\subsubsection{Subsystems perpendicular the boost}
We now consider that the equal length subsystems (separated by the distance $x$) are in a direction perpendicular to the boost. The mutual complexity then reads
\begin{eqnarray}\label{MCPER}
	\Delta\mathcal{C}^{\perp}&=&\frac{V_{d-2}}{8\pi G_{d+1}}\Bigg[-\frac{2^{d-2}\pi^{\frac{d-1}{2}}}{(d-1)^{2}}\left(\frac{\Gamma(\frac{d}{2d-2})}{\Gamma(\frac{1}{(2d-2)})}\right)^{d-3}\left(\frac{2}{l^{d-2}}+\frac{1}{x^{d-2}}-\frac{1}{(2l+x)^{d}}\right)\nonumber\\
	&-&\frac{1}{z_{I}^{d}(d-1)(2b_{0})^{2}}\left[c_{2}-c_{0}(d-1)+\frac{I_{l}(d-2)}{2(d-1)b_{0}^{2}}\right]\left(2l^{2}+x^{2}-(2l+x)^{2}\right)\Bigg]~.
\end{eqnarray}
The changes in mutual complexity due to involvement of boost can be written down with the help of the above expressions. For the parallel set up this reads
\begin{eqnarray}
	\delta \mathcal{C}^{\parallel}&=&-\frac{V_{d-2}}{8\pi G_{d+1}}\frac{1}{z_{I}^{d}(2b_{0})^{2}(d-1)}\Bigg[c_{2}-c_{0}d+\frac{(d-2)\pi}{2(d-1)b_{0}}\left(\frac{I_{l}}{b_{0}}-\frac{b_{1}}{b_{0}}\right)\Bigg]\left(2l^{2}+x^{2}-(2l+x)^{2}\right)~.
\end{eqnarray}
On the other hand, for the perpendicular set up, it reads
\begin{eqnarray}
	\delta \mathcal{C}^{\perp}
	&=&-\frac{V_{d-2}}{8\pi G_{d+1}}\frac{1}{z_{I}^{d}(d-1)(2b_{0})^{2}}\left[c_{2}-c_{0}(d-1)+\frac{I_{l}(d-2)}{2(d-1)b_{0}^{2}}\right]\left(2l^{2}+x^{2}-(2l+x)^{2}\right)~.
\end{eqnarray}

\section{Conclusion}\label{conc}
We now summarize our findings. In this paper, we have holographically computed and studied various information theoretic measures to quantify quantum correlation for mixed states. By choosing a proper gravity dual, we also investigate the effect of boost on these quantities. This amounts to looking at the effect of IR deformations of the quantum information theoretic quantities in the boundary CFT. The bulk geometry we have considered in this paper is the boosted black brane geometry with the direction of boost compactified on a circle. One of the motivations to consider this geometry lies in the fact that this type of geometry leads to Kaluza-Klein gauge charges. The presence of the boost in a particular direction provides an opportunity to choose the subsystems either in the direction of boost or in the perpendicular orientation with respect to the direction of boost. By considering this geometry, we first holographically compute the entanglement entropy for strip-like subsystems both along the boost and perpendicular to the boost. We have performed the mentioned computations under the thin strip approximation as we are interested in the leading order change (next to pure AdS) in the information theoretic quantities due to the inclusion of boost. We show that as long we confine ourselves to the leading order change in the HEE, both the HRT prescription and RT prescription produces the same result. The reason for this is that the static minimal surface emerges as the dominating piece of the HRT surface if one neglects the higher order terms. We also show that in both parallel and perpendicular scenario, one can relate the leading order change in the HEE (due to boost) to the CFT excitation energy and pressure. This relation has an appearance of a thermodynamics like law as it introduces the notion of an entanglement temperature. The boost on the bulk spacetime can be thought as a charged excitation in the CFT side. It is important to observe that the HEE increases with increase in the boost parameter for both the perpendicular as well as the parallel cases. A possible reason for this increase in the entanglement entropy can be an increase in the strip area of the subsystem living on the CFT side. Further, it is also observed that the entanglement entropy in the perpendicular case is greater than the parallel cases. This asymmetry arises due to the difference in the `pressure' of the CFT in the perpendicular and the parallel case, the pressure being more in the parallel case due to the effect of boost.\\
 We then proceed to holographically compute two different possible candidates to quantify quantum correlation of mixed states, namely, the minimal area of the entanglement wedge cross-section (EWCS) and entanglement negativity. We also compute the mutual information which determines the state of connectivity (connected or disconnected phase) of EWCS. Similar to the computation of HEE, we obtain these quantities for subsystems both along the boost and perpendicular to the boost.\\  
Further, inclusion of the boost in the dual field theory (or in other words, inclusion of excitations in the bulk) leads to a smaller value of the critical separation length $x_c$, at which the mutual information vanishes. This in turn also means that the connected entanglement wedge gets disconnected earlier, in comparison to the $\beta=0$ case. We also observe that the value of $x_c$ decreases with the increase in the $\beta$ parameter. This has been observed in both the parallel and perpendicular set ups. These observations have also been shown graphically. In case of entanglement negativity, we have considered two different set ups. Firstly, we consider two adjacent strip-like subsystems with different lengths and then we consider two disjoint strip-like subsystems with different lengths. The mentioned configurations have been considered for both parallel and perpendicular scenarios. We observe that in case of the disjoint set up, the entanglement negativity vanishes at a particular value of the separation distance $x=x_c^{\prime}$ and $x_c^{\prime}$ decreases with the increase in the $\beta$ parameter. We have shown this graphically. We have also computed the changes in the mentioned entanglement measures due to the boost, with respect to the pure $AdS_{d+1}$.\\
 Remarkably we observe that the leading order changes in HEE, EWCS (for adjacent subsystems of equal length) and entanglement negativity (for adjacent subsystems of equal length) are proportional to each other upto a constant factor. This in turn helps us to relate the leading order changes in EWCS and entanglement negativity to the excitation energy and pressure of boundary CFT. This we show by using the generalized first law of entanglement. Furthermore, by using these changes, we have shown the subsystem independent asymmetry ratio of these information theoretic quantities. These asymmetry ratios are bounded from the above where the bound is obtained by taking the $\beta\rightarrow1$ limit.\\
In addition to the measures of quantum correlation, we also compute the subregion complexity for boosted black brane, from which we evaluate the mutual complexity. We observe that the mutual complexity is superadditive in this case. We also provide a thermodynamics like law for HSC for both parallel and perpendicular strip. Finally, we consider the AdS wave geometry which is obtained by taking appropriate limits of the boosted black brane geometry. We compute the mentioned information theoretic quantities for this geometry once again.\\
We now make a final remark on the entanglement asymmetry. We observe that entanglement asymmetry is maximum for AdS wave geometry, which corresponds to the zero temperature CFT at the boundary. This can be explained by virtue of the entanglement pressure. The wave like excitation in the zero temperature CFT creates finite entanglement pressure along the direction of the wave, hence pressure in the transverse direction vanishes. But in case of finite temperature CFT at the boundary (which in turn implies including a black brane geometry in the bulk), there is also an entanglement pressure in the transverse direction. Thus, asymmetry in the entanglement entropy exists if there is an uniform wave like excitation or uniform flow in the CFT. The boosted black brane systems are used here since there are the only known examples to study asymmetric systems. It would also be interesting to explore other systems like Bianchi models having more generic asymmetry. 
\section*{Acknowledgments}
ARC would like to thank SNBNCBS for the Junior Research Fellowship. AS would like to acknowledge the support by Council of Scientific and Industrial Research (CSIR, Govt. of India) for the Senior Research Fellowship. 

\section*{Data availability information}
There is no associated data.

\section*{Funding and/or Conflicts of interests/Competing interests}
There is no conflict of interest and no funding.
 
 \section{Appendix A: List of beta function identities}\label{8}
 In this appendix, we will give some useful integrals which have been used in this paper.
 \begin{eqnarray}
 b_{0}&=&\int_{0}^{1}~dt~t^{d-1}\frac{1}{\sqrt{R}}=\frac{1}{2(d-1)}B\left(\frac{d}{2d-2},\frac{1}{2}\right)=\frac{\sqrt{\pi}\Gamma(\frac{d}{2d-2})}{\Gamma(\frac{1}{2d-2})}\nonumber\\
 b_{1}&=&\int_{0}^{1}~dt~t^{2d-1}\frac{1}{\sqrt{R}}=
 \frac{1}{2(d-1)}B\left(\frac{d}{d-1},\frac{1}{2}\right)=\frac{\sqrt{\pi}\Gamma(\frac{d}{d-1})}{(d+1)\Gamma(\frac{1}{2}+\frac{1}{d-1})}\nonumber\\
 b_{2}&=&\int_{0}^{1}~dt~t^{3d-1}\frac{1}{\sqrt{R}}=\frac{1}{2(d-1)}B\left(\frac{3d}{2d-2},\frac{1}{2}\right)\nonumber\\
 I_{l}&=&\int_{0}^{1}~dt~t^{d-1}(1-t^{d})\frac{1}{R^{\frac{3}{2}}}=\frac{d+1}{d-1}b_{1}-\frac{1}{d-1}b_{0}\nonumber\\
 c_{0}&=&\int_{0}^{1}~dt~\frac{t^{d}}{\sqrt{R}}=\frac{1}{2(d-1)}B\left(\frac{d+1}{2(d-1)},\frac{1}{2}\right)=\frac{\pi}{2(d^{2}-1)b_{1}}\nonumber\\
 c_{1}&=&\int_{0}^{1}~dt~\frac{t^{2d}}{\sqrt{R}}=\frac{1}{2(d-1)}B\left(\frac{2d+1}{2(d-1)},\frac{1}{2}\right)=\frac{\pi}{2(d-1)(2d+1)b_{2}}\nonumber\\
 c_{2}&=&\int_{0}^{1}~dt~\frac{(1-t^{d})}{R^{\frac{3}{2}}}=\frac{2}{d-1}c_{0}+\frac{d-2}{2(d-1)^{2}}B\left(\frac{1}{2(d-1)},\frac{1}{2}\right)=\frac{\pi}{(d+1)(d-1)^{2}b_{1}}+\frac{\pi(d-2)}{2(d-1)^{2}b_{0}}\nonumber\\
 c_{3}&=&\int~dt~\frac{t^{d}(1-t^{d})}{R^{\frac{3}{2}}}=\frac{2}{d-1}c_{0}+\frac{d+2}{d-1}c_{1}\int_{0}^{1}\frac{dt}{\sqrt{1-t^{2(d-1)}}}=\frac{\pi}{2(d-1)b_{0}}\nonumber\\
 J_{l}&=&\int_{0}^{1}~dt~t^{d-1}\left(\frac{\beta^{2}\gamma^{2}}{4}t^{d}+\beta^{4}\gamma^{4}\left(\frac{3(1-t^{d})}{8(1-t^{2(d-1)})}-\frac{1}{2}\right)\right)\frac{(1-t^{d})}{R^{\frac{3}{2}}}=\beta^{2}\gamma^{2}J_{1}+\beta^{4}\gamma^{4}J_{2}\nonumber\\
 K_{l}&=&\int_{0}^{1}~dt~t^{d-1}\left(\frac{\beta^{2}\gamma^{2}}{4}t^{d}+\beta^{4}\gamma^{4}\left(\frac{3(1-t^{d})}{8(1-t^{2(d-1)})}-\frac{1}{2}\right)\right)\frac{(1-t^{d})}{R^{\frac{3}{2}}}=\beta^{2}\gamma^{2}K_{1}+\beta^{4}\gamma^{4}K_{2}
 \end{eqnarray}
where $R=1-t^{2d-2}$ and $J_{1}, J_{2}, K_{1}, K_{2}$ are given by
 \begin{eqnarray}
 J_{1}&=&\frac{1}{4(d-1)}\left((2d+1)b_{2}-(d+1)b_{1}\right)\nonumber\\
 J_{2}&=&\frac{1}{8(d-1)^{2}}\left((3-2d)b_{0}-2(d+1)(3-d)b_{1}+3(2d+1)b_{2}\right)-\frac{I_{l}}{2}\nonumber\\
 K_{1}&=&\frac{1}{2(d-1)}c_{0}+\frac{d+2}{4(d-1)}c_{1}\nonumber\\
 K_{2}&=&-\frac{d-4}{2(d-1)^{2}}c_{0}+\frac{(d+2)(d-4)}{8(d-1)^{2}}c_{1}+\frac{d}{8(d-1)}c_{2}~.
 \end{eqnarray}
 Note that $B(m,n)=\frac{\Gamma(m)\Gamma(n)}{\Gamma(m+n)}$ is the standard beta function and we have used the identity
 \begin{equation}
 B(x,\frac{1}{2})B(x+\frac{1}{2},\frac{1}{2})=\frac{\pi}{x}~.
 \end{equation}
 Some more integrals are as follows
 \begin{eqnarray}
 a_{0}&=&\int_{0}^{1}~dt~t^{-d+1}\frac{1}{\sqrt{R}}=\frac{1}{2(d-1)}B\left(\frac{1-(d/2)}{d-1},\frac{1}{2}\right)\nonumber\\
 a_{1}&=&\int_{0}^{1}~dt~t^{-d+1}\frac{t^{d}}{\sqrt{R}}=\frac{1}{2(d-1)}B\left(\frac{1}{d-1},\frac{1}{2}\right)\nonumber\\
 a_{2}&=&\int_{0}^{1}~dt~t^{-d+1}\frac{t^{2d}}{\sqrt{R}}=\frac{1}{2(d-1)}B\left(\frac{1+(d/2)}{d-1},\frac{1}{2}\right)\nonumber\\
 I_{a}&=&\int_{0}^{1}~dt~t^{d-1}(1-t^{2d})\frac{1}{R^{\frac{3}{2}}}=\frac{2d+1}{d-1}b_{0}-\frac{1}{d-1}b_{0}~.
 \end{eqnarray}
We record some more identities given as 
 \begin{eqnarray}
 b_{0}=(2-d)a_{0}~,~~b_{1}=\frac{2}{d+1}a_{1}~,~b_{2}=\frac{2+d}{2d+1}a_{2} ~~.
 \end{eqnarray}
\section{Appendix B: Expressions of $\lambda_1$, $\lambda_2$, $\lambda_3$ and $\lambda_4$}\label{9}
In this appendix, we provide explicit expressions of $\lambda_1$, $\lambda_2$, $\lambda_3$ and $\lambda_4$ which have appeared in our analysis. This reads
\begin{eqnarray}
	\lambda_1&=& \left[\frac{4}{b_1(d+1)}\right] \left[\frac{\frac{1}{2}-\left(1+\frac{2\beta^{2}\gamma^{2}}{d-1}\right)\frac{b_{1}}{b_{0}}+\frac{\beta^{2}\gamma^{2}}{d-1}}{\frac{d-1}{d+1}+\frac{2\beta^{2}\gamma^{2}}{d+1}}\right]\\
	\lambda_2&=& - \left[\frac{3}{d+1}\right]\left[\frac{\frac{d-1}{2}+\beta^{2}\gamma^{2}}{\frac{d-1}{d+1}+\frac{2\beta^{2}\gamma^{2}}{d+1}}\right]\\
	\lambda_3 &=& \left[\frac{4}{b_1(d+1)}\right]\left[\frac{\frac{1}{2}\left(1+\beta^{2}\gamma^{2}\right)-\left(\frac{b_{1}}{b_{0}}+\beta^{2}\gamma^{2}\frac{I_{l}}{b_{0}}\right)}{\frac{d-1}{d+1}+\beta^{2}\gamma^{2}}\right]\\
	\lambda_4 &=& - \frac{3}{2}~~.
\end{eqnarray}
\bibliographystyle{hephys}  
\bibliography{Reference}

\begin{thebibliography}{100}
\newcommand{\enquote}[1]{``#1''}

\bibitem{Maldacena:1997re}
J.~M. Maldacena, \enquote{{The Large N limit of superconformal field theories
  and supergravity}},
  \href{http://dx.doi.org/10.1023/A:1026654312961}{\emph{Int. J. Theor. Phys.}
  \textbf{38} (1999) 1113}, \href{http://arxiv.org/abs/hep-th/9711200}{{\tt
  arXiv:hep-th/9711200}}.

\bibitem{Witten:1998qj}
E.~Witten, \enquote{{Anti-de Sitter space and holography}},
  \href{http://dx.doi.org/10.4310/ATMP.1998.v2.n2.a2}{\emph{Adv. Theor. Math.
  Phys.} \textbf{2} (1998) 253},
  \href{http://arxiv.org/abs/hep-th/9802150}{{\tt arXiv:hep-th/9802150}}.

\bibitem{Aharony:1999ti}
O.~Aharony, S.~S. Gubser, J.~M. Maldacena, H.~Ooguri and Y.~Oz, \enquote{{Large
  N field theories, string theory and gravity}},
  \href{http://dx.doi.org/10.1016/S0370-1573(99)00083-6}{\emph{Phys. Rept.}
  \textbf{323} (2000) 183}, \href{http://arxiv.org/abs/hep-th/9905111}{{\tt
  arXiv:hep-th/9905111}}.

\bibitem{Chuang:2000}
M.~A. Nielsen and I.~L. Chuang, Quantum Computation and Quantum Information,
  Cambridge University Press 2000.

\bibitem{Ryu:2006bv}
S.~Ryu and T.~Takayanagi, \enquote{{Holographic derivation of entanglement
  entropy from AdS/CFT}},
  \href{http://dx.doi.org/10.1103/PhysRevLett.96.181602}{\emph{Phys. Rev.
  Lett.} \textbf{96} (2006) 181602},
  \href{http://arxiv.org/abs/hep-th/0603001}{{\tt arXiv:hep-th/0603001}}.

\bibitem{Ryu:2006ef}
S.~Ryu and T.~Takayanagi, \enquote{{Aspects of Holographic Entanglement
  Entropy}}, \href{http://dx.doi.org/10.1088/1126-6708/2006/08/045}{\emph{JHEP}
  \textbf{08} (2006) 045}, \href{http://arxiv.org/abs/hep-th/0605073}{{\tt
  arXiv:hep-th/0605073}}.

\bibitem{Nishioka:2009un}
T.~Nishioka, S.~Ryu and T.~Takayanagi, \enquote{{Holographic Entanglement
  Entropy: An Overview}},
  \href{http://dx.doi.org/10.1088/1751-8113/42/50/504008}{\emph{J. Phys. A}
  \textbf{42} (2009) 504008}, \href{http://arxiv.org/abs/0905.0932}{{\tt
  arXiv:0905.0932 [hep-th]}}.

\bibitem{Srednicki:1993im}
M.~Srednicki, \enquote{{Entropy and area}},
  \href{http://dx.doi.org/10.1103/PhysRevLett.71.666}{\emph{Phys. Rev. Lett.}
  \textbf{71} (1993) 666}, \href{http://arxiv.org/abs/hep-th/9303048}{{\tt
  arXiv:hep-th/9303048}}.

\bibitem{Terhal_2002}
B.~M. Terhal, M.~Horodecki, D.~W. Leung and D.~P. DiVincenzo, \enquote{The
  entanglement of purification},
  \href{http://dx.doi.org/10.1063/1.1498001}{\emph{Journal of Mathematical
  Physics} \textbf{43[9]} (2002) 4286–4298}.

\bibitem{Takayanagi:2017knl}
T.~Takayanagi and K.~Umemoto, \enquote{{Entanglement of purification through
  holographic duality}},
  \href{http://dx.doi.org/10.1038/s41567-018-0075-2}{\emph{Nature Phys.}
  \textbf{14[6]} (2018) 573}, \href{http://arxiv.org/abs/1708.09393}{{\tt
  arXiv:1708.09393 [hep-th]}}.

\bibitem{Nguyen:2017yqw}
P.~Nguyen, T.~Devakul, M.~G. Halbasch, M.~P. Zaletel and B.~Swingle,
  \enquote{{Entanglement of purification: from spin chains to holography}},
  \href{http://dx.doi.org/10.1007/JHEP01(2018)098}{\emph{JHEP} \textbf{01}
  (2018) 098}, \href{http://arxiv.org/abs/1709.07424}{{\tt arXiv:1709.07424
  [hep-th]}}.

\bibitem{Vidal:2002zz}
G.~Vidal and R.~Werner, \enquote{{Computable measure of entanglement}},
  \href{http://dx.doi.org/10.1103/PhysRevA.65.032314}{\emph{Phys. Rev. A}
  \textbf{65} (2002) 032314}, \href{http://arxiv.org/abs/quant-ph/0102117}{{\tt
  arXiv:quant-ph/0102117}}.

\bibitem{PhysRevA.58.883}
K.~\ifmmode~\dot{Z}\else \.{Z}\fi{}yczkowski, P.~Horodecki, A.~Sanpera and
  M.~Lewenstein, \enquote{Volume of the set of separable states},
  \href{http://dx.doi.org/10.1103/PhysRevA.58.883}{\emph{Phys. Rev. A}
  \textbf{58} (1998) 883}.

\bibitem{PhysRevA.60.3496}
K.~\ifmmode~\dot{Z}\else \.{Z}\fi{}yczkowski, \enquote{Volume of the set of
  separable states. II},
  \href{http://dx.doi.org/10.1103/PhysRevA.60.3496}{\emph{Phys. Rev. A}
  \textbf{60} (1999) 3496}.

\bibitem{Jain:2017aqk}
P.~Jain, V.~Malvimat, S.~Mondal and G.~Sengupta, \enquote{{Holographic
  entanglement negativity conjecture for adjacent intervals in $AdS_3/CFT_2$}},
  \href{http://dx.doi.org/10.1016/j.physletb.2019.04.037}{\emph{Phys. Lett. B}
  \textbf{793} (2019) 104}, \href{http://arxiv.org/abs/1707.08293}{{\tt
  arXiv:1707.08293 [hep-th]}}.

\bibitem{PhysRevLett.122.141601}
K.~Tamaoka, \enquote{Entanglement Wedge Cross Section from the Dual Density
  Matrix}, \href{http://dx.doi.org/10.1103/PhysRevLett.122.141601}{\emph{Phys.
  Rev. Lett.} \textbf{122} (2019) 141601}.

\bibitem{Ghasemi:2021jiy}
M.~Ghasemi, A.~Naseh and R.~Pirmoradian, \enquote{{Odd entanglement entropy and
  logarithmic negativity for thermofield double states}},
  \href{http://dx.doi.org/10.1007/JHEP10(2021)128}{\emph{JHEP} \textbf{10}
  (2021) 128}, \href{http://arxiv.org/abs/2106.15451}{{\tt arXiv:2106.15451
  [hep-th]}}.

\bibitem{Basak:2022gcv}
J.~K. Basak, H.~Chourasiya, V.~Raj and G.~Sengupta, \enquote{{Odd entanglement
  entropy in Galilean conformal field theories and flat holography}},
  \href{http://arxiv.org/abs/2203.03902}{{\tt arXiv:2203.03902 [hep-th]}}.

\bibitem{Dutta:2019gen}
S.~Dutta and T.~Faulkner, \enquote{{A canonical purification for the
  entanglement wedge cross-section}},
  \href{http://dx.doi.org/10.1007/JHEP03(2021)178}{\emph{JHEP} \textbf{03}
  (2021) 178}, \href{http://arxiv.org/abs/1905.00577}{{\tt arXiv:1905.00577
  [hep-th]}}.

\bibitem{Chu:2019etd}
J.~Chu, R.~Qi and Y.~Zhou, \enquote{{Generalizations of Reflected Entropy and
  the Holographic Dual}},
  \href{http://dx.doi.org/10.1007/JHEP03(2020)151}{\emph{JHEP} \textbf{03}
  (2020) 151}, \href{http://arxiv.org/abs/1909.10456}{{\tt arXiv:1909.10456
  [hep-th]}}.

\bibitem{Basak:2022cjs}
J.~K. Basak, H.~Chourasiya, V.~Raj and G.~Sengupta, \enquote{{Reflected entropy
  in Galilean conformal field theories and flat holography}},
  \href{http://arxiv.org/abs/2202.01201}{{\tt arXiv:2202.01201 [hep-th]}}.

\bibitem{S.Arora:2009book}
S.~Arora and B.~Barak, \enquote{Computational complexity: A modern approach,
  Cambridge University Press (2009)}, .

\bibitem{C.Moore:2011book}
C.~Moore and S.~Mertens, \enquote{The nature of computation, Oxford University
  Press (2011)}, .

\bibitem{Nielsen1}
M.A.Nielsen, \enquote{{A geometric approach to quantum circuit lower bound}},
  \href{http://arxiv.org/abs/0502070}{{\tt arXiv:0502070 [quant-ph]}}.

\bibitem{Nielsen2}
M.~G. A. C.~D. Michael A.~Nielsen, Mark R.~Dowling, \enquote{{ Quantum
  Computation as Geometry}},
  \href{http://dx.doi.org/10.1126/science.1121541}{\emph{Science} },
  \href{http://arxiv.org/abs/0603161}{{\tt arXiv:0603161 [quant-ph]}}.

\bibitem{Susskind:2018pmk}
L.~Susskind, \href{http://dx.doi.org/10.1007/978-3-030-45109-7}{\enquote{{Three
  Lectures on Complexity and Black Holes}}, }SpringerBriefs in Physics,
  Springer 2018, \href{http://arxiv.org/abs/1810.11563}{{\tt arXiv:1810.11563
  [hep-th]}}.

\bibitem{Ali:2018fcz}
T.~Ali, A.~Bhattacharyya, S.~Shajidul~Haque, E.~H. Kim and N.~Moynihan,
  \enquote{{Time Evolution of Complexity: A Critique of Three Methods}},
  \href{http://dx.doi.org/10.1007/JHEP04(2019)087}{\emph{JHEP} \textbf{04}
  (2019) 087}, \href{http://arxiv.org/abs/1810.02734}{{\tt arXiv:1810.02734
  [hep-th]}}.

\bibitem{Alishahiha:2018lfv}
M.~Alishahiha, K.~Babaei~Velni and M.~R. Mohammadi~Mozaffar, \enquote{{Black
  hole subregion action and complexity}},
  \href{http://dx.doi.org/10.1103/PhysRevD.99.126016}{\emph{Phys. Rev. D}
  \textbf{99[12]} (2019) 126016}, \href{http://arxiv.org/abs/1809.06031}{{\tt
  arXiv:1809.06031 [hep-th]}}.

\bibitem{PhysRevD.99.086016}
E.~C\'aceres, J.~Couch, S.~Eccles and W.~Fischler, \enquote{Holographic
  purification complexity},
  \href{http://dx.doi.org/10.1103/PhysRevD.99.086016}{\emph{Phys. Rev. D}
  \textbf{99} (2019) 086016}.

\bibitem{Caceres:2019pgf}
E.~Caceres, S.~Chapman, J.~D. Couch, J.~P. Hernandez, R.~C. Myers and S.-M.
  Ruan, \enquote{{Complexity of Mixed States in QFT and Holography}},
  \href{http://dx.doi.org/10.1007/JHEP03(2020)012}{\emph{JHEP} \textbf{03}
  (2020) 012}, \href{http://arxiv.org/abs/1909.10557}{{\tt arXiv:1909.10557
  [hep-th]}}.

\bibitem{Agon:2018zso}
C.~A. Ag\'on, M.~Headrick and B.~Swingle, \enquote{{Subsystem Complexity and
  Holography}}, \href{http://dx.doi.org/10.1007/JHEP02(2019)145}{\emph{JHEP}
  \textbf{02} (2019) 145}, \href{http://arxiv.org/abs/1804.01561}{{\tt
  arXiv:1804.01561 [hep-th]}}.

\bibitem{Ruan:2021wep}
S.-M. Ruan, {Circuit Complexity of Mixed States}, Ph.D. thesis, Waterloo U.
  2021.

\bibitem{Mateos:2011ix}
D.~Mateos and D.~Trancanelli, \enquote{{The anisotropic N=4 super Yang-Mills
  plasma and its instabilities}},
  \href{http://dx.doi.org/10.1103/PhysRevLett.107.101601}{\emph{Phys. Rev.
  Lett.} \textbf{107} (2011) 101601},
  \href{http://arxiv.org/abs/1105.3472}{{\tt arXiv:1105.3472 [hep-th]}}.

\bibitem{Bhatta:2019eog}
A.~Bhatta, S.~Chakrabortty, S.~Dengiz and E.~Kilicarslan, \enquote{{High
  temperature behavior of non-local observables in boosted strongly coupled
  plasma: A holographic study}},
  \href{http://dx.doi.org/10.1140/epjc/s10052-020-8206-1}{\emph{Eur. Phys. J.
  C} \textbf{80[7]} (2020) 663}, \href{http://arxiv.org/abs/1909.03088}{{\tt
  arXiv:1909.03088 [hep-th]}}.

\bibitem{SHURYAK200564}
E.~Shuryak, \enquote{What RHIC experiments and theory tell us about properties
  of quark–gluon plasma?},
  \href{http://dx.doi.org/https://doi.org/10.1016/j.nuclphysa.2004.10.022}{\emph{Nuclear
  Physics A} \textbf{750[1]} (2005) 64}, quark-Gluon Plasma. New Discoveries at
  RHIC: Case for the Strongly Interacting Quark-Gluon Plasma. Contributions
  from the RBRC Workshop held May 14-15, 2004,
  \href{https://www.sciencedirect.com/science/article/pii/S0375947404011340}{{\tt
  URL}}.

\bibitem{BACK200528}
J.~et~al, \enquote{The PHOBOS perspective on discoveries at RHIC},
  \href{http://dx.doi.org/https://doi.org/10.1016/j.nuclphysa.2005.03.084}{\emph{Nuclear
  Physics A} \textbf{757[1]} (2005) 28}.

\bibitem{SHURYAK200948}
E.~Shuryak, \enquote{Physics of strongly coupled quark–gluon plasma},
  \href{http://dx.doi.org/https://doi.org/10.1016/j.ppnp.2008.09.001}{\emph{Progress
  in Particle and Nuclear Physics} \textbf{62[1]} (2009) 48},
  \href{https://www.sciencedirect.com/science/article/pii/S0146641008000732}{{\tt
  URL}}.

\bibitem{Shuryak:2014zxa}
E.~Shuryak, \enquote{{Strongly coupled quark-gluon plasma in heavy ion
  collisions}},
  \href{http://dx.doi.org/10.1103/RevModPhys.89.035001}{\emph{Rev. Mod. Phys.}
  \textbf{89} (2017) 035001}, \href{http://arxiv.org/abs/1412.8393}{{\tt
  arXiv:1412.8393 [hep-ph]}}.

\bibitem{PhysRevLett.103.232001}
M.~Panero, \enquote{Thermodynamics of the QCD Plasma and the Large-$N$ Limit},
  \href{http://dx.doi.org/10.1103/PhysRevLett.103.232001}{\emph{Phys. Rev.
  Lett.} \textbf{103} (2009) 232001}.

\bibitem{Bhaseen:2013ypa}
M.~J. Bhaseen, B.~Doyon, A.~Lucas and K.~Schalm, \enquote{{Far from equilibrium
  energy flow in quantum critical systems}},
  \href{http://dx.doi.org/10.1038/nphys3220}{\emph{Nature Phys.} \textbf{11}
  (2015) 5}, \href{http://arxiv.org/abs/1311.3655}{{\tt arXiv:1311.3655
  [hep-th]}}.

\bibitem{Mishra:2018tzj}
R.~Mishra and H.~Singh, \enquote{{Entanglement entropy at higher orders for the
  states of $a = 3$ Lifshitz theory}},
  \href{http://dx.doi.org/10.1016/j.nuclphysb.2018.11.012}{\emph{Nucl. Phys. B}
  \textbf{938} (2019) 307}, \href{http://arxiv.org/abs/1804.01361}{{\tt
  arXiv:1804.01361 [hep-th]}}.

\bibitem{Balasubramanian:1999re}
V.~Balasubramanian and P.~Kraus, \enquote{{A Stress tensor for Anti-de Sitter
  gravity}}, \href{http://dx.doi.org/10.1007/s002200050764}{\emph{Commun. Math.
  Phys.} \textbf{208} (1999) 413},
  \href{http://arxiv.org/abs/hep-th/9902121}{{\tt arXiv:hep-th/9902121}}.

\bibitem{Kraus:1999di}
P.~Kraus, F.~Larsen and R.~Siebelink, \enquote{{The gravitational action in
  asymptotically AdS and flat space-times}},
  \href{http://dx.doi.org/10.1016/S0550-3213(99)00549-0}{\emph{Nucl. Phys. B}
  \textbf{563} (1999) 259}, \href{http://arxiv.org/abs/hep-th/9906127}{{\tt
  arXiv:hep-th/9906127}}.

\bibitem{Mishra:2015cpa}
R.~Mishra and H.~Singh, \enquote{{Perturbative entanglement thermodynamics for
  AdS spacetime: Renormalization}},
  \href{http://dx.doi.org/10.1007/JHEP10(2015)129}{\emph{JHEP} \textbf{10}
  (2015) 129}, \href{http://arxiv.org/abs/1507.03836}{{\tt arXiv:1507.03836
  [hep-th]}}.

\bibitem{Mishra:2016yor}
R.~Mishra and H.~Singh, \enquote{{Entanglement asymmetry for boosted black
  branes and the bound}},
  \href{http://dx.doi.org/10.1142/S0217751X17500919}{\emph{Int. J. Mod. Phys.
  A} \textbf{32[16]} (2017) 1750091},
  \href{http://arxiv.org/abs/1603.06058}{{\tt arXiv:1603.06058 [hep-th]}}.

\bibitem{hubeny2007covariant}
V.~E. Hubeny, M.~Rangamani and T.~Takayanagi, \enquote{A covariant holographic
  entanglement entropy proposal}, \emph{Journal of High Energy Physics}
  \textbf{2007[07]} (2007) 062.

\bibitem{mishra2018study}
R.~Mishra, Study of Geometrical Aspects of Holographic Entanglement Entropy and
  First Law, Ph.D. thesis, HOMI BHABHA NATIONAL INSTITUTE 2018.

\bibitem{Maulik:2020tzm}
S.~Maulik and H.~Singh, \enquote{{Entanglement entropy and the first law at
  third order for boosted black branes}},
  \href{http://dx.doi.org/10.1007/JHEP04(2021)065}{\emph{JHEP} \textbf{04}
  (2021) 065}, \href{http://arxiv.org/abs/2012.09530}{{\tt arXiv:2012.09530
  [hep-th]}}.

\bibitem{Jokela:2019ebz}
N.~Jokela and A.~P\"onni, \enquote{{Notes on entanglement wedge cross
  sections}}, \href{http://dx.doi.org/10.1007/JHEP07(2019)087}{\emph{JHEP}
  \textbf{07} (2019) 087}, \href{http://arxiv.org/abs/1904.09582}{{\tt
  arXiv:1904.09582 [hep-th]}}.

\bibitem{BabaeiVelni:2019pkw}
K.~Babaei~Velni, M.~R. Mohammadi~Mozaffar and M.~Vahidinia, \enquote{{Some
  Aspects of Entanglement Wedge Cross-Section}},
  \href{http://dx.doi.org/10.1007/JHEP05(2019)200}{\emph{JHEP} \textbf{05}
  (2019) 200}, \href{http://arxiv.org/abs/1903.08490}{{\tt arXiv:1903.08490
  [hep-th]}}.

\bibitem{Saha:2021kwq}
A.~Saha and S.~Gangopadhyay, \enquote{{Holographic study of entanglement and
  complexity for mixed states}},
  \href{http://dx.doi.org/10.1103/PhysRevD.103.086002}{\emph{Phys. Rev. D}
  \textbf{103[8]} (2021) 086002}, \href{http://arxiv.org/abs/2101.00887}{{\tt
  arXiv:2101.00887 [hep-th]}}.

\bibitem{Chowdhury:2021idy}
A.~R. Chowdhury, A.~Saha and S.~Gangopadhyay, \enquote{{Entanglement wedge
  cross-section for noncommutative Yang-Mills theory}},
  \href{http://dx.doi.org/10.1007/JHEP02(2022)192}{\emph{JHEP} \textbf{02}
  (2022) 192}, \href{http://arxiv.org/abs/2106.04562}{{\tt arXiv:2106.04562
  [hep-th]}}.

\bibitem{Sahraei:2021wqn}
M.~Sahraei, M.~J. Vasli, M.~R.~M. Mozaffar and K.~B. Velni,
  \enquote{{Entanglement wedge cross section in holographic excited states}},
  \href{http://dx.doi.org/10.1007/JHEP08(2021)038}{\emph{JHEP} \textbf{08}
  (2021) 038}, \href{http://arxiv.org/abs/2105.12476}{{\tt arXiv:2105.12476
  [hep-th]}}.

\bibitem{Liu:2021rks}
P.~Liu, C.~Niu, Z.-J. Shi and C.-Y. Zhang, \enquote{{Entanglement wedge minimum
  cross-section in holographic massive gravity theory}},
  \href{http://dx.doi.org/10.1007/JHEP08(2021)113}{\emph{JHEP} \textbf{08}
  (2021) 113}, \href{http://arxiv.org/abs/2104.08070}{{\tt arXiv:2104.08070
  [hep-th]}}.

\bibitem{Basu:2021awn}
D.~Basu, A.~Chandra, V.~Raj and G.~Sengupta, \enquote{{Entanglement Wedge in
  Flat Holography and Entanglement Negativity}},
  \href{http://arxiv.org/abs/2106.14896}{{\tt arXiv:2106.14896 [hep-th]}}.

\bibitem{Kudler-Flam:2018qjo}
J.~Kudler-Flam and S.~Ryu, \enquote{{Entanglement negativity and minimal
  entanglement wedge cross sections in holographic theories}},
  \href{http://dx.doi.org/10.1103/PhysRevD.99.106014}{\emph{Phys. Rev. D}
  \textbf{99[10]} (2019) 106014}, \href{http://arxiv.org/abs/1808.00446}{{\tt
  arXiv:1808.00446 [hep-th]}}.

\bibitem{Kusuki:2019zsp}
Y.~Kusuki, J.~Kudler-Flam and S.~Ryu, \enquote{{Derivation of holographic
  negativity in AdS$_3$/CFT$_2$}},
  \href{http://dx.doi.org/10.1103/PhysRevLett.123.131603}{\emph{Phys. Rev.
  Lett.} \textbf{123[13]} (2019) 131603},
  \href{http://arxiv.org/abs/1907.07824}{{\tt arXiv:1907.07824 [hep-th]}}.

\bibitem{Blanco:2013joa}
D.~D. Blanco, H.~Casini, L.-Y. Hung and R.~C. Myers, \enquote{{Relative Entropy
  and Holography}},
  \href{http://dx.doi.org/10.1007/JHEP08(2013)060}{\emph{JHEP} \textbf{08}
  (2013) 060}, \href{http://arxiv.org/abs/1305.3182}{{\tt arXiv:1305.3182
  [hep-th]}}.

\bibitem{Chaturvedi:2016rft}
P.~Chaturvedi, V.~Malvimat and G.~Sengupta, \enquote{{Entanglement negativity,
  Holography and Black holes}},
  \href{http://dx.doi.org/10.1140/epjc/s10052-018-5969-8}{\emph{Eur. Phys. J.
  C} \textbf{78[6]} (2018) 499}, \href{http://arxiv.org/abs/1602.01147}{{\tt
  arXiv:1602.01147 [hep-th]}}.

\bibitem{Chaturvedi:2016rcn}
P.~Chaturvedi, V.~Malvimat and G.~Sengupta, \enquote{{Holographic Quantum
  Entanglement Negativity}},
  \href{http://dx.doi.org/10.1007/JHEP05(2018)172}{\emph{JHEP} \textbf{05}
  (2018) 172}, \href{http://arxiv.org/abs/1609.06609}{{\tt arXiv:1609.06609
  [hep-th]}}.

\bibitem{Jain:2017xsu}
P.~Jain, V.~Malvimat, S.~Mondal and G.~Sengupta, \enquote{{Holographic
  entanglement negativity for adjacent subsystems in AdS$_{d+1}$/CFT$_{d}$}},
  \href{http://dx.doi.org/10.1140/epjp/i2018-12113-0}{\emph{Eur. Phys. J. Plus}
  \textbf{133[8]} (2018) 300}, \href{http://arxiv.org/abs/1708.00612}{{\tt
  arXiv:1708.00612 [hep-th]}}.

\bibitem{Jain:2017uhe}
P.~Jain, V.~Malvimat, S.~Mondal and G.~Sengupta, \enquote{{Covariant
  holographic entanglement negativity for adjacent subsystems in AdS$_3$
  /CFT$_2$}},
  \href{http://dx.doi.org/10.1016/j.nuclphysb.2019.114683}{\emph{Nucl. Phys. B}
  \textbf{945} (2019) 114683}, \href{http://arxiv.org/abs/1710.06138}{{\tt
  arXiv:1710.06138 [hep-th]}}.

\bibitem{Malvimat:2018izs}
V.~Malvimat, H.~Parihar, B.~Paul and G.~Sengupta, \enquote{{Entanglement
  Negativity in Galilean Conformal Field Theories}},
  \href{http://dx.doi.org/10.1103/PhysRevD.100.026001}{\emph{Phys. Rev. D}
  \textbf{100[2]} (2019) 026001}, \href{http://arxiv.org/abs/1810.08162}{{\tt
  arXiv:1810.08162 [hep-th]}}.

\bibitem{Malvimat:2018cfe}
V.~Malvimat, S.~Mondal and G.~Sengupta, \enquote{{Time Evolution of
  Entanglement Negativity from Black Hole Interiors}},
  \href{http://dx.doi.org/10.1007/JHEP05(2019)183}{\emph{JHEP} \textbf{05}
  (2019) 183}, \href{http://arxiv.org/abs/1812.04424}{{\tt arXiv:1812.04424
  [hep-th]}}.

\bibitem{Rogerson:2022yim}
D.~Rogerson, F.~Pollmann and A.~Roy, \enquote{{Entanglement entropy and
  negativity in the Ising model with defects}},
  \href{http://arxiv.org/abs/2204.03601}{{\tt arXiv:2204.03601 [hep-th]}}.

\bibitem{Matsumura:2022ide}
A.~Matsumura, \enquote{{Role of coherence in entanglement due to gravity}},
  \href{http://arxiv.org/abs/2204.00324}{{\tt arXiv:2204.00324 [quant-ph]}}.

\bibitem{Bertini:2022fnr}
B.~Bertini, K.~Klobas and T.-C. Lu, \enquote{{Entanglement Negativity and
  Mutual Information after a Quantum Quench: Exact Link from Space-Time
  Duality}}, \href{http://arxiv.org/abs/2203.17254}{{\tt arXiv:2203.17254
  [quant-ph]}}.

\bibitem{Roik:2022gbb}
J.~Roik, K.~Bartkiewicz, A.~\v{C}ernoch and K.~Lemr, \enquote{{Entanglement
  quantification from collective measurements processed by machine learning}},
  \href{http://arxiv.org/abs/2203.01607}{{\tt arXiv:2203.01607 [quant-ph]}}.

\bibitem{Dong:2021oad}
X.~Dong, S.~McBride and W.~W. Weng, \enquote{{Replica Wormholes and Holographic
  Entanglement Negativity}}, \href{http://arxiv.org/abs/2110.11947}{{\tt
  arXiv:2110.11947 [hep-th]}}.

\bibitem{Bhattacharya:2021dnd}
A.~Bhattacharya, A.~Bhattacharyya, P.~Nandy and A.~K. Patra, \enquote{{Partial
  islands and subregion complexity in geometric secret-sharing model}},
  \href{http://dx.doi.org/10.1007/JHEP12(2021)091}{\emph{JHEP} \textbf{12}
  (2021) 091}, \href{http://arxiv.org/abs/2109.07842}{{\tt arXiv:2109.07842
  [hep-th]}}.

\bibitem{Hejazi:2021yhz}
K.~Hejazi and H.~Shapourian, \enquote{{Symmetry protected entanglement in
  random mixed states}}, \href{http://arxiv.org/abs/2112.00032}{{\tt
  arXiv:2112.00032 [quant-ph]}}.

\bibitem{Afrasiar:2021hld}
M.~Afrasiar, J.~Kumar~Basak, V.~Raj and G.~Sengupta, \enquote{{Holographic
  Entanglement Negativity for Disjoint Subsystems in Conformal Field Theories
  with a Conserved Charge}}, \href{http://arxiv.org/abs/2106.14918}{{\tt
  arXiv:2106.14918 [hep-th]}}.

\bibitem{Basu:2022nds}
D.~Basu, H.~Parihar, V.~Raj and G.~Sengupta, \enquote{{Entanglement negativity,
  reflected entropy and anomalous gravitation}},
  \href{http://arxiv.org/abs/2202.00683}{{\tt arXiv:2202.00683 [hep-th]}}.

\bibitem{Malvimat:2018ood}
V.~Malvimat, S.~Mondal, B.~Paul and G.~Sengupta, \enquote{{Covariant
  holographic entanglement negativity for disjoint intervals in
  $AdS_3/CFT_2$}},
  \href{http://dx.doi.org/10.1140/epjc/s10052-019-7032-9}{\emph{Eur. Phys. J.
  C} \textbf{79[6]} (2019) 514}, \href{http://arxiv.org/abs/1812.03117}{{\tt
  arXiv:1812.03117 [hep-th]}}.

\bibitem{KumarBasak:2020viv}
J.~Kumar~Basak, H.~Parihar, B.~Paul and G.~Sengupta, \enquote{{Holographic
  entanglement negativity for disjoint subsystems in
  $\mathrm{AdS_{d+1}/CFT_d}$}}, \href{http://arxiv.org/abs/2001.10534}{{\tt
  arXiv:2001.10534 [hep-th]}}.

\bibitem{Maldacena:2013xja}
J.~Maldacena and L.~Susskind, \enquote{{Cool horizons for entangled black
  holes}}, \href{http://dx.doi.org/10.1002/prop.201300020}{\emph{Fortsch.
  Phys.} \textbf{61} (2013) 781}, \href{http://arxiv.org/abs/1306.0533}{{\tt
  arXiv:1306.0533 [hep-th]}}.

\bibitem{Susskind:2014moa}
L.~Susskind, \enquote{{Entanglement is not enough}},
  \href{http://dx.doi.org/10.1002/prop.201500095}{\emph{Fortsch. Phys.}
  \textbf{64} (2016) 49}, \href{http://arxiv.org/abs/1411.0690}{{\tt
  arXiv:1411.0690 [hep-th]}}.

\bibitem{Brown:2015lvg}
A.~R. Brown, D.~A. Roberts, L.~Susskind, B.~Swingle and Y.~Zhao,
  \enquote{{Complexity, action, and black holes}},
  \href{http://dx.doi.org/10.1103/PhysRevD.93.086006}{\emph{Phys. Rev. D}
  \textbf{93[8]} (2016) 086006}, \href{http://arxiv.org/abs/1512.04993}{{\tt
  arXiv:1512.04993 [hep-th]}}.

\bibitem{Brown:2015bva}
A.~R. Brown, D.~A. Roberts, L.~Susskind, B.~Swingle and Y.~Zhao,
  \enquote{{Holographic Complexity Equals Bulk Action?}},
  \href{http://dx.doi.org/10.1103/PhysRevLett.116.191301}{\emph{Phys. Rev.
  Lett.} \textbf{116[19]} (2016) 191301},
  \href{http://arxiv.org/abs/1509.07876}{{\tt arXiv:1509.07876 [hep-th]}}.

\bibitem{Cai:2016xho}
R.-G. Cai, S.-M. Ruan, S.-J. Wang, R.-Q. Yang and R.-H. Peng, \enquote{{Action
  growth for AdS black holes}},
  \href{http://dx.doi.org/10.1007/JHEP09(2016)161}{\emph{JHEP} \textbf{09}
  (2016) 161}, \href{http://arxiv.org/abs/1606.08307}{{\tt arXiv:1606.08307
  [gr-qc]}}.

\bibitem{Goto:2018iay}
K.~Goto, H.~Marrochio, R.~C. Myers, L.~Queimada and B.~Yoshida,
  \enquote{{Holographic Complexity Equals Which Action?}},
  \href{http://dx.doi.org/10.1007/JHEP02(2019)160}{\emph{JHEP} \textbf{02}
  (2019) 160}, \href{http://arxiv.org/abs/1901.00014}{{\tt arXiv:1901.00014
  [hep-th]}}.

\bibitem{Alishahiha:2018tep}
M.~Alishahiha, A.~Faraji~Astaneh, M.~R. Mohammadi~Mozaffar and A.~Mollabashi,
  \enquote{{Complexity Growth with Lifshitz Scaling and Hyperscaling
  Violation}}, \href{http://dx.doi.org/10.1007/JHEP07(2018)042}{\emph{JHEP}
  \textbf{07} (2018) 042}, \href{http://arxiv.org/abs/1802.06740}{{\tt
  arXiv:1802.06740 [hep-th]}}.

\bibitem{PhysRevD.100.086004}
M.~Alishahiha and A.~Faraji~Astaneh, \enquote{Complexity of hyperscaling
  violating theories at finite cutoff},
  \href{http://dx.doi.org/10.1103/PhysRevD.100.086004}{\emph{Phys. Rev. D}
  \textbf{100} (2019) 086004}.

\bibitem{Carmi:2017jqz}
D.~Carmi, S.~Chapman, H.~Marrochio, R.~C. Myers and S.~Sugishita, \enquote{{On
  the Time Dependence of Holographic Complexity}},
  \href{http://dx.doi.org/10.1007/JHEP11(2017)188}{\emph{JHEP} \textbf{11}
  (2017) 188}, \href{http://arxiv.org/abs/1709.10184}{{\tt arXiv:1709.10184
  [hep-th]}}.

\bibitem{Couch:2016exn}
J.~Couch, W.~Fischler and P.~H. Nguyen, \enquote{{Noether charge, black hole
  volume, and complexity}},
  \href{http://dx.doi.org/10.1007/JHEP03(2017)119}{\emph{JHEP} \textbf{03}
  (2017) 119}, \href{http://arxiv.org/abs/1610.02038}{{\tt arXiv:1610.02038
  [hep-th]}}.

\bibitem{Alishahiha:2015rta}
M.~Alishahiha, \enquote{{Holographic Complexity}},
  \href{http://dx.doi.org/10.1103/PhysRevD.92.126009}{\emph{Phys. Rev. D}
  \textbf{92[12]} (2015) 126009}, \href{http://arxiv.org/abs/1509.06614}{{\tt
  arXiv:1509.06614 [hep-th]}}.

\bibitem{Emparan:2021hyr}
R.~Emparan, A.~M. Frassino, M.~Sasieta and M.~Toma\v{s}evi\'c,
  \enquote{{Holographic complexity of quantum black holes}},
  \href{http://dx.doi.org/10.1007/JHEP02(2022)204}{\emph{JHEP} \textbf{02}
  (2022) 204}, \href{http://arxiv.org/abs/2112.04860}{{\tt arXiv:2112.04860
  [hep-th]}}.

\bibitem{Bhattacharyya:2022ren}
A.~Bhattacharyya, G.~Katoch and S.~R. Roy, \enquote{{Complexity of warped
  conformal field theory}}, \href{http://arxiv.org/abs/2202.09350}{{\tt
  arXiv:2202.09350 [hep-th]}}.

\bibitem{Alishahiha:2021thv}
M.~Alishahiha, S.~Banerjee, J.~Kames-King and E.~Loos, \enquote{{Complexity as
  a holographic probe of strong cosmic censorship}},
  \href{http://dx.doi.org/10.1103/PhysRevD.105.026001}{\emph{Phys. Rev. D}
  \textbf{105[2]} (2022) 026001}, \href{http://arxiv.org/abs/2106.14578}{{\tt
  arXiv:2106.14578 [hep-th]}}.

\bibitem{Alishahiha:2019cib}
M.~Alishahiha, K.~Babaei~Velni and M.~Reza~Tanhayi, \enquote{{Complexity and
  near extremal charged black branes}},
  \href{http://dx.doi.org/10.1016/j.aop.2021.168398}{\emph{Annals Phys.}
  \textbf{425} (2021) 168398}, \href{http://arxiv.org/abs/1901.00689}{{\tt
  arXiv:1901.00689 [hep-th]}}.

\bibitem{Engelhardt:2021mju}
N.~Engelhardt and r.~Folkestad, \enquote{{General bounds on holographic
  complexity}}, \href{http://dx.doi.org/10.1007/JHEP01(2022)040}{\emph{JHEP}
  \textbf{01} (2022) 040}, \href{http://arxiv.org/abs/2109.06883}{{\tt
  arXiv:2109.06883 [hep-th]}}.

\bibitem{Mounim:2021bba}
A.~Mounim and W.~M\"uck, \enquote{{Reparameterization dependence is useful for
  holographic complexity}},
  \href{http://dx.doi.org/10.1007/JHEP07(2021)010}{\emph{JHEP} \textbf{07}
  (2021) 010}, \href{http://arxiv.org/abs/2101.10909}{{\tt arXiv:2101.10909
  [hep-th]}}.

\bibitem{Ghanbarian:2020cdv}
N.~Ghanbarian and M.~R. Tanhayi, \enquote{{\textquoteleft{}Mutual
  complexity\textquoteright{} in hyperscaling violating background}},
  \href{http://dx.doi.org/10.1142/S0218271821500139}{\emph{Int. J. Mod. Phys.
  D} \textbf{30[02]} (2021) 2150013}.

\bibitem{Borvayeh:2020yip}
Z.~Borvayeh, M.~R. Tanhayi and S.~Rafibakhsh, \enquote{{Holographic Complexity
  of Subregions in the Hyperscaling Violating Theories}},
  \href{http://dx.doi.org/10.1142/S0217732320501916}{\emph{Mod. Phys. Lett. A}
  \textbf{35[23]} (2020) 2050191}, \href{http://arxiv.org/abs/2006.08478}{{\tt
  arXiv:2006.08478 [hep-th]}}.

\bibitem{HosseiniMansoori:2017tsm}
S.~A. Hosseini~Mansoori and M.~M. Qaemmaqami, \enquote{{Complexity growth,
  butterfly velocity and black hole thermodynamics}},
  \href{http://dx.doi.org/10.1016/j.aop.2020.168244}{\emph{Annals Phys.}
  \textbf{419} (2020) 168244}, \href{http://arxiv.org/abs/1711.09749}{{\tt
  arXiv:1711.09749 [hep-th]}}.

\bibitem{Akhavan:2019zax}
A.~Akhavan and F.~Omidi, \enquote{{On the Role of Counterterms in Holographic
  Complexity}}, \href{http://dx.doi.org/10.1007/JHEP11(2019)054}{\emph{JHEP}
  \textbf{11} (2019) 054}, \href{http://arxiv.org/abs/1906.09561}{{\tt
  arXiv:1906.09561 [hep-th]}}.

\bibitem{Omidi:2020oit}
F.~Omidi, \enquote{{Regularizations of Action-Complexity for a Pure BTZ Black
  Hole Microstate}},
  \href{http://dx.doi.org/10.1007/JHEP07(2020)020}{\emph{JHEP} \textbf{07}
  (2020) 020}, \href{http://arxiv.org/abs/2004.11628}{{\tt arXiv:2004.11628
  [hep-th]}}.

\bibitem{Hashemi:2019aop}
S.~S. Hashemi, G.~Jafari and A.~Naseh, \enquote{{First law of holographic
  complexity}},
  \href{http://dx.doi.org/10.1103/PhysRevD.102.106008}{\emph{Phys. Rev. D}
  \textbf{102[10]} (2020) 106008}, \href{http://arxiv.org/abs/1912.10436}{{\tt
  arXiv:1912.10436 [hep-th]}}.

\bibitem{Doroudiani:2019llj}
M.~Doroudiani, A.~Naseh and R.~Pirmoradian, \enquote{{Complexity for Charged
  Thermofield Double States}},
  \href{http://dx.doi.org/10.1007/JHEP01(2020)120}{\emph{JHEP} \textbf{01}
  (2020) 120}, \href{http://arxiv.org/abs/1910.08806}{{\tt arXiv:1910.08806
  [hep-th]}}.

\bibitem{Akhavan:2018wla}
A.~Akhavan, M.~Alishahiha, A.~Naseh and H.~Zolfi, \enquote{{Complexity and
  Behind the Horizon Cut Off}},
  \href{http://dx.doi.org/10.1007/JHEP12(2018)090}{\emph{JHEP} \textbf{12}
  (2018) 090}, \href{http://arxiv.org/abs/1810.12015}{{\tt arXiv:1810.12015
  [hep-th]}}.

\bibitem{Alishahiha:2017hwg}
M.~Alishahiha, A.~Faraji~Astaneh, A.~Naseh and M.~H. Vahidinia, \enquote{{On
  complexity for F(R) and critical gravity}},
  \href{http://dx.doi.org/10.1007/JHEP05(2017)009}{\emph{JHEP} \textbf{05}
  (2017) 009}, \href{http://arxiv.org/abs/1702.06796}{{\tt arXiv:1702.06796
  [hep-th]}}.

\bibitem{Karar:2019bwy}
S.~Karar, R.~Mishra and S.~Gangopadhyay, \enquote{{Holographic complexity of
  boosted black brane and Fisher information}},
  \href{http://dx.doi.org/10.1103/PhysRevD.100.026006}{\emph{Phys. Rev. D}
  \textbf{100[2]} (2019) 026006}, \href{http://arxiv.org/abs/1904.13090}{{\tt
  arXiv:1904.13090 [hep-th]}}.

\bibitem{Ghodrati:2019hnn}
M.~Ghodrati, X.-M. Kuang, B.~Wang, C.-Y. Zhang and Y.-T. Zhou, \enquote{{The
  connection between holographic entanglement and complexity of purification}},
  \href{http://dx.doi.org/10.1007/JHEP09(2019)009}{\emph{JHEP} \textbf{09}
  (2019) 009}, \href{http://arxiv.org/abs/1902.02475}{{\tt arXiv:1902.02475
  [hep-th]}}.

\end{thebibliography}

\end{document}